\documentclass[12pt]{article}
\usepackage[colorlinks=true, linkcolor=black, citecolor=blue, urlcolor=black, pdfpagemode={UseNone}]{hyperref} 
\usepackage{graphicx}
\usepackage[round,authoryear]{natbib}
\usepackage{multirow}
\usepackage{dcolumn,booktabs}
\usepackage{algorithm,algorithmicx,algpseudocode,algcompatible}
\usepackage{authblk} 
\usepackage{url} 
\usepackage{enumerate}
\usepackage[table]{xcolor}
\usepackage{xspace}
\usepackage[font=small]{caption} 
\usepackage{amsmath,amssymb,bm,amsthm,mathtools} 
\usepackage{placeins} 
\usepackage{silence} 
\usepackage{anyfontsize} 

\graphicspath{{figures}}

\DeclareMathOperator{\MD}{\mbox{MD}}
\DeclareMathOperator{\MSE}{\mbox{MSE}}
\DeclareMathOperator{\E}{\mathbb{E}}

\DeclareMathOperator*{\argmin}{\mbox{argmin}}

\DeclareMathOperator*{\diag}{diag}

\newcommand{\p}{p} 
\newcommand{\m}{{\mbox{\scriptsize{con}}}}
\newcommand{\Sym}{{\mbox{Sym}}}
\newcommand{\penalt}{q} 

\newcommand{\numd}{k} 
\newcommand{\eig}{\lambda} 

\newcommand{\new}{\mbox{\footnotesize new}}

\newcommand{\AN}{\mbox{\footnotesize A09}}
\newcommand{\ALYZ}{\mbox{\footnotesize ALYZ}}

\newcommand{\eps}{\varepsilon}

\newcommand{\bzero}{\boldsymbol{0}}
\newcommand{\bone}{\boldsymbol{1}}
\newcommand{\bcdot}{\boldsymbol{\cdot}}

\newcommand{\bx}{\boldsymbol{x}}
\newcommand{\by}{\boldsymbol{y}}
\newcommand{\bz}{\boldsymbol{z}}

\newcommand{\bW}{\boldsymbol{W}}
\newcommand{\bZ}{\boldsymbol{Z}}
\newcommand{\bX}{\boldsymbol{X}}
\newcommand{\bbeta}{\boldsymbol{\beta}}
\newcommand{\bgamma}{\boldsymbol{\gamma}}

\newcommand{\bmu}{\boldsymbol{\mu}}
\newcommand{\bSigma}{\boldsymbol{\Sigma}}

\newcommand{\btop}{\boldsymbol{\top}}

\newcommand{\hx}{\widehat{x}}
\newcommand{\hy}{\widehat{y}}
\newcommand{\hbeta}{\boldsymbol{\widehat{\beta}}}
\newcommand{\hmu}{\widehat{\mu}}

\newcommand{\hsigma}{\widehat{\sigma}}
\newcommand{\halpha}{\widehat{\alpha}}

\newcommand{\tth}{\widetilde{h}}
\newcommand{\tn}{\widetilde{n}}

\newcommand{\ty}{\widetilde{y}}
\newcommand{\tE}{\widetilde{E}}
\newcommand{\tH}{\widetilde{H}}

\newcommand{\tY}{\widetilde{Y}}
\newcommand{\ttH}{\widetilde{H}}

\newcommand{\bhmu}{\boldsymbol{\widehat{\mu}}}
\newcommand{\bhSigma}{\boldsymbol{\widehat{\Sigma}}}
\newcommand{\bhbeta}{\boldsymbol{\widehat{\beta}}}
\newcommand{\bhgamma}{\boldsymbol{\widehat{\gamma}}}

\newcommand{\btx}{\boldsymbol{\widetilde{x}}}
\newcommand{\bty}{\boldsymbol{\widetilde{y}}}

\newcommand{\btX}{\boldsymbol{\widetilde{X}}}

\newcommand{\Ximp}{\boldsymbol{\widehat{X}}}
\newcommand{\ximp}{\boldsymbol{\widehat{x}}}

\begin{document}

\def\spacingset#1{\renewcommand{\baselinestretch}
{#1}\small\normalsize} \spacingset{1}


\title{\bf Least Trimmed Squares Regression with\\
   \vspace{2mm} Missing Values and Cellwise Outliers\\
   \vspace{5mm}}

\author[1]{Jakob Raymaekers}

\author[2]{Peter J. Rousseeuw}

\affil[1]{Department of Mathematics, 
  University of Antwerp, Belgium,\linebreak
  jakob.raymaekers@uantwerpen.be}

\affil[2]{Section of Statistics and Data Science,
          KU Leuven, Belgium,\linebreak
          peter@rousseeuw.net}
\setcounter{Maxaffil}{0}
\renewcommand\Affilfont{\itshape\small}

\date{August 3, 2026} 
\maketitle

\bigskip
\begin{abstract}
Regression is the workhorse of statistics, and is 
often faced with real data that contain outliers.
When these are casewise outliers, that is, cases
that are entirely wrong or belong to a different
population, the issue can be remedied by existing
casewise robust regression methods. It is another
matter when cellwise outliers occur, that is,
suspicious individual entries in the data matrix 
containing the regressors and the response.
We propose a new regression method that is robust
to both casewise and cellwise outliers, and 
handles missing values as well. Its construction
allows for skewed distributions. We show that it
obeys the first breakdown result for cellwise
robust regression. It is also the first such
method that is geared to making robust 
out-of-sample predictions. Its performance is 
studied by simulation, and it is illustrated on 
a substantial real dataset.
\end{abstract}

\vspace{5mm}

\newpage
\spacingset{1.5}

\section{Introduction} \label{sec:intro}

Real data often contains outliers. These outlying 
data deviate from the structure suggested by the 
majority of the data, thereby raising concern 
that they may have been generated by a different 
mechanism. Whether these outlying data are errors 
or informative pieces of information, it is 
essential to detect them.

Our preferred approach for detecting outliers is 
that of robust statistics. This approach first 
fits a model to the majority of the data, and 
then finds outliers by measuring their distance 
from the robust fit. 
Many proposals for robust regression have been 
made, such as the least trimmed squares (LTS) 
method \citep{rousseeuw1984least}. LTS minimizes 
the sum of squared residuals on a subset of the 
cases, rather than on all of them as in OLS. 
By focusing on a subset, LTS finds the central 
part of the data that fits most tightly. 
Suspicious points are then detected by looking
at the residuals with respect to this fit. 
For a book-length treatment see \cite{RL1987}.

What all these methods have in common, is that
they consider casewise outliers. This assumes 
that a case is either entirely clean, or 
entirely suspicious. More recently, a different 
paradigm has gained traction in the development of 
robust statistics. Instead of assuming outlying 
cases, it assumes instead that some cells in the 
data matrix (containing the predictors and the
response) have been replaced by suspicious 
values. This cellwise paradigm was formalized 
by \cite{alqallaf2009}. The challenges this
poses are reviewed by \cite{challenges}, with a 
summary of statistical settings in which 
progress has been made, such as cellwise robust 
estimation of covariance matrices.

Cellwise robust regression methods are 
substantially less established. The first proposal
was by \cite{leung2016robust}. They proposed to 
first estimate a cellwise robust covariance 
matrix on the predictors and response combined. 
Next, they inverted the submatrix corresponding 
with the predictors and multiplied it with the 
column corresponding with the covariance between 
predictors and response. This results in 
an estimate of the regression coefficients that 
performs quite well, but depends strongly on the 
assumption that the uncontaminated data are close 
to Gaussian. A second proposal is the Shooting S 
estimator by \cite{ollerer2016shooting}. It 
adopts the coordinate descent algorithm for linear 
regression and incorporates robustness in each 
of the fits. By working per coordinate, a 
sparsity penalty can be integrated 
\citep{bottmer2022sparse}. 
Finally, a recent proposal is the method 
of \cite{CRlasso}. It optimizes an 
elegant objective function inspired by mean-shift 
penalized objectives for casewise robust 
regression \citep{she2011outlier}.

We propose a new method called 
{\it cellwise least trimmed squares} 
(cellLTS) that generalizes the casewise 
LTS regression estimator. It is robust 
to both cellwise and casewise outliers, 
and handles missing values in a natural 
way. It can also deal with skewed data, by
incorporating symmetrization in its 
construction. Moreover, it obeys the first 
cellwise breakdown result for regression.

\begin{figure}[!ht]
\centering
\includegraphics[width = 0.95\columnwidth]
  {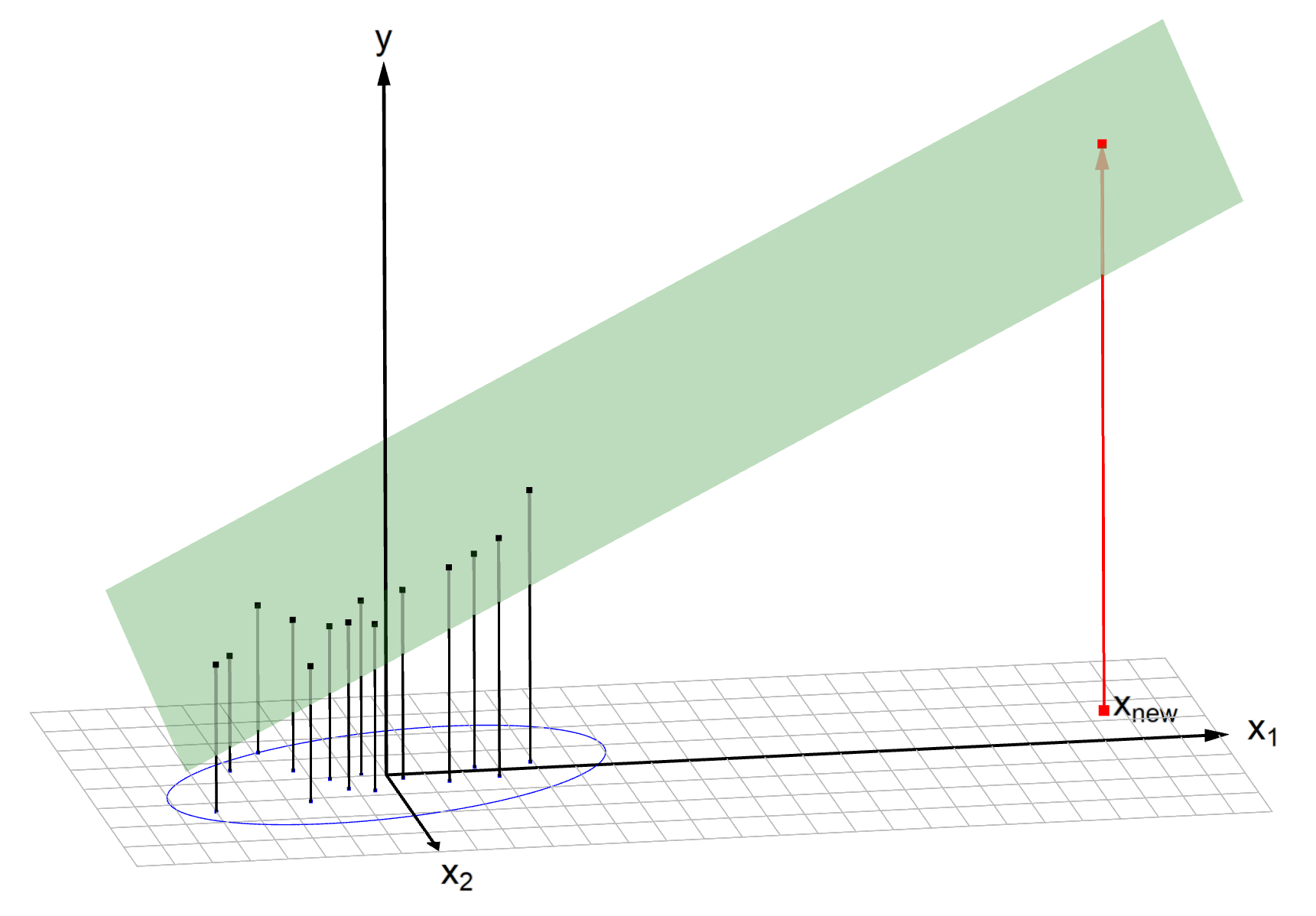}\\
\caption{A toy example to illustrate the 
issue of out-of-sample prediction in 
contaminated data.}
\label{fig:toy}
\end{figure}

Let us look at the small toy example in 
Figure~\ref{fig:toy}, with two regressors 
$x_1$ and $x_2$ and the response $y$. The 
black points are the training data, that
nicely follow a linear model, and their
$(x_1,x_2)$ all lie inside the horizontal 
tolerance ellipse and are well-behaved. 
So here it is easy to fit a regression, 
shown as the light green plane.

But what about prediction? 
In-sample prediction is easy, as it is a 
byproduct of estimation. But out-of-sample 
prediction on new data is another matter, 
because the new data may be 
contaminated itself. This happens with the
new regressor point $\bx_{\new}$ (in red)
at the right hand side of the $(x_1,x_2)$ 
horizontal plane, that has a huge 
cellwise outlier in its first coordinate. 
How can we compute a prediction $\hy$ in 
this $\bx_{\new}$? Just plugging in the 
regression coefficients would follow the 
vertical red arrow upward until it meets
the regression plane, yielding a very 
high $\hy$ that is outside the range of 
the training data. The existing cellwise 
robust regression methods do not address 
this issue, but we will do so in 
what follows. We will come back to this 
example.

The remainder of the paper is organized as 
follows. Section~\ref{sec:prelim} describes 
preliminaries including symmetrization and a 
recent cellwise robust covariance estimator. 
Section~\ref{sec:methodology} introduces the 
proposed two-step regression method, proves 
its good breakdown behavior, and addresses
out-of-sample prediction.
Section~\ref{sec:algo} discusses the 
optimization of the regression objective 
and faster symmetrization. 
Section~\ref{sec:simulation} contains an
empirical study, followed by a real data 
application in Section~\ref{sec:realdata}. 
Section~\ref{sec:conclusion} concludes.

\section{Preliminaries} \label{sec:prelim}
In this section we introduce the setup and some 
preliminary tools needed to construct our new 
method. 

\subsection{Data symmetrization}
\label{sec:symmetrization}

We assume that the uncontaminated data is 
generated by the linear model
\begin{equation}\label{eq:linmodel}
  Y = \alpha + \bbeta^{\top} \bX  + E
\end{equation}
where the random variable $Y$ is the response, 
$\alpha$ is the unknown intercept, $\bbeta$ 
is the vector of unknown slope parameters, 
$\bX$ is the random vector of the regressors with 
location $\bmu_{\bX}$ and scatter matrix 
$\bSigma_{\bX}$, and $E$ is an i.i.d. error term.

For reasons that will be explained later, 
it may be useful to consider a 
transformation of the regression 
model~\eqref{eq:linmodel}. We can consider 
an independent copy $(\bX',E',Y')$ of
$(\bX,E,Y)$, from which it follows that
\begin{equation*}
  Y-Y' = \bbeta^{\top} (\bX-\bX') + E-E'\;.
\end{equation*}
If we denote the variables $\tY := Y-Y'$\,,
$\btX := \bX - \bX'$\, and $\tE := E-E'$ 
we thus obtain
\begin{equation}
  \tY = \bbeta^{\top} \btX + \tE
\end{equation}
in which all variables are symmetric. Note that
the slope parameter vector remains the same, and
that the intercept  $\alpha$ vanishes (but it 
can easily be estimated afterward). 
Because the symmetrized variables $\tY$, $\btX$ 
and $\tE$ are defined as convolutions, they tend 
to be a bit closer to Gaussian than the raw 
variables.

If the second moments exist, we can see 
that the transformed covariates $\btX$ now 
have the property that $\E[\btX] = \bzero$ 
and $\E[\btX\,\btX^\top] = 2 \E[\bX \bX^\top]$. 
In other words, the covariance matrix of 
the symmetrized data is twice the 
covariance matrix of the original $\bX$. 
On the other hand, the information of the 
location of $\bX$ is lost in the 
symmetrization.

The above reasoning is for populations. 
When we have a finite univariate sample 
$\by = \{y_1,\ldots,y_n\}$ of $Y$ we 
construct the symmetrized dataset 
$\Sym(\by) = \{y_{\ell} -y_i\,;\,
i\neq \ell\}$
that contains $n(n-1)$ entries. 
It is well-known that the empirical 
distribution of $\Sym(\by)$ is a good
approximation of the distribution of
$Y - Y'$. We also symmetrize 
$\{\bx_1,\ldots,\bx_n\}$ to
$\{\bx_{\ell} - \bx_i\,;\, i \neq \ell\}$.

\subsection{The cellwise robust MCD method}
\label{sec:cellMCD}
The proposed regression estimator will make use 
of the cellwise minimum covariance determinant 
(cellMCD) estimator by \cite{cellMCD}. The 
cellMCD method estimates the location and the 
covariance matrix of a multivariate dataset, and
flags outlying cells. 
It works as follows.

Suppose we observe $n$ cases 
$\bx_1, \ldots, \bx_n$ from a $\p$-variate 
population with unknown center $\bmu_{\bX}$ and 
covariance matrix $\bSigma_{\bX}$. Denote by 
$\bW$ an unknown $n \times \p$ binary matrix 
in which $\bW_{ij} = 1$ indicates the clean 
cells, and $\bW_{ij} = 0$ the contaminated 
cells. The cellMCD method then estimates 
$\bmu_{\bX}$, $\bSigma_{\bX}$, and $\bW$ by 
maximizing the observed likelihood of the 
clean cells, under a penalty ensuring $\bW$ 
does not contain too many zeros. 
More specifically, cellMCD minimizes the 
objective 
\begin{equation} \label{eq:cellMCD}
 \sum_{i=1}^{n} {\Big(
 \ln |\bSigma^{(\bW_{i\bcdot})}| + 
 d(\bW_{i\bcdot})\ln(2\pi) +
 \MD^2(\bx_i,\bW_{i\bcdot}\,,\bmu,\bSigma)
 \,\Big)} + \sum_{j=1}^\p \penalt_j
 ||\bone_\p - \bW_{\bcdot j}||_0
\end{equation}
with respect to $(\bmu, \bSigma, \bW)$, 
under the constraints 
$\eig_\p(\bSigma) \geqslant a$ and 
$||\bW_{\bcdot j}||_0 \geqslant h$
for all $j=1,\ldots,\p$.
The first sum of this objective function is 
minus twice the log likelihood of the clean 
cells, in which the row $\bW_{i\bcdot}$ 
indicates that the log-determinant, the 
dimension $d(\bW_{i\bcdot})$ and the 
Mahalanobis distance are 
calculated only on the variables $j$ with
$\bW_{ij} = 1$. The second sum ensures that 
not too many cells are flagged as outlying. 
Indeed, the notation 
$||\bone_\p - \bW_{\bcdot j}||_0$ stands 
for the number of nonzero entries in the 
vector $\bone_\p - \bW_{\bcdot j}$ which is 
the number of zero weights in column $j$ of 
$\bW$, so the number of flagged cells in 
column $j$ of $\bX$. The constants 
$\penalt_j$ for $j=1,\ldots,\p$ are computed 
from the desired percentage (by default 1\%) 
of flagged cells in the absence of 
contamination. At the same time we keep the 
robustness constraints that 
$||\bW_{\bcdot j}||_0 \geqslant h \approx 0.75n$
and $\eig_\p(\bSigma) \geqslant a> 0$ which
lower bounds the smallest eigenvalue of the 
covariance matrix, with default
$a = 10^{-4}$.

The cellMCD estimator minimizes the objective 
of~\eqref{eq:cellMCD} through block-coordinate 
descent. More precisely, we iterate between 
updating $(\bmu, \bSigma)$ and $\bW$ while 
keeping the other quantity fixed. Updating 
$\bW$ for fixed $(\bmu, \bSigma)$ boils down 
to a thresholding rule based on the 
contribution of each cell to the 
objective~\eqref{eq:cellMCD}. Updating 
$(\bmu,\bSigma)$ for 
fixed $\bW$ works by optimizing the observed 
likelihood, as is done in missing value 
problems. It has been proved that each step
of the algorithm lowers the objective, so
the algorithm always converges. The cellMCD
was applied to cluster analysis
\citep{zaccaria2025}
and to outlier explanation 
\citep{mayrhofer2026}.

\subsection{Least trimmed squares regression}
\label{sec:LTS}

In addition the the cellMCD method, we will 
also apply least trimmed squares (LTS)
regression \citep{rousseeuw1984least}. 
Suppose we want to fit a regression model
\begin{equation} \label{eq:linmod}
  y_i = \alpha + \bx_i^\top \bbeta  + 
  \mbox{noise}_i
\end{equation}
to $n$ data points 
$(\bx_1\,,y_1), \ldots, (\bx_n\,,y_n)$
where the $\bx_i$ are $d$-variate.
The classical least squares (OLS) estimator 
estimates $\alpha$ and $\bhbeta$ by minimizing 
the sum of all squared residuals $r_i^2$\,, 
where $r_i=y_i-\halpha - \bhbeta^{\,\top}\bx_i$\,. 
OLS is not robust because the effect of a case
$(\bx_i\,,y_i)$ on the objective is unbounded. 
LTS regression instead minimizes a trimmed 
sum of squared residuals
\begin{equation*}
 (\halpha, \bhbeta)_{\mbox{\tiny LTS}} =
 \argmin_{\alpha, \bbeta} 
 \sum_{i=1}^{h} r_{(i)}^2
\end{equation*}
where $r_{(1)}^2 \leqslant r_{(2)}^2 \leqslant
\ldots \leqslant r_{(n)}^2$ denote the ordered 
squared residuals. A typical choice of $h$
is $0.75n$, so that the LTS fit can withstand 
up to 25\% of outlying cases. Also this 
algorithm works by alternating between 
updating the $h$-subset and the regression 
coefficients.

\section{Methodology}\label{sec:methodology}

\subsection{Estimating the regression coefficients}
\label{sec:estimation}

The proposed regression estimator consists 
of two steps. 
In the first step we clean the data matrix of 
the regressors, by imputing missing values as
well as outlying cells. This is done without
using information about the response variable.
In the second step we regress the response on 
this cleaned data matrix, to obtain estimates
of $\alpha$ and $\bbeta$. This regression is
carried out by the robust LTS method, to 
mitigate the effect of potentially outlying 
response values. The resulting cellwise robust 
version of LTS we call cellLTS.

We rewrite the regression 
model~\eqref{eq:linmod} in matrix form as
\begin{equation} \label{eq:linmodmat}
 \by = \alpha + \bX \bbeta + \mbox{noise}
\end{equation}
where the $n \times 1$ column vector 
$\by$ contains the response, 
and the $n \times \p$ matrix $\bX$
has the regressors. The $\p \times 1$
column vector $\bbeta$ contains the 
regression slopes, and $\alpha$ is the 
intercept. In this formulation the 
matrix $\bX$ does not contain a
column of 1's because the intercept is
a separate term.
We proceed by describing the steps of
the proposed cellLTS method. 

\vspace{3mm}
\noindent \textbf{Step 1.} Cleaning the
data matrix of regressors.

\begin{enumerate}[(a)]
\item Run the cellMCD algorithm described
in Section~\ref{sec:cellMCD} to obtain a 
cellwise robust estimate $\bhSigma_X$ of the 
scatter matrix of the predictors. Instead of 
running cellMCD on $\bX$, however, we apply 
it to the symmetrized data $\Sym(\bX)$ 
obtained by symmetrizing each of its columns, 
and afterward we divide the resulting scatter 
matrix by 2. For this estimation we run 
cellMCD with $\tth := h(h-1)$ and center fixed 
at the origin, since we know that $\Sym(\bX)$ 
is symmetric about $\bzero$. 
Interestingly, the choice $h \approx n/2$
yields $\tth \approx \tn/4$. This is in line
with the use of first quartile of the 
symmetrized data 
in the $Q_n$ estimator \citep{Qn}.
An advantage of symmetrization is that
it typically brings the distribution closer 
to Gaussianity and increases statistical 
efficiency (albeit at a computational 
cost, that we will address in 
Section~\ref{sec:fastsymm}).

\item Obtain $\bhmu_X$ by applying a robust 
  univariate location estimate to each
  column of $\bX$.
\item Obtain $\bW_{\bX}$. We first compute 
the matrix
$\bZ \coloneqq (\diag(\bhSigma_X))^{-1/2}
(\bX - \bhmu_X \bone_n)$. Its cells can
be seen as robust $z$-scores of the cells 
in $\bX$. Next we initialize $\bW_{\bX}$
by setting $(\bW^{(0)}_{\bX})_{ij}$ to 
zero whenever $|z_{ij}| >
\sqrt{\chi^2_{1,0.99}} \approx 2.57$\,,
and to 1 otherwise. Starting from 
$\bW^{(0)}_{\bX}$ we obtain the final 
$\bW_{\bX}$ by iterations that optimize 
the cellMCD objective with respect to 
$\bW_{\bX}$\,, while keeping $\bhmu_{\bX}$ 
and $\bhSigma_{\bX}$ fixed.
\item Obtain $\Ximp$ by imputing the cells
with $(\bW_{\bX})_{ij} = 0$ with their 
best linear prediction according to 
$(\bhmu_X$, $\bhSigma_X)$. More precisely,
\begin{equation}\label{eq:Ximp}
  \hx_{ij} = \begin{cases}
  x_{ij} \qquad \mbox{ if } \;\;
  \bW_{ij} = 1\\
  (\hmu_X)_j + (\bhSigma_X)_{j,(\bW_{i\bcdot})}
  \left(\bhSigma_X^{(\bW_{i\bcdot})}\right)^{-1}
  (\bx_i^{(\bW_{i\bcdot})} - 
  \bhmu_X^{(\bW_{i\bcdot})})\;\; 
  \mbox{ if } \;\; \bW_{ij} = 0\;.\\
  \end{cases}
\end{equation}
  This minimizes the $i$-th term of the
  negative log likelihood in~\eqref{eq:cellMCD} 
  for given \mbox{$(\bhmu_X$, $\bhSigma_X)$,} 
  as in the E-step of the EM algorithm. 

\end{enumerate}

\noindent Some remarks on this approach are in order:
\begin{itemize}
\item We could refine the robust location estimate of 
step (b) by using it as a starting value for cellMCD 
iteration steps with fixed $\bhSigma_X$. This requires
more computation, but in our experiments it did not 
improve the final estimator.
\item The justification for step (c)
is twofold. One, we need an imputation for $\bX$ 
and not $\Sym(\bX)$, so we cannot use the output
of step (a) directly. Two, we want to flag outliers 
in-sample and out-of-sample in a uniform way, so 
we use a start that is the same in both situations.
\end{itemize}

\noindent \textbf{Step 2.} To obtain robust estimates 
of the regression coefficients $\alpha$ and $\bbeta$, 
we do the following.
\begin{enumerate}[(a)]
\setcounter{enumi}{4}
\item We symmetrize the response variable $\by$
  to $\Sym(\by)$, which helps when the response 
  is skewed. We then standardize $\Sym(\by)$ by 
  dividing it by the scale estimate of its 
  univariate MCD with the same $\tth = h(h-1)$
  as before, and denote the result as $\bty$.\\
  We also symmetrize the imputed 
  $\Ximp$ of \eqref{eq:Ximp} to $\Sym(\Ximp)$.
  We do not include a column of 1's because the 
  symmetrized model has no intercept.
  We standardize $\Sym(\Ximp)$ by dividing
  each column $j$ by 
  $\sqrt{(\bhSigma_{\bX})_{jj}}$\,, and denote
  the resulting regressor matrix as $\btX$.

\item Both $\btX$ and $\bty$ have 
$\tn := n(n-1)$ 
rows, and we will consider $\tth$-subsets 
$\ttH$ of $\{1,\ldots,\tn\}$ with 
$\tth = h(h-1)$ elements.
The cellLTS estimator of $\bbeta$ is defined as
\begin{equation}\label{eq:ridgeLTS}
  \bhbeta = \argmin_{\bbeta(\ttH)} \Big(
  \sum_{\ell \in \ttH} r^2(\ty_\ell, 
  \btx_{\ell}, \bbeta) +
  \lambda ||\bbeta||_2^2 \Big)
\end{equation}
with residuals 
$r(\ty_{\ell},\btx_{\ell},\bbeta)
:= \ty_{\ell} - \btx_{\ell}^\top\bbeta$.
The second term is a ridge penalty with
a small $\lambda$, to avoid collinearity
problems. Penalizing $||\bbeta||_2$ 
makes sense because the components of
$\bbeta$ are in similar units. This 
is due to the scaling of the variables
in step (e), which made them unitless and 
gave them similar scales. This also makes
it possible to set a default value for
$\lambda$, like $10^{-4}$.
\item The intercept of the regression of
the unsymmetrized $\by$ on $\Ximp$ 
is estimated as follows. First
compute the pseudo residuals 
$\bm{r}^* = \by - \Ximp\,\bhbeta$.
Then apply the univariate robust MCD
to them, and set the intercept $\halpha$
equal to its location estimate.
\item Transform the regression coefficients
to undo the standardization in step (e).
\end{enumerate}

\subsection{Computing out-of-sample predictions}
\label{sec:OOS}
In section~\ref{sec:estimation} we have 
described the two-step procedure to obtain 
robust regression coefficients. We now discuss 
how to obtain in-sample and out-of-sample 
predictions.

The in-sample predictions are straightforward 
once the algorithm has run. For the $i$-th
observation, the predicted $\hy_i$ is given by 
$\hy_i = \halpha + \ximp_i \bhbeta$ where the
imputed $\ximp_i$ was computed earlier in 
step~(d).

Now suppose we want to obtain the out-of-sample 
prediction at a new $\bx$. Note that this new
$\bx$ may itself contain missing values and/or
outlying cells! Therefore we cannot just 
use $\halpha + \bx^\top \bhbeta$, as we saw
in Figure~\ref{fig:toy}.
Instead we first need to compute the 
corresponding cleaned case $\ximp$. 
We begin by flagging cells of $\bx$
by optimizing the cellMCD objective 
of~\eqref{eq:cellMCD} in which we keep the 
estimates $(\bhmu_{\bX}, \bhSigma_{\bX})$ 
fixed. We initiate this iterative 
optimization by applying the marginal 
detection rule to 
$\bz \coloneqq (\diag(\bhSigma_X))^{-1/2}
(\bx - \bhmu_X)$. Then we 
obtain $\ximp$ by applying the imputation 
rule~\eqref{eq:Ximp} to $\bx$ instead of
$\bx_i$\,. Finally, the out-of-sample 
prediction is given by 
$\hy = \halpha + \ximp^\top \bhbeta$. The 
ability to compute reasonable predictions 
for partly corrupted out-of-sample inputs 
is new, and constitutes a major advantage 
of the cellLTS approach.

Note that, because the imputation in a
new point $\bx$ starts from the same marginal 
detection rule that we used for $\bW_{\bX}$, 
we treat in-sample prediction and out-of-sample
prediction consistently. In other words, if we
input one of the in-sample $\bx_i$ as a new 
$\bx$, the prediction will be the same as
before.

\begin{figure}[!ht]
\centering
\includegraphics[width = 0.90\columnwidth]
  {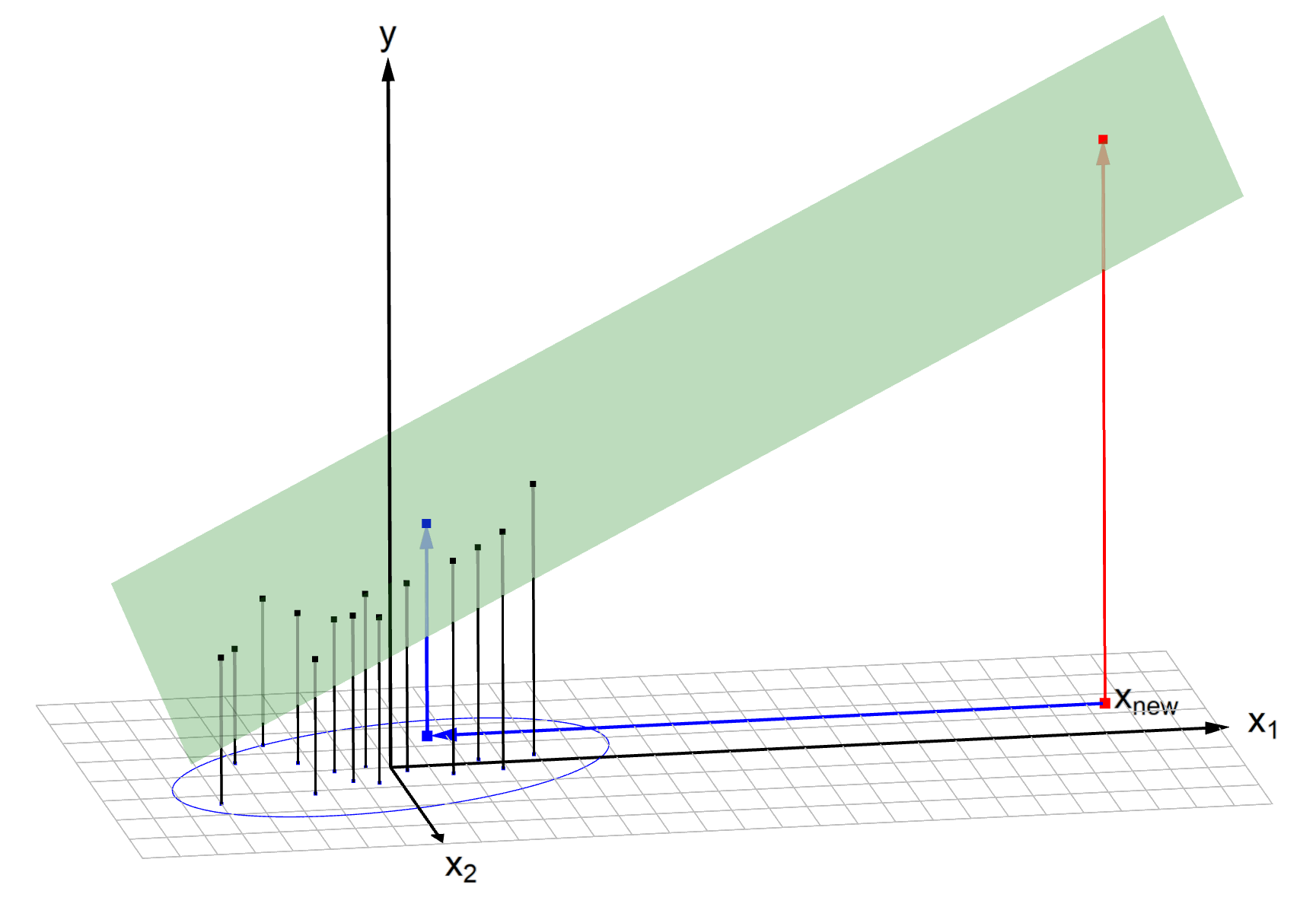}\\
\caption{In the setting of Figure~\ref{fig:toy},
  the cellLTS out-of-sample prediction in
  $\bx_{\new}$ first cleans the cellwise outlier
  along the horizontal blue arrow, and then 
  follows the vertical blue arrow upward.}
\label{fig:toy2}
\end{figure}

\vspace{-2mm}
Let us return to the toy example in 
Figure~\ref{fig:toy} where we wanted an 
out-of-sample prediction in the contaminated
point $\bx_{\new}$\,. Merely plugging its cells
$x_1$ and $x_2$ into the regression formula
would follow the red arrow upward, yielding
a prediction far outside the range of the 
in-sample data. Instead cellLTS flags the first 
coordinate, and imputes it along the horizontal  
blue arrow in the way the EM algorithm would. 
And finally it makes a prediction from this
cleaned point by following the vertical 
blue arrow upward, yielding a prediction
in the data range.

\subsection{Breakdown} \label{sec:bdv}

For the breakdown value of the cellLTS
regression estimator we use  
model~\eqref{eq:linmodel} but write it 
a little differently as $\by = 
[\bone_n\,;\bX]\,\bgamma + \mbox{noise}$ 
where $\by = (y_1,\ldots,y_n)$ is the 
$n \times 1$ column vector of responses.
The $(\p+1) \times 1$ vector
$\bgamma = (\alpha,\beta_1,\ldots, 
\beta_\p)^{\btop}$ combines
the intercept and the slopes. 
We also combine the actual data 
in the $n \times (\p+1)$ matrix 
$\bZ = [\by\;;\bX]$ whose $i$-th row 
is $(y_i,x_{i1},\ldots,x_{i\p})$.

We now contaminate the dataset $\bZ$.
Replacing at most $m$ cells in each 
column of $\bZ$ by arbitrary values
yields a contaminated matrix
denoted as $\bZ^\m$.
We define the {\it cellwise  
finite-sample breakdown value} of the 
regression estimator $\bhgamma$ at 
the dataset $\bZ$ as 
\begin{equation} \label{eq:bdvreg}
  \varepsilon^*_n(\bhgamma, \bZ) :=
  \min \left\{\frac{m}{n}:\;
	\sup_{\bZ^\m}{\left|\left
	|\bhgamma(\bZ^\m) -
	\bhgamma(\bZ)\right|\right|} =
	\infty\right\}.
\end{equation}
That is, it is the smallest fraction
$m/n$ that can move the vector of
regression coefficients 
$\bhgamma(\bZ^\m)$ arbitrarily far 
away.

The {\it casewise} breakdown value is 
defined similarly, but instead of 
replacing at most $m$ cells in each 
column, it replaces at most $m$ entire
rows of $\bZ$.
Casewise breakdown values of regression
methods have been studied extensively
in the past, see e.g. \cite{RL1987}.
As in the casewise setting, we 
need to make a geometric assumption on 
the data in order to formally prove the
breakdown value. We say that $\bX$ is 
in general position iff no more than 
$\p(\p-1)$ of the $\bx_k - \bx_i$ 
belong to the same affine hyperplane 
in $\mathbb{R}^\p$\,. This also 
implies that no more than $\p$ points
$\bx_i$ belong to the same hyperplane.
When the data are
generated by a continuous distribution,
it is in general position with
probability 1. When the data is not
quite in general position that doesn't 
mean that cellLTS cannot be used, it 
only means that the breakdown value
stated below is no longer guaranteed.

\vspace{3mm}
\noindent \textbf{Proposition.}
{\it If $\bX$ is in general position and 
$h \geqslant [(n + \p + 1)/2]$,
then the breakdown value of cellLTS with
$\tth = h(h-1)$ is 
$\eps^*_n(\bhgamma, \bZ) = (n-h+1)/n$.}

\vspace{3mm}
This is the first time a breakdown result
was obtained for a \mbox{cellwise} robust 
regression method. 
With the choice $h = [(n + \p + 1)/2]$ 
we attain the maximal breakdown
value, which is about 50\%. However,
unless the data is massively contaminated,
the more typical choice for cellMCD and
LTS is $h = 0.75n$.

The following result shows that we also 
have robustness for casewise outliers.

\vspace{3mm}
\noindent \textbf{Corollary.} {\it The casewise 
breakdown value of cellLTS equals 
its cellwise breakdown value.}

\vspace{3mm}
Therefore, cellLTS has the same casewise
breakdown value as the casewise LTS.
Both proofs can be found in 
section~\ref{supp:proofs} of the
Supplementary Material.

\vspace{3mm}
The proposed method could  easily be 
extended by replacing the LTS in step (f) by 
another regression estimator with high casewise 
breakdown value, provided the penalty term is
preserved. This would change the algorithm but 
not the cellwise breakdown value. Likewise we 
could replace the $||\bbeta||_2^2$ in 
\eqref{eq:ridgeLTS} by $||\bbeta||_1$\;. 

\section{More about the algorithm}\label{sec:algo}

\subsection{Optimizing the regression objective
function} \label{sec:ridgeLTS}

Note that~\eqref{eq:ridgeLTS} can be rewritten 
by extending the dataset $[\bty\;;\btX]$
by adding $\p$ rows of length $\p+1$, namely
$(0,\sqrt{\lambda},0,\ldots,0)$ up to 
$(0,0,\ldots,0,\sqrt{\lambda})$. The regressor 
columns of this extra $\p \times (\p+1)$ 
matrix are the basis vectors
multiplied by $\sqrt{\lambda}$, and 
their responses are all zero.
If we now denote 
$\ttH^* = \{\tn+1,\ldots,\tn+\p\}$
we can write the optimization problem in
\eqref{eq:ridgeLTS} as
\begin{equation}\label{eq:ridgeLTS2}
  \bhbeta = \argmin_{\bbeta(\ttH)} 
	\sum_{\ell \in \ttH \cup \ttH^*} 
    r^2(\ty_\ell, \btx_\ell, \bbeta)
\end{equation}
where $\ttH$ is any $\tth$-subset of 
$\{1,\ldots,\tn\}$ and $\ttH^*$ stays fixed.
In this way the penalty term becomes part
of an LTS objective function, with the
constraint that the last $\p$ cases must 
belong to the subset. This allows us to
carry out the optimization by a  
modification of the FastLTS 
algorithm of \cite{FastLTS} 
in which the C-step is constrained to
contain the last $\p$ cases, so it is only 
allowed to replace the other $\tth$ cases.
This C-step still lowers the objective
in every step, until it converges.
Afterward we assign weights to the actual
$n$ cases just as in FastLTS, and then 
carry out a weighted least squares fit 
with the same ridge penalty as 
in~\eqref{eq:ridgeLTS}.

\subsection{Faster symmetrization}
\label{sec:fastsymm}

The symmetrization strategy laid out previously 
renders the data more symmetric, which improves 
the performance of estimators such as cellMCD. 
In principle we can compute all $n(n-1)$ pairwise 
differences and work with that large set. 
Alternatively, we can consider the smaller set 
of differences
\begin{equation}
  \tY_N = \{y_\ell - y_i\,;\, 1\leqslant i
          < \ell \leqslant n\}
\end{equation}
of size $N = n(n-1)/2$, so the set of all
differences equals $\tY_N\,\cup\,-\tY_N$\,.
We can define $\btX_N$ analogously, so the
symmetrized regression dataset becomes
$(\btX_N, \tY_N)\cup(-\btX_N,-\tY_N)$. Most 
estimators $\bhbeta$ of the slope vector of 
a regression without intercept, including 
LTS, satisfy the natural condition that 
$$\bhbeta((\bX,Y)\cup(-\bX,-Y)) =
\bhbeta(\bX,Y)$$ so that it suffices to compute
$\bhbeta(\bX_N,Y_N)$ on $N$ datapoints instead of 
$2N$. Analogously, most estimators $\bhSigma$ of a
covariance matrix with center $\bzero$, including
cellMCD, satisfy 
$$\bhSigma(\bX\cup-\bX) = \bhSigma(\bX)$$
so it suffices to compute $\bhSigma(\btX_N)$.

Still, $N$ is of the order $O(n^2)$, which creates
issues for large sample sizes $n$. It would be
better if $N$ were less than quadratic in $n$.
In the setting of casewise robust M-estimation
of covariance, \cite{dumbgen2024} studied a 
smaller set of pairwise differences. 
In our notation it is
\begin{equation}
 \tY_{n,\numd} =
 \{y_{i+j} - y_i\,;\, 1\leqslant i 
 \leqslant n\,,\, 1\leqslant j 
 \leqslant \numd\}
\end{equation}
for some $1 \leqslant k < n$. Here $y_{i+j}$ is 
interpreted in a cyclic way,
meaning that $y_{n+s} := y_s$ for
$1 \leqslant s \leqslant n$. To avoid 
systematic effects, they first randomly permute
the indices $1,\ldots,n$. A major advantage of
the set $\tY_{n,\numd}$ is that each case $i$
occurs in the same number ($2\numd$) of pairwise 
differences, so each case gets the same weight.
The set $\tY_{n,\numd}$ has $nk$ elements, so 
for fixed $\numd$ it is $O(n)$.

Instead of $\tY_{n,\numd}$ we propose to draw 
$\numd$ random permutations $\tY^j$ and each 
time generate the $n$ pairs in $\tY_{n,1}^j$\,, 
and then use the combined set of pairs
\begin{equation} \label{eq:oursymm}
\tY_{n,1}^1\cup \ldots \cup\tY_{n,1}^\numd\;.
\end{equation}
This set has the same number of pairs and
each case still has the same weight, but it
is likely to yield more stable results 
because an unfortunate permutation will only
have a small effect.
\cite{dumbgen2024} found empirically that when
replacing $\btX_N$ by $\btX_{n,\numd}$ the result 
of their estimator was sufficiently close already 
for $\numd=20$. The next section will confirm 
that this is also true for~\eqref{eq:oursymm} in
cellLTS.

\section{Simulation study}\label{sec:simulation}

As the cellwise regression setup does not 
allow for affine equivariance, and the 
performance of regression methods depends 
on the distribution of the regressors, we 
have to consider a variety of simulation 
setups.

For the predictors, we consider the pairs 
$(n,\p) = (100, 10)$ and $(400, 20)$ of 
sample size and dimension, in line with the 
simulation study of \cite{cellMCD} for
covariance estimators. For the distribution 
of the clean data we consider the multivariate 
Gaussian distribution centered at the origin 
with covariance matrix $\bSigma_{\AN}$ given 
by $(\bSigma_{\AN})_{ij} = (-0.9)^{|i-j|}$. 
We also use the random covariance matrix 
$\bSigma_{\ALYZ}$ constructed by 
\cite{Agostinelli2015}. In addition, we 
generate two setups with skewed predictors 
by first sampling independent 
centered exponential variables with rate 1, 
or standard lognormal variables. Then we 
linearly transform these data so that their
classical covariance matrix becomes 
$\bSigma_{\AN}$ or $\bSigma_{\ALYZ}$.

To generate a response, we set the true 
coefficients equal to $\bbeta = 
[\p \;\;\; \p-1 \;\;\ldots \;\;
2 \;\;\; 1]$ and 
set the variance of the error terms to be
$$\sigma^2 = \frac{\bbeta^\top
  \bSigma_X \bbeta}{1/R^2-1}\;,$$
where we choose $R^2 = 0.9$ by default. 
Then we generate 
$y_i = \bx_i^\top \bbeta + e_i$ where 
$e_i$ follows a Gaussian distribution with 
mean zero and variance $\sigma^2$.

To generate cellwise outliers, we start by 
drawing a fraction $\eps = 20\%$ of cells 
at random in each column of the clean data 
matrix, including the response variable. 
Each contaminated cell is then put at a 
value $\gamma = 1,\ldots,10$\,. 
Results for $\eps = 10\%$ are shown in 
Section~\ref{supp:sim} 
of the Supplementary Material, as well as 
results for $\p=10$. They are all 
qualitatively similar.

\subsection{Accuracy of coefficients and
predictions}
In our simulations we compare the proposed 
cellLTS procedure with OLS and the three 
competitors, and use the following 
abbreviations in the figures:
\begin{itemize}
\item \textbf{OLS}: least squares regression;
\item \textbf{3SGS}: the three-step 
  generalized S-estimator 
  of \cite{leung2016robust};
\item \textbf{Shooting S}: the Shooting S 
  estimator as implemented in
  \cite{bottmer2022sparse};
\item \textbf{STMW}: the algorithm by 
 Su, Tarr, Muller and Wang (2024). 
\end{itemize}

To compare the different estimators of $\bbeta$ 
we compute the distance
\begin{equation*}
  \MD = \MD(\bhbeta, \bbeta, \sigma^2 
  \bSigma_{\bX}^{-1}) = \sqrt{
  (\bhbeta - \bbeta)^{\top} \bSigma_{\bX}
  (\bhbeta - \bbeta)/\sigma^2}
\end{equation*}
that is inspired by the asymptotic distribution 
of the OLS estimator on clean data.

\begin{figure}[!hb]
\vspace{-5mm}
\centering
\includegraphics[width = 0.45\columnwidth]
{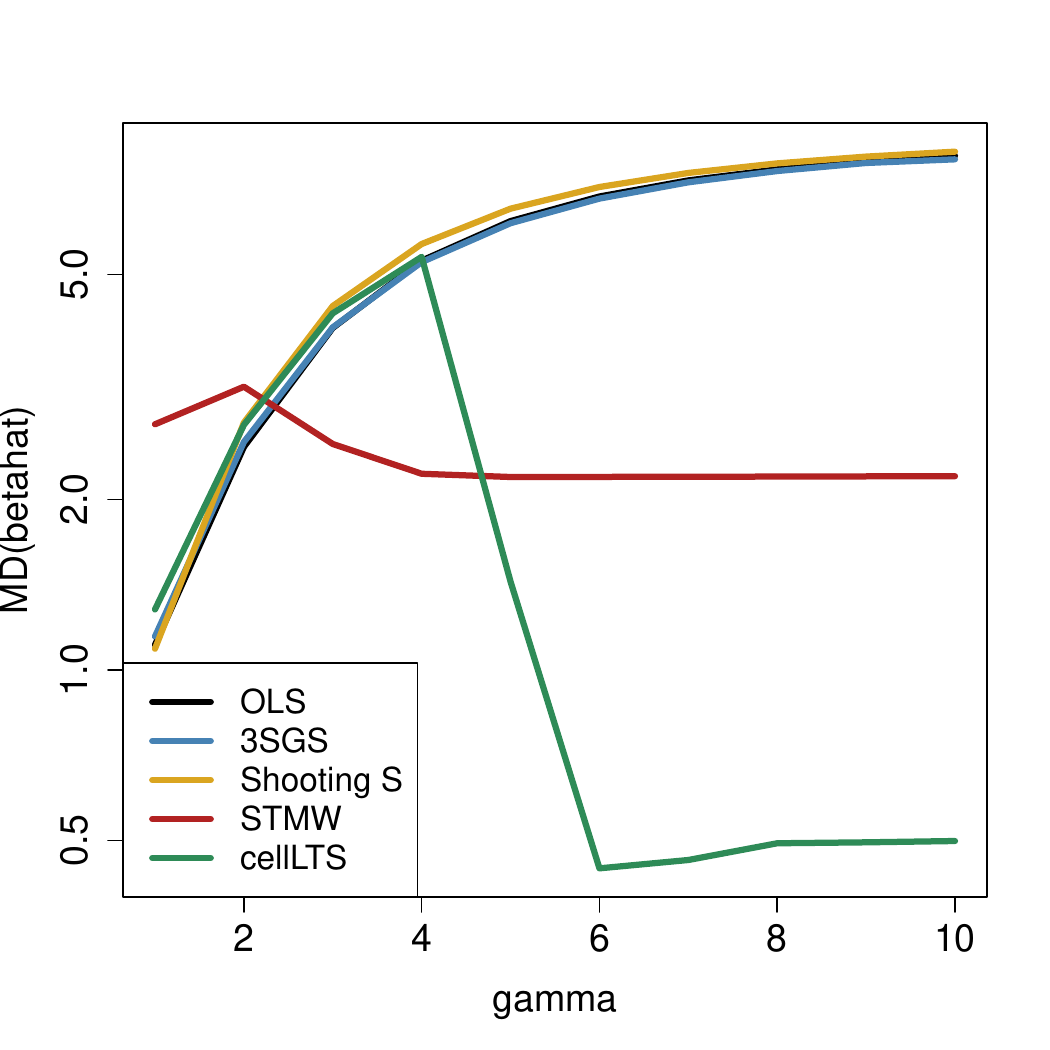}
\includegraphics[width = 0.45\columnwidth]
{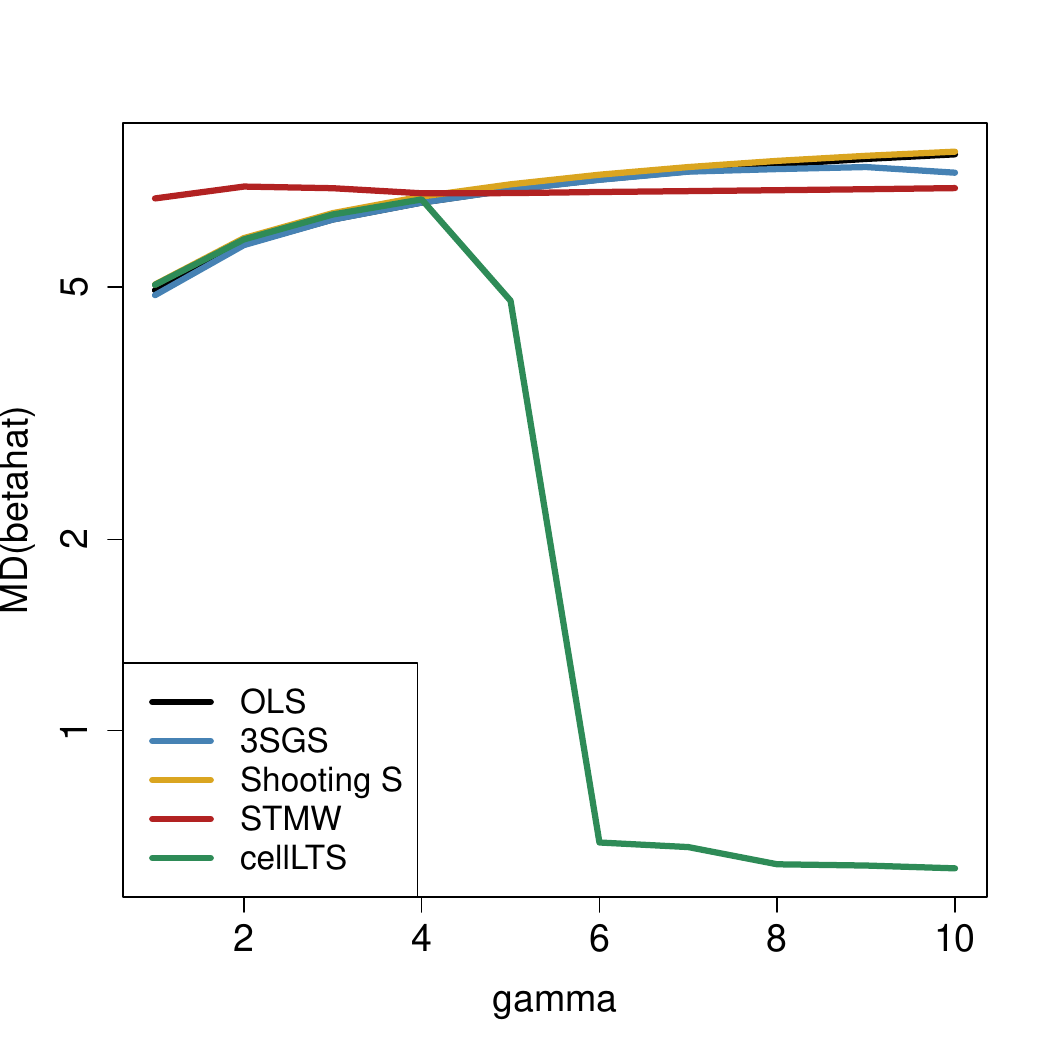}\\
\vspace{-6mm}
\includegraphics[width = 0.45\columnwidth]
{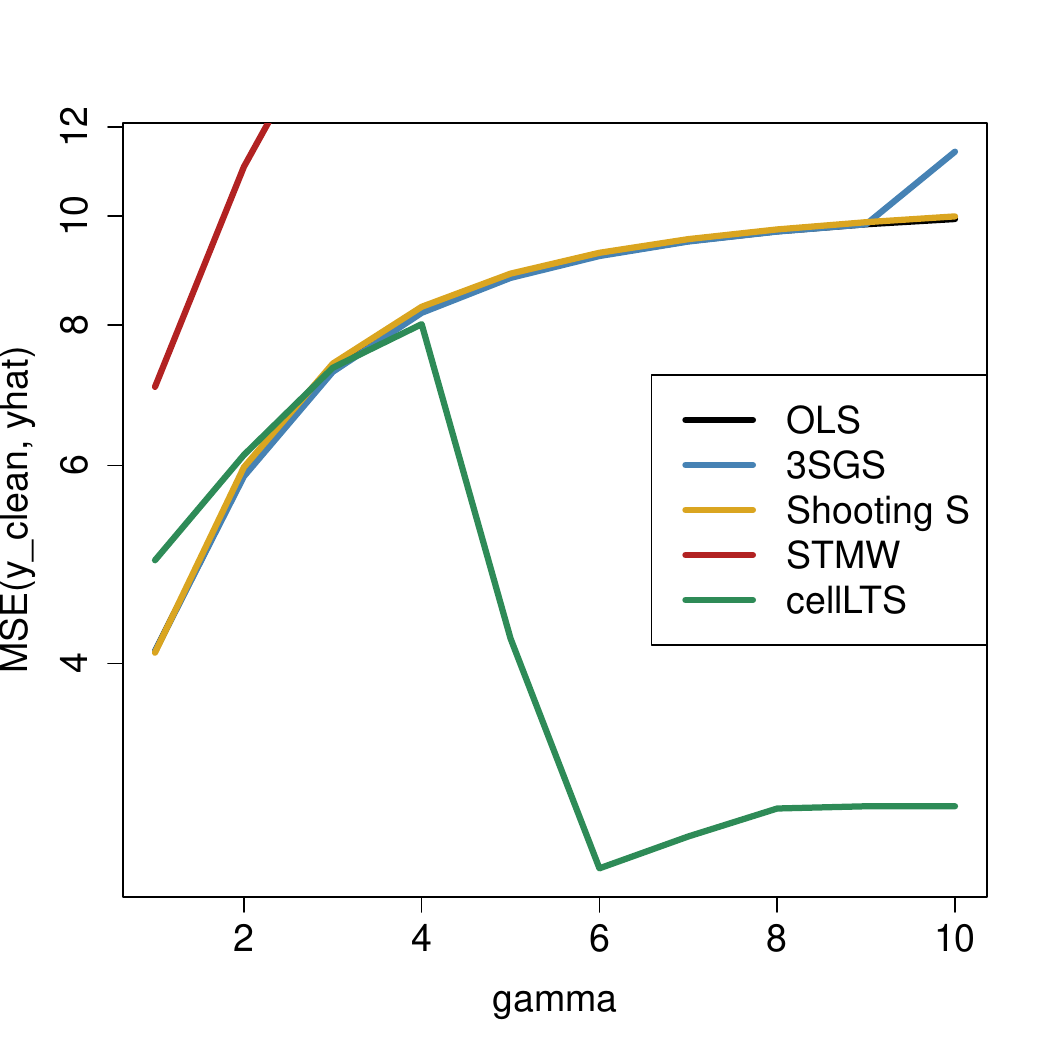}
\includegraphics[width = 0.45\columnwidth]
{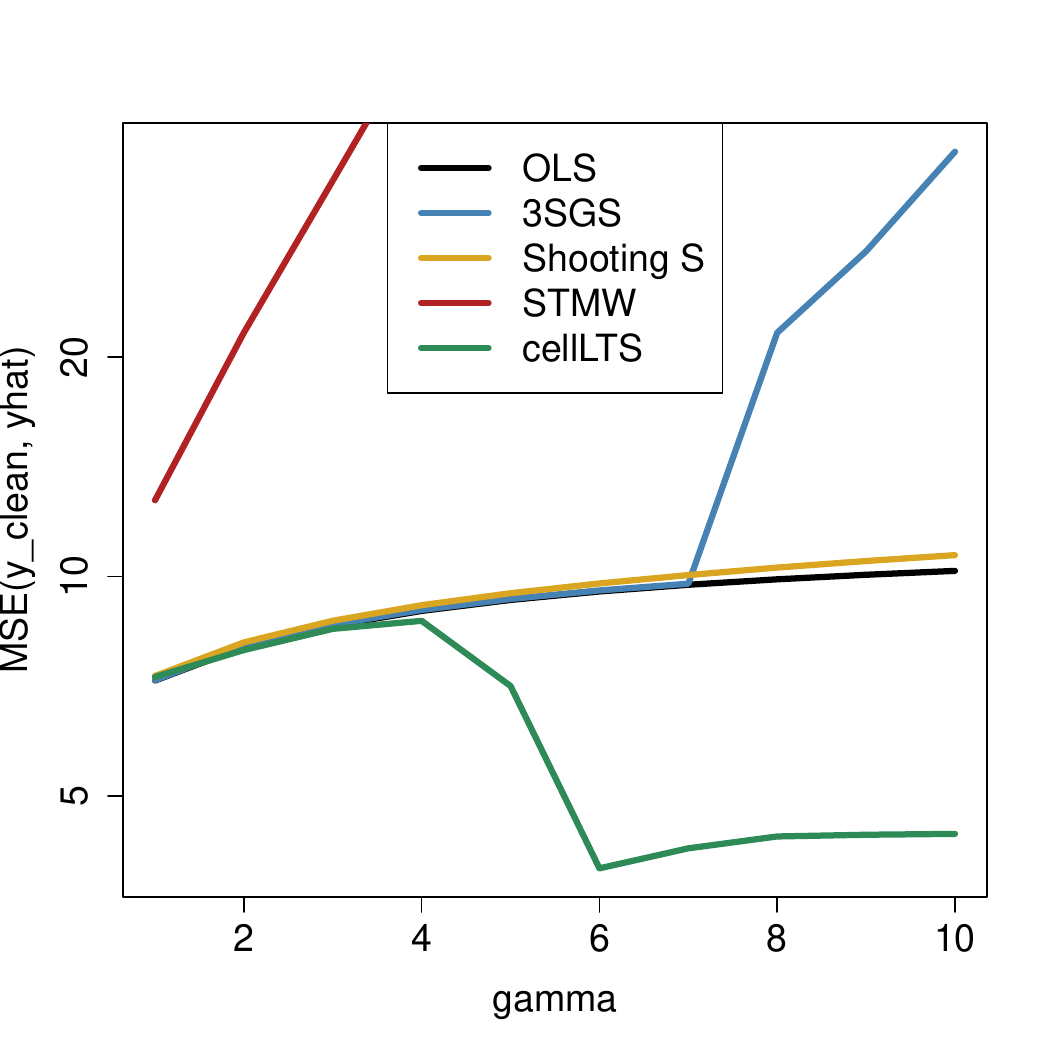}\\
\caption{Top: average MD (on log scale) of the 
estimated coefficients for $n = 400$, $\p = 20$, 
$\eps = 20\%$ of cellwise outliers, and 
$\bSigma = \bSigma_{\ALYZ}$ (left) or 
$\bSigma = \bSigma_{\AN}$ (right), for normal
predictors.
Bottom: corresponding MSE, also on log scale.}
\label{fig:MD_MSE_normal}
\end{figure}

In the top row of 
Figure~\ref{fig:MD_MSE_normal} we see that
the MD distance of the coefficients grows
with $\gamma$ for 3SGS, OLS and Shooting S,
both for $\bSigma = \bSigma_{\ALYZ}$ (left) 
and $\bSigma = \bSigma_{\AN}$ (right).
Note that MD is shown on a log scale.
The curve for STMW is more stable but at a
high level. The MD of cellLTS grows for
small $\gamma$, when the contaminated cells 
are not far enough to be considered truly
outlying, and then comes down for higher
$\gamma$, indicating high accuracy. 

We also generate clean out-of-sample regressor
matrices and their clean responses $\by^*$,
after which we contaminate the regressor 
matrices $\bX^*$ exactly as in the training
data. We then compute the out-of sample 
predictions $\widehat{\by}^{\,*}$, and their 
MSE
\begin{equation*}
  \MSE = \frac{1}{n} \sum_{i=1}^n 
  (\widehat{y}_i^{\,*} - y_i^*)^2\;.
\end{equation*}

\begin{figure}[!ht]
\vspace{-5mm}
\centering
\includegraphics[width = 0.45\columnwidth]
  {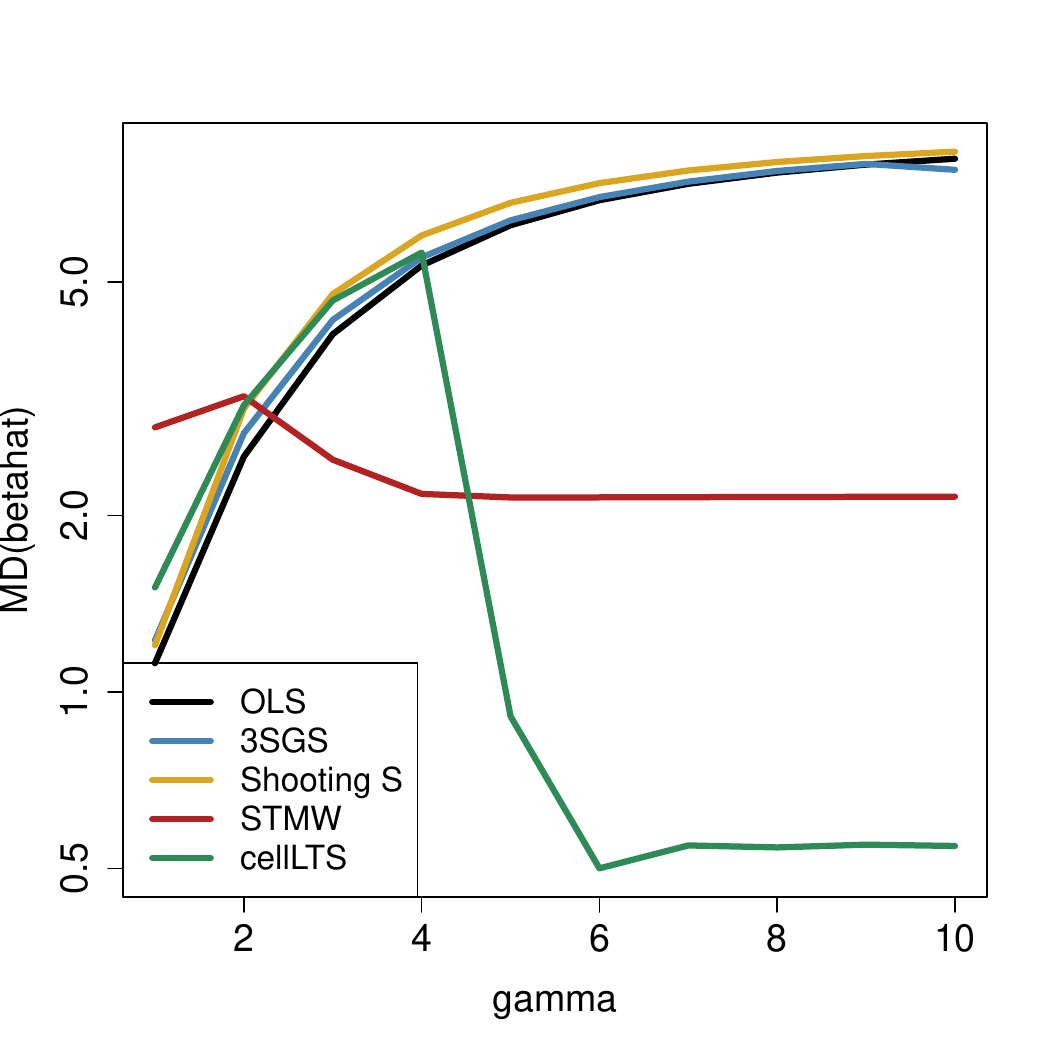}
\includegraphics[width = 0.45\columnwidth]
  {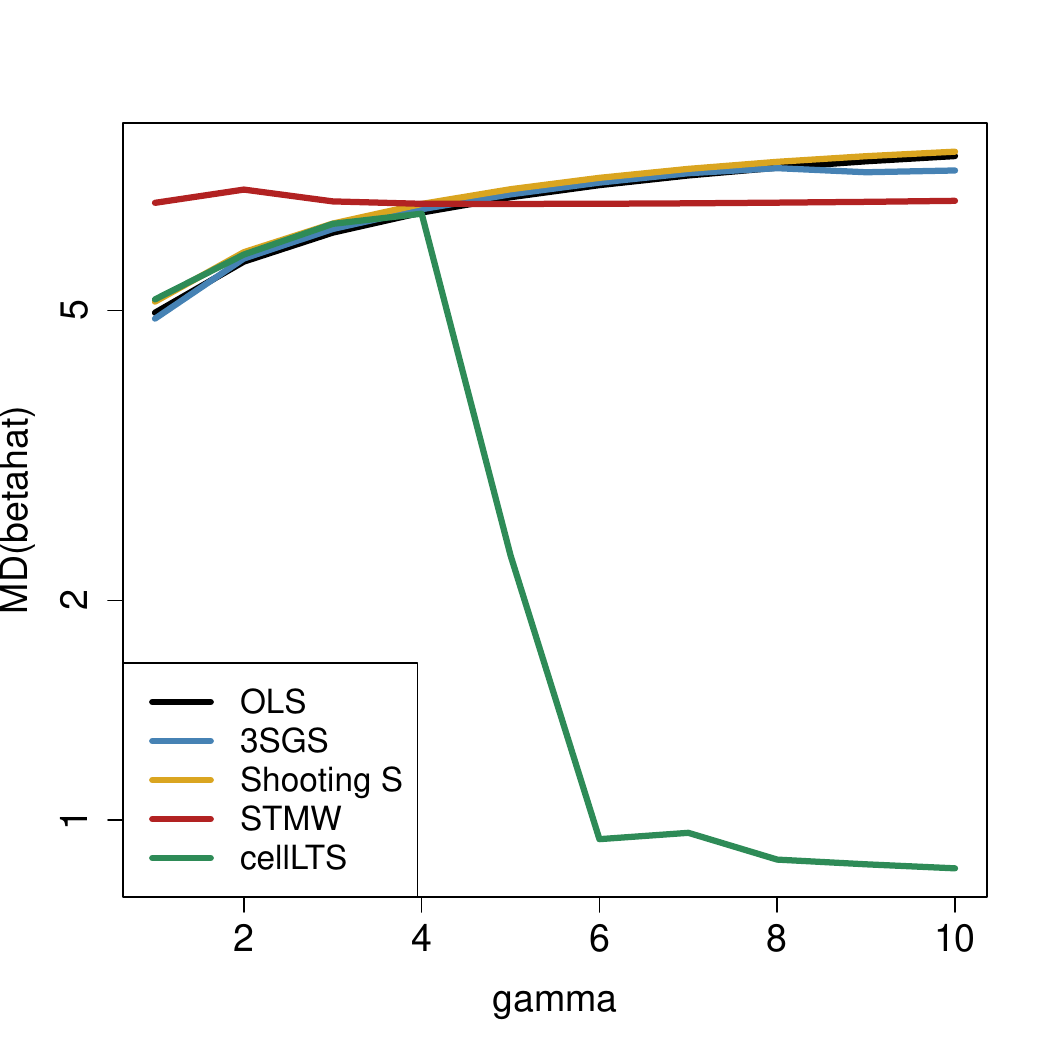}\\
\vspace{-6mm}
\includegraphics[width = 0.45\columnwidth]
  {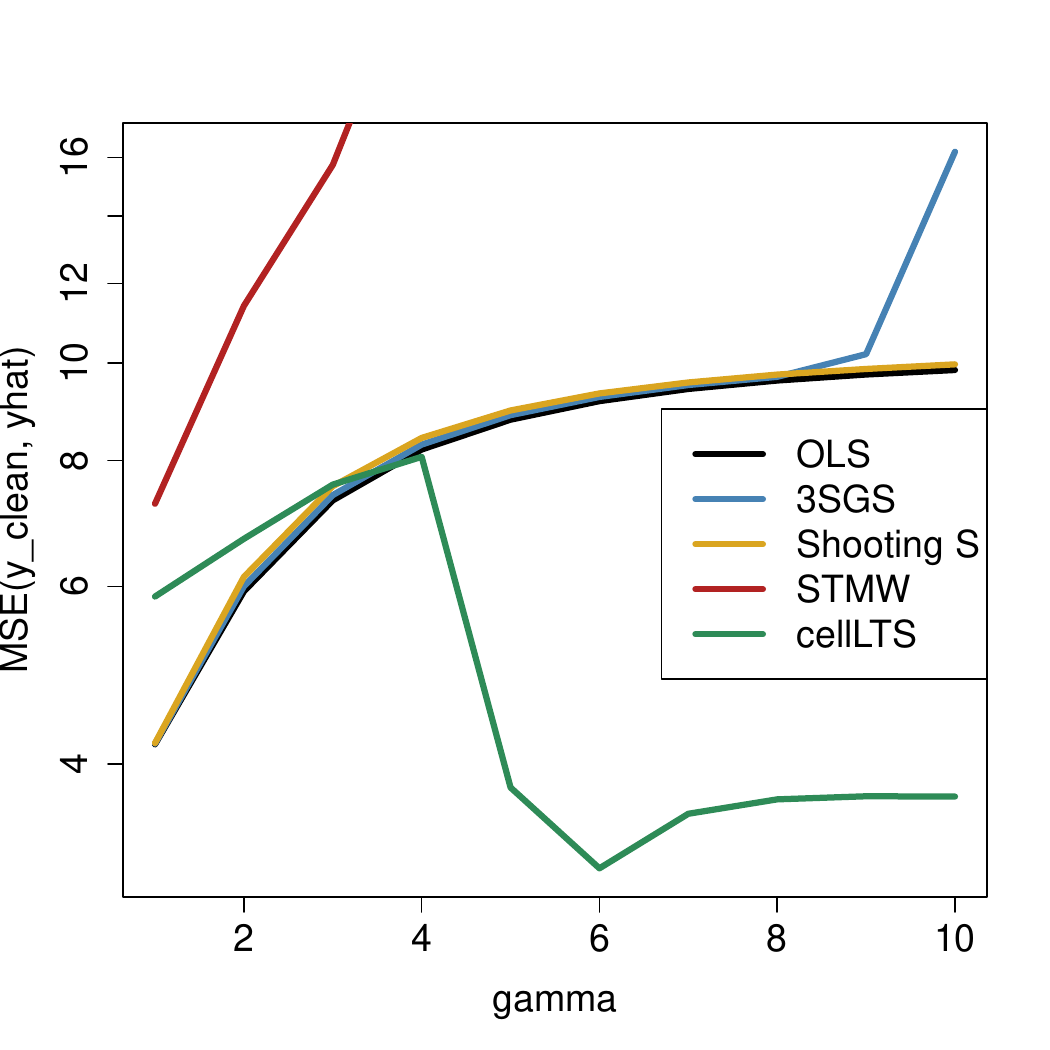}
\includegraphics[width = 0.45\columnwidth]
  {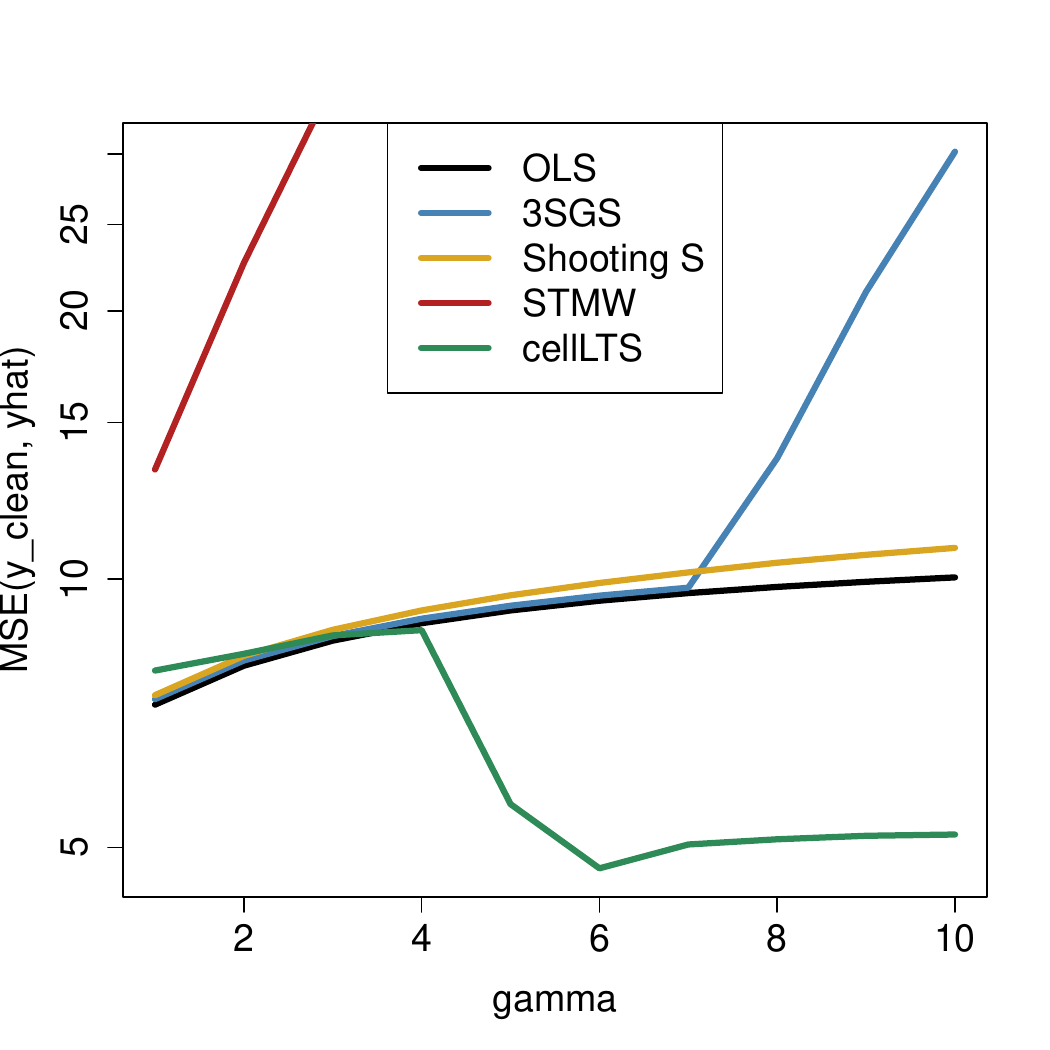}\\
\caption{Like Figure~\ref{fig:MD_MSE_normal}, 
but for exponential predictors.}
\label{fig:MD_MSE_exponential}
\end{figure}

The MSE of the out-of-sample predictions 
in the bottom row of 
Figure~\ref{fig:MD_MSE_normal} looks 
different from the MD. Here STMW is the most 
affected, followed by 3SGS that breaks away 
from OLS and Shooting S at $\bSigma_{\AN}$\,.
In contrast, cellLTS continues to perform
well. This is because cellLTS robustly
imputes outlying cells in the 
out-of-sample $\bX^*$ before computing 
its predictions.

\begin{figure}[!ht]
\vspace{-5mm}
\centering
\includegraphics[width = 0.45\columnwidth]
  {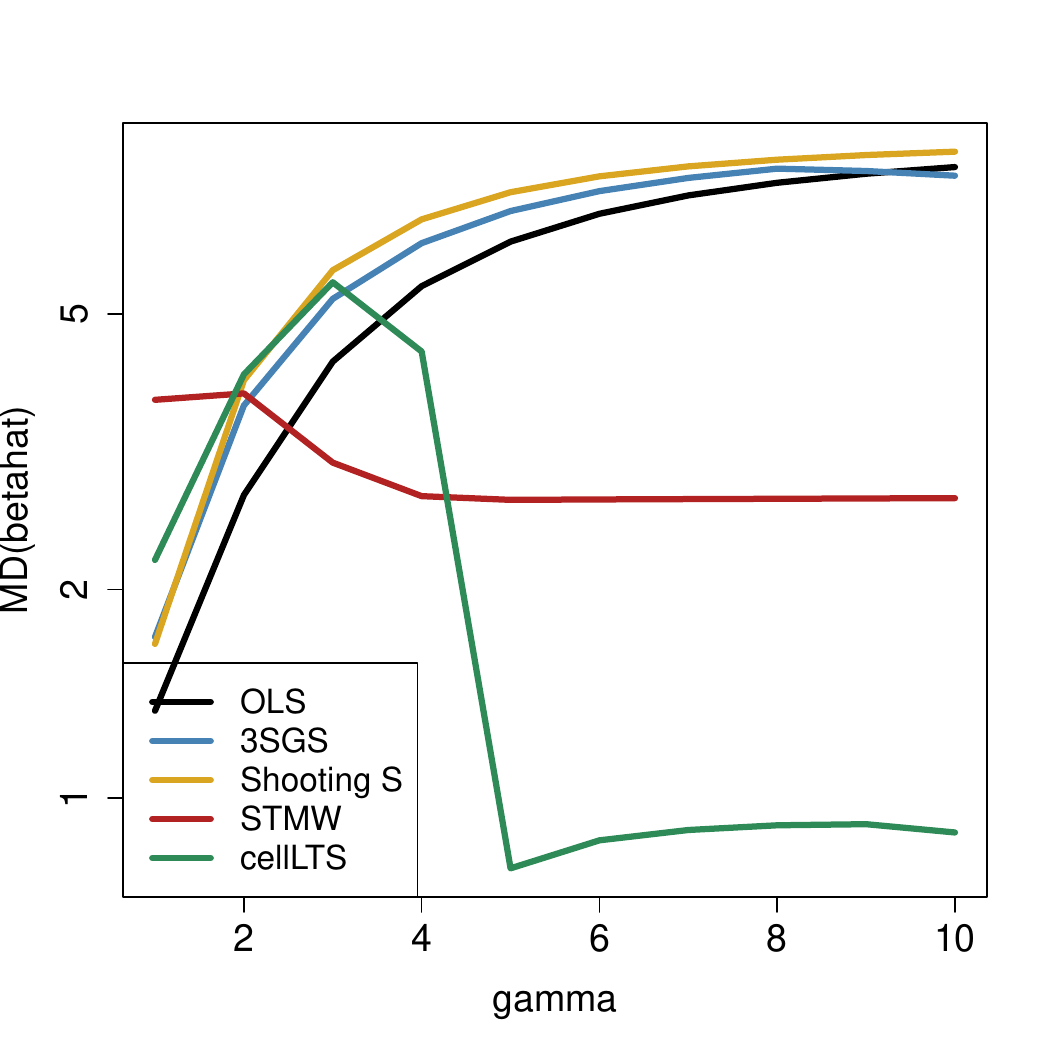}
\includegraphics[width = 0.45\columnwidth]
  {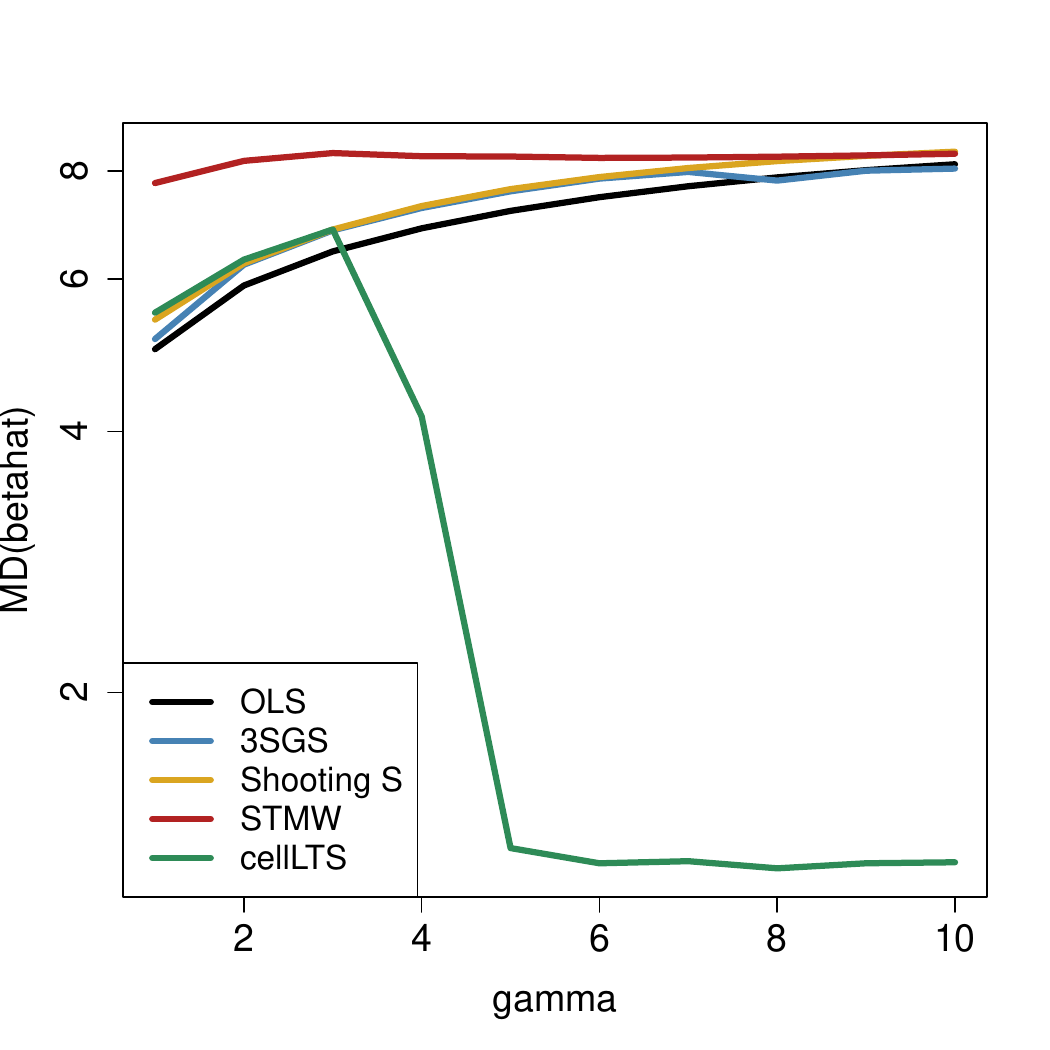}\\
\vspace{-6mm}
\includegraphics[width = 0.45\columnwidth]
  {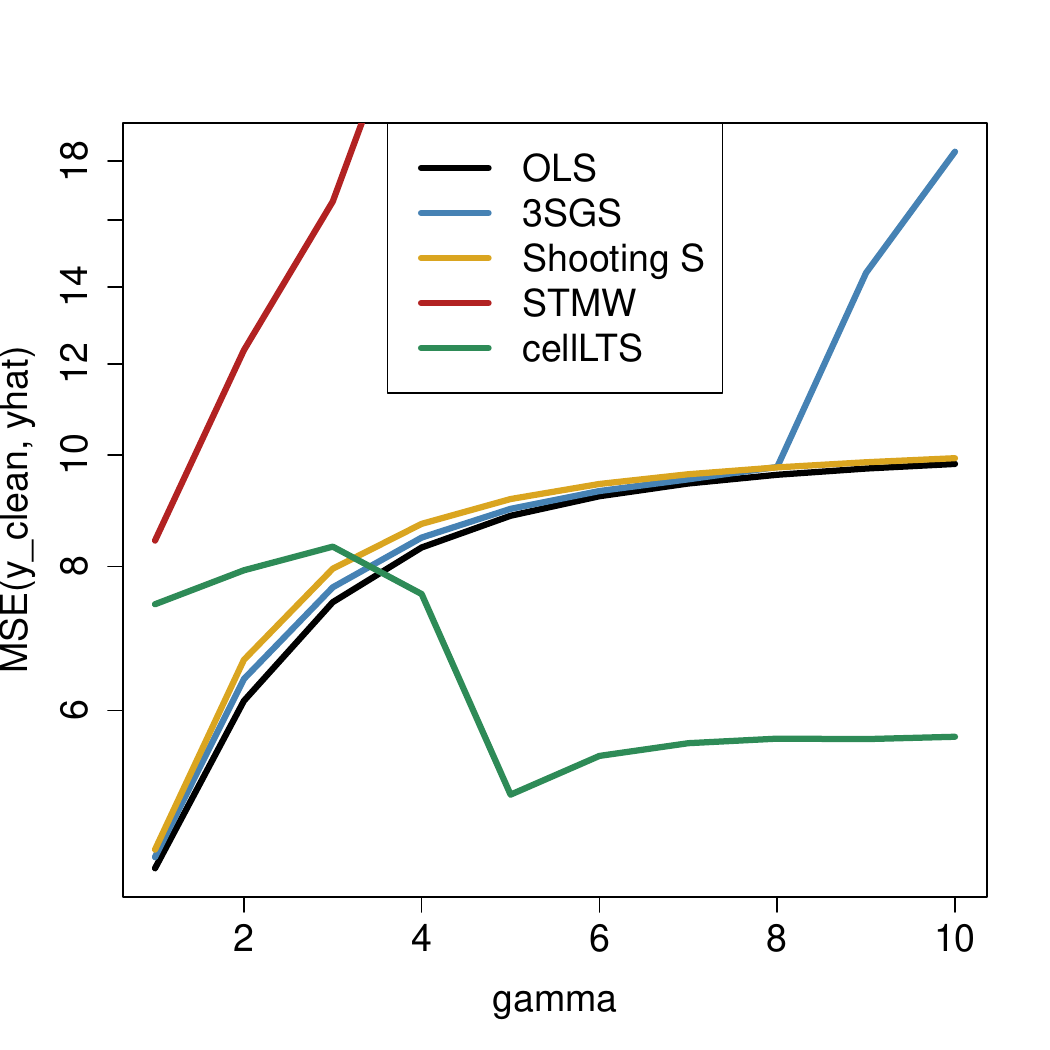}
\includegraphics[width = 0.45\columnwidth]
  {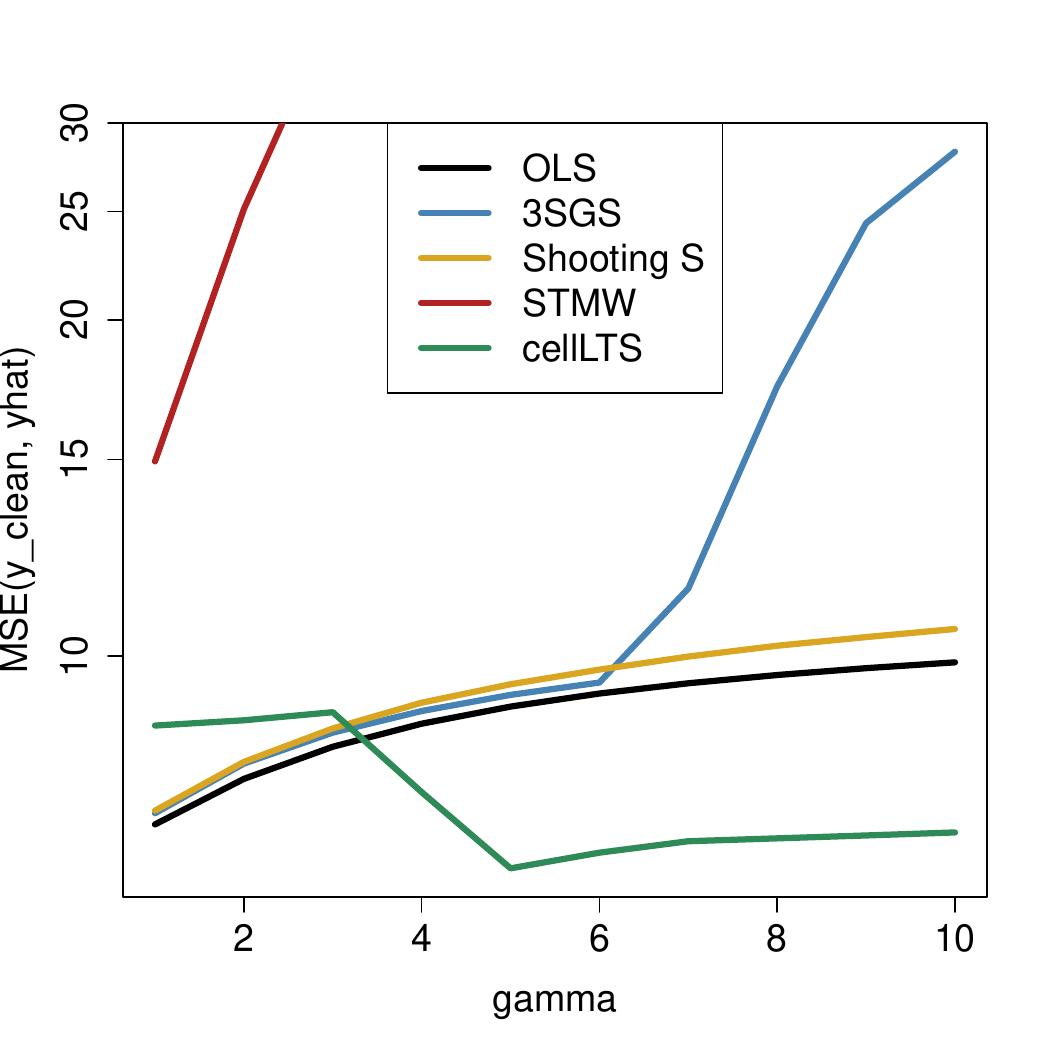}\\
\caption{Like Figure~\ref{fig:MD_MSE_normal}, 
but for lognormal predictors.}
\label{fig:MD_MSE_lognormal}
\end{figure}

Figure \ref{fig:MD_MSE_exponential} repeats
the simulation for skewed regressors $\bX$
derived from the exponential distribution.
The results are qualitatively similar to 
those in Figure \ref{fig:MD_MSE_normal},
where $\bX$ was normally distributed before
its contamination by outlying cells. The
same can be said about skewed $\bX$
based on the lognormal distribution in
Figure \ref{fig:MD_MSE_lognormal}. We
conclude that cellLTS outperforms the
earlier methods over a range of settings, 
both in terms of coefficient accuracy and 
out-of-sample prediction.

\subsection{Simulation of symmetrization}
\begin{figure}[!ht]
\centering
\includegraphics[width = 0.42\columnwidth]{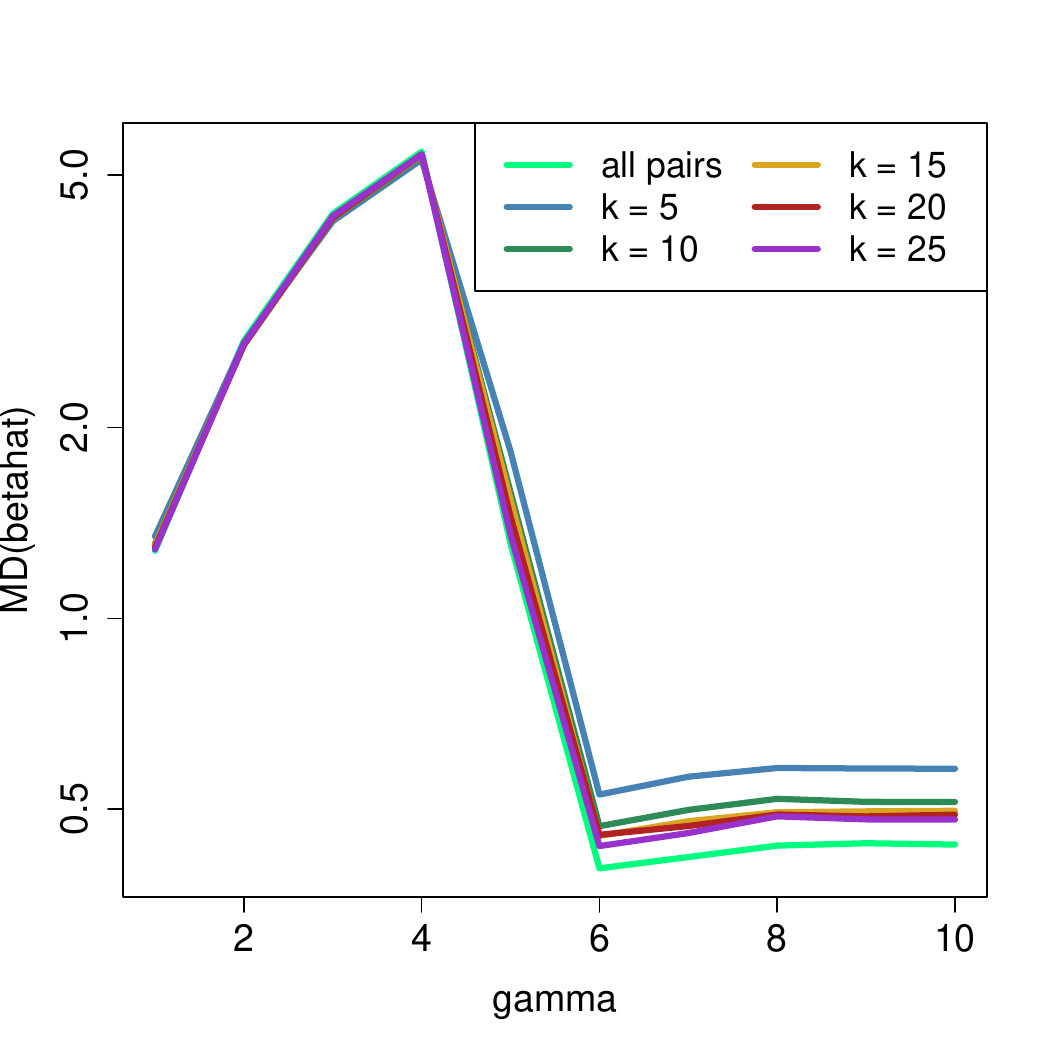}
\includegraphics[width = 0.42\columnwidth]{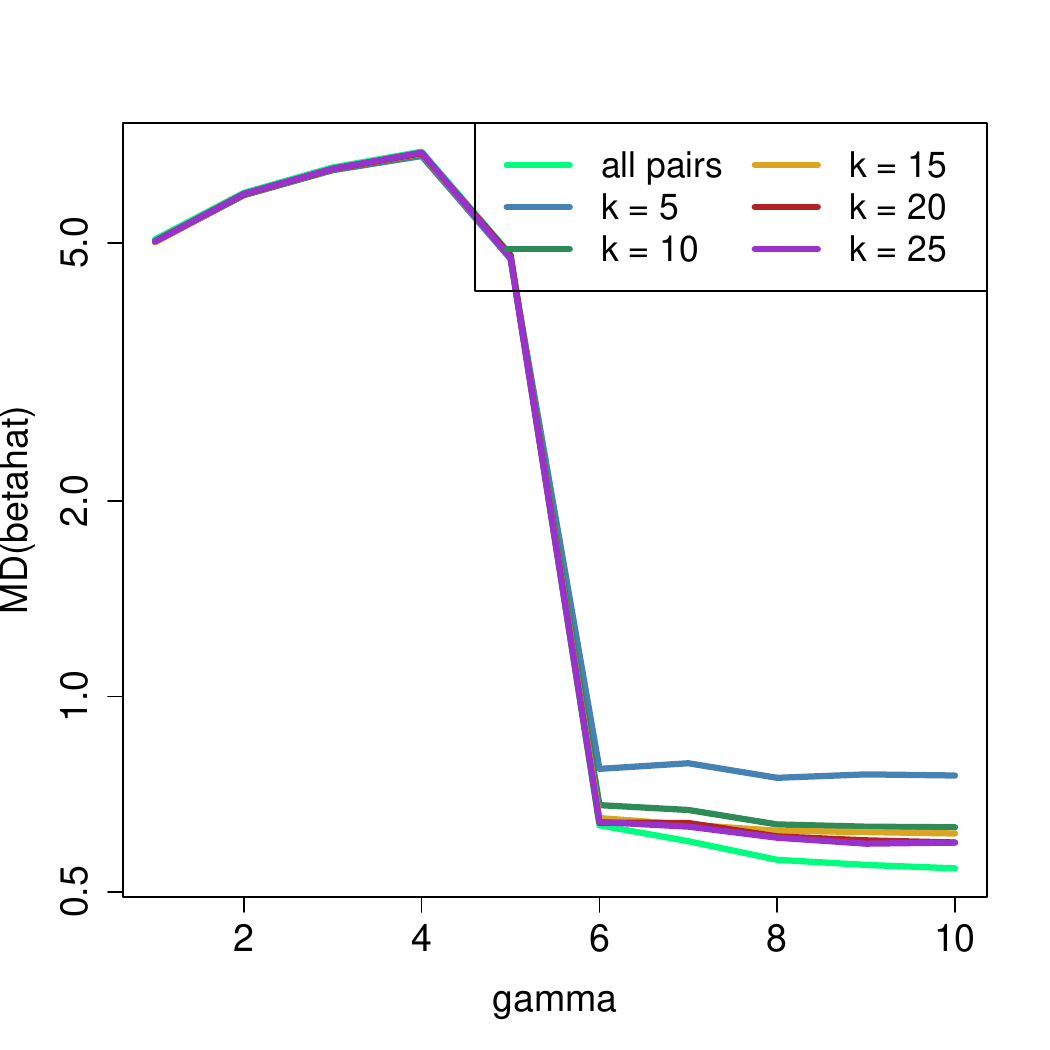}\\
\vspace{-6mm}
\includegraphics[width = 0.42\columnwidth]{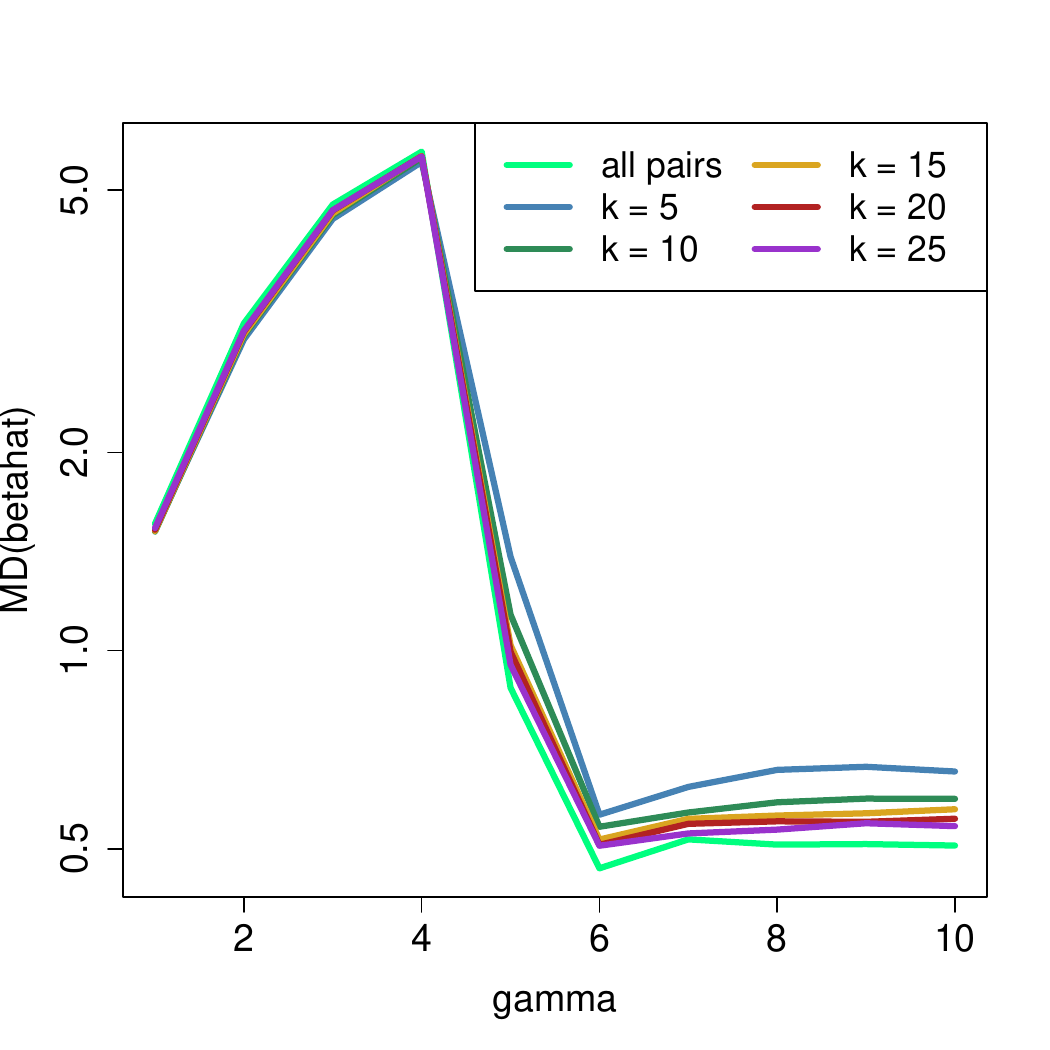}
\includegraphics[width = 0.42\columnwidth]{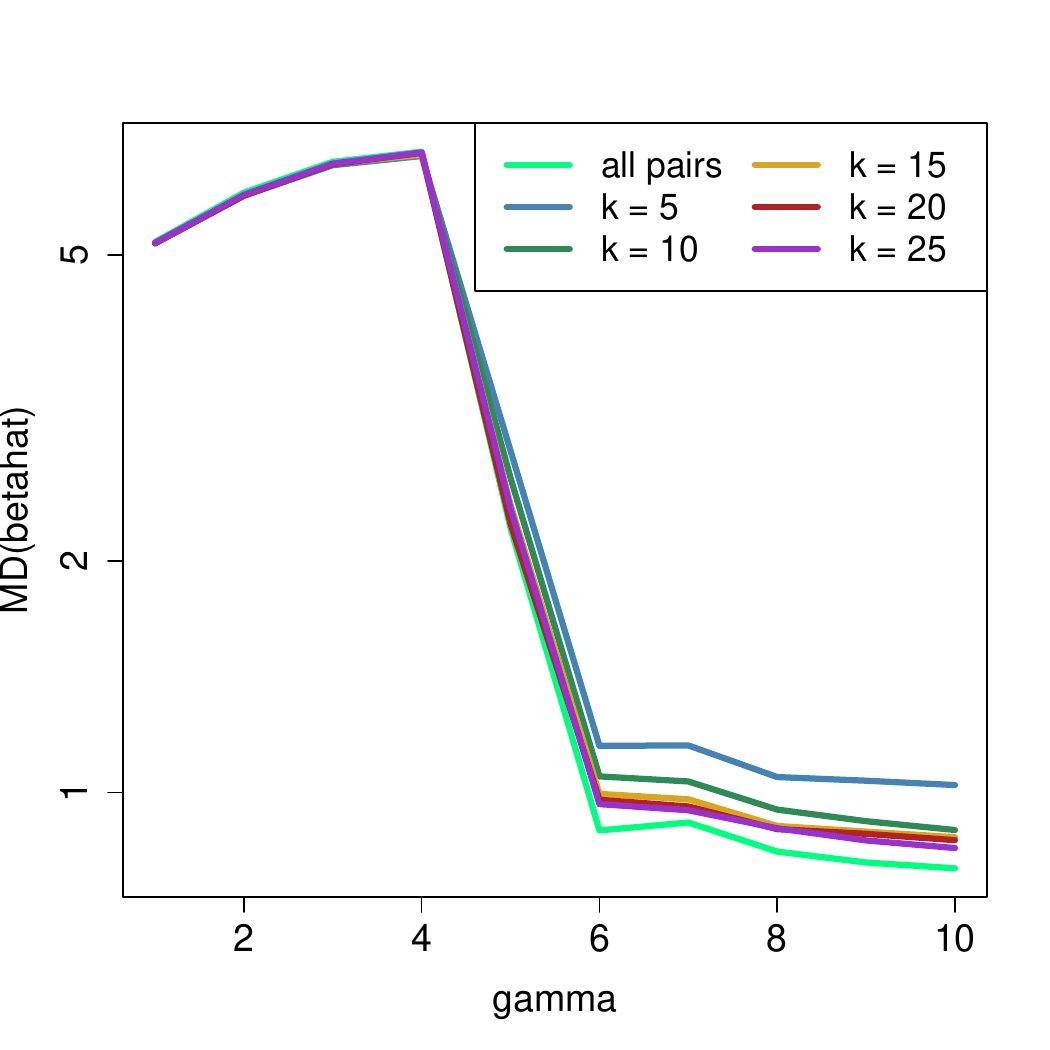}\\
\vspace{-6mm}
\includegraphics[width = 0.42\columnwidth]{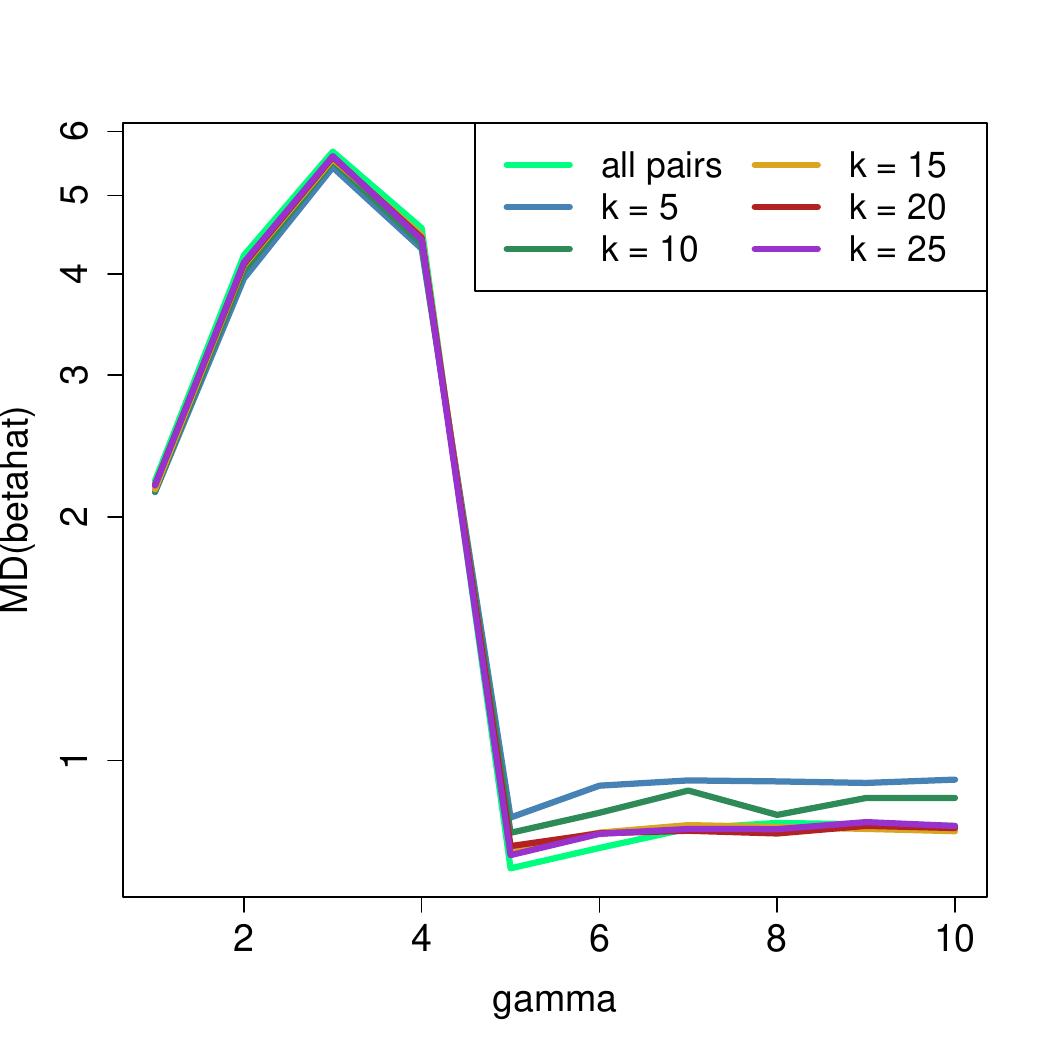}
\includegraphics[width = 0.42\columnwidth]{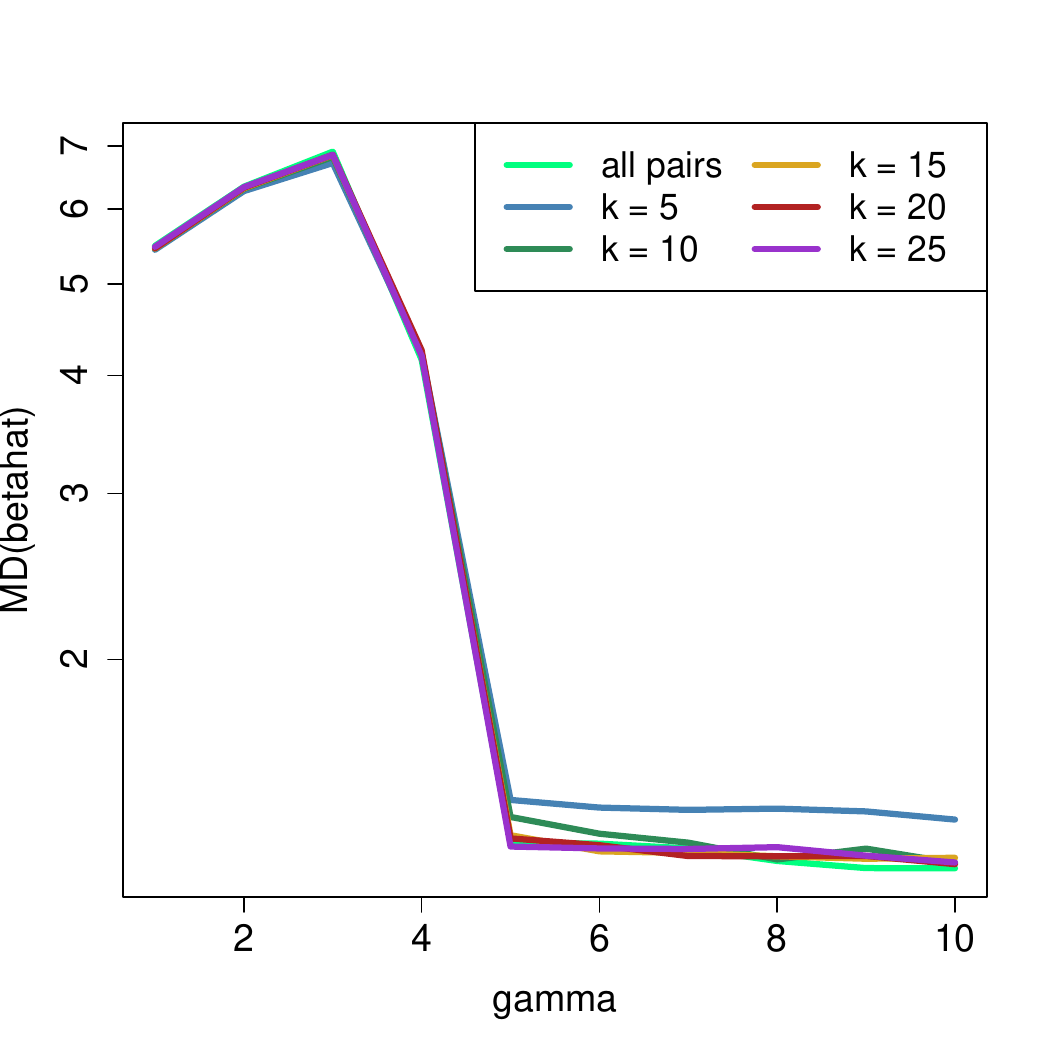}\\
\caption{Top row: average MD (on log scale) of the 
estimated coefficients for different symmetrization 
strategies and normal predictors. 
The data has dimension $\p = 20$, 
$\eps = 20\%$ of cellwise outliers, and 
$\bSigma = \bSigma_{\ALYZ}$ (left) or 
$\bSigma = \bSigma_{\AN}$ (right).
Middle row: same for exponential predictors.
Bottom row: same for lognormal predictors.}
\label{fig:symmetrization}
\end{figure}

In this section we verify that symmetrization 
with $k = 20$ random permutations as described 
in Section~\ref{sec:fastsymm} performs about as 
well as taking all $\mathcal{O}(n^2)$ pairs. 
The data is generated in the same way as before.
Figure~\ref{fig:symmetrization}
shows the accuracy of the estimated regression 
coefficients for normal, exponential, and 
lognormal data. We again show the results for 
$n = 400$, $\p = 20$, and $\varepsilon = 20\%$. 
The results for the other settings are 
qualitatively similar and can be found in 
Section~\ref{supp:sim} 
of the Supplementary Material. 
As we can see, working with all $n(n-1)/2$ 
pairs generally performs best, but the 
differences are rather small. Whereas $k=5$ 
and $k=10$ still perform visibly worse, 
especially for higher values of $\gamma$, 
the difference becomes tiny for higher values 
of $k$. This supports our default choice 
of $k=20$.

\section{Real data example: Modeling cancer 
mortality}\label{sec:realdata}

To illustrate cellLTS on real data, we analyze
the US Cancer dataset containing information on 
demographics and cancer statistics for counties 
in the US. The data was compiled from different 
sources, most notably the U.S. Census Bureau 
and the National Cancer Institute, and collects 
figures from the period 2010-2016. The data is 
publicly available on Kaggle at
\url{https://www.kaggle.com/datasets/varunraskar/cancer-regression/data}. 
The dataset contains 3047 rows, corresponding 
to the counties, and 33 columns. The response 
variable is \texttt{target\_deathrate}, which 
is the death rate due to cancer (in cases per 
100 000). The other features include 
information such as median income, median age, 
population estimates, degree of schooling, 
birth rates, and others.

For illustrative purposes, we build a 
regression model using five covariates:\linebreak 
\texttt{incidencerate}, \texttt{medincome}, 
\texttt{medianage}, \texttt{pcths18\_24} and 
\texttt{pctemployed16\_over}. 
The meaning of these five regressors is 
given in Table \ref{tab:USCancersmall_features}.
(An analysis involving all regressors can be 
found in Section~\ref{supp:uscancer} 
of the Supplementary Material.)

\begin{table}[ht]
\centering
\vspace{2mm}
\begin{tabular}{p{3.8cm}|p{11cm}}
\hline
feature & explanation\\
\hline
\texttt{incidencerate}& Incidence rate of 
  cancer (per 100 000).\\
\texttt{medincome}& Median income in the 
  county (in thousands of dollars).\\
\texttt{medianage}& Median age in the county.\\
\texttt{pcths18\_24}& Percentage of residents 
  aged 18-24 whose highest\\ & 
  attained education is a high school diploma.\\
\texttt{pctemployed16\_over}& Percentage of 
  people aged 16 and over who are employed.\\
\hline
\end{tabular}
\caption{The explanatory features for 
   predicting cancer mortality. }
\label{tab:USCancersmall_features}
\end{table}

\begin{figure}[!ht]
\centering
\includegraphics[width=0.75\linewidth]
  {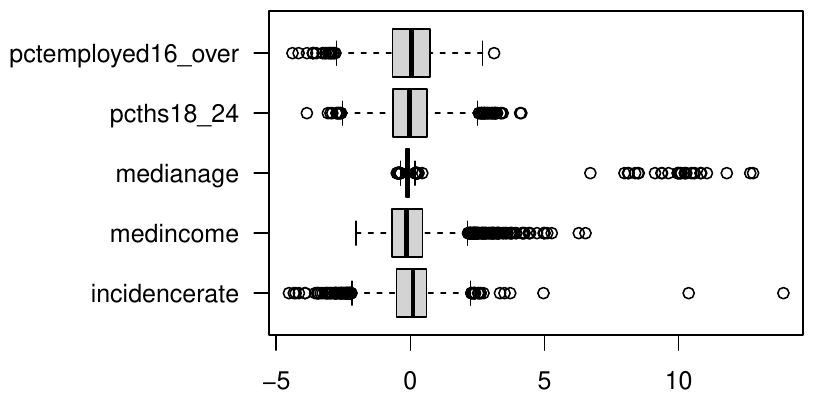}\\
\caption{Boxplots of the standardized regressors 
   for predicting cancer mortality.}
\label{fig:USCancersmall_boxplot}
\end{figure}

\phantom{abc}\\
In Figure~\ref{fig:USCancersmall_boxplot} we
see that these five features are roughly 
symmetric in the middle, but there are some 
extreme outliers in the median age and 
the incidence rate.

To start the analysis, we fit a linear model 
with OLS and with the proposed cellLTS method 
to model the death rate as a linear function 
of the five predictors. For OLS we imputed 
the 5\% missing values in 
\texttt{pctemployed16\_over} with the mean,
whereas cellLTS did not require any 
preprocessing because it is designed to
handle missing data. The resulting 
coefficients are given in the first two 
columns of 
Table~\ref{tab:USCancersmall_coefficients}. 
We see that the coefficients are fairly similar 
overall, with the exception of the coefficient 
for \texttt{medianage}. It is -0.73 for 
cellLTS, whereas it is almost zero for OLS. The 
explanation is found in the extreme outliers of 
the \texttt{medianage} variable. When these 
cases are removed by hand, we obtain an OLS 
coefficient of -0.61 which is much more in 
line with the -0.73 of cellLTS.

\begin{table}[!ht]
\vspace{5mm}
\centering
\begin{tabular}{lrrr}
\hline
 & OLS & cellLTS & OLS of cleaned \\ 
\hline
\texttt{intercept} & 122.50 & 157.16 & 147.27 \\ 
\texttt{incidencerate} & 0.23 & 0.24 & 0.23 \\ 
\texttt{medincome} & -0.66 & -0.75 & -0.65 \\ 
\texttt{medianage} & -0.01 & -0.73 & -0.61 \\ 
\texttt{pcths18\_24} & 0.47 & 0.62 & 0.55 \\ 
\texttt{pctemployed16}\_over & -0.56 & -0.83 & -0.62 \\ 
\hline
\end{tabular}
\caption{Regression coefficients of the US 
Cancer data, fit by OLS, cellLTS, and OLS 
after removing cases with extreme outliers 
in the median age.}
\label{tab:USCancersmall_coefficients}
\end{table}

\begin{figure}[!ht]
\centering
\includegraphics[width=0.47\linewidth]
{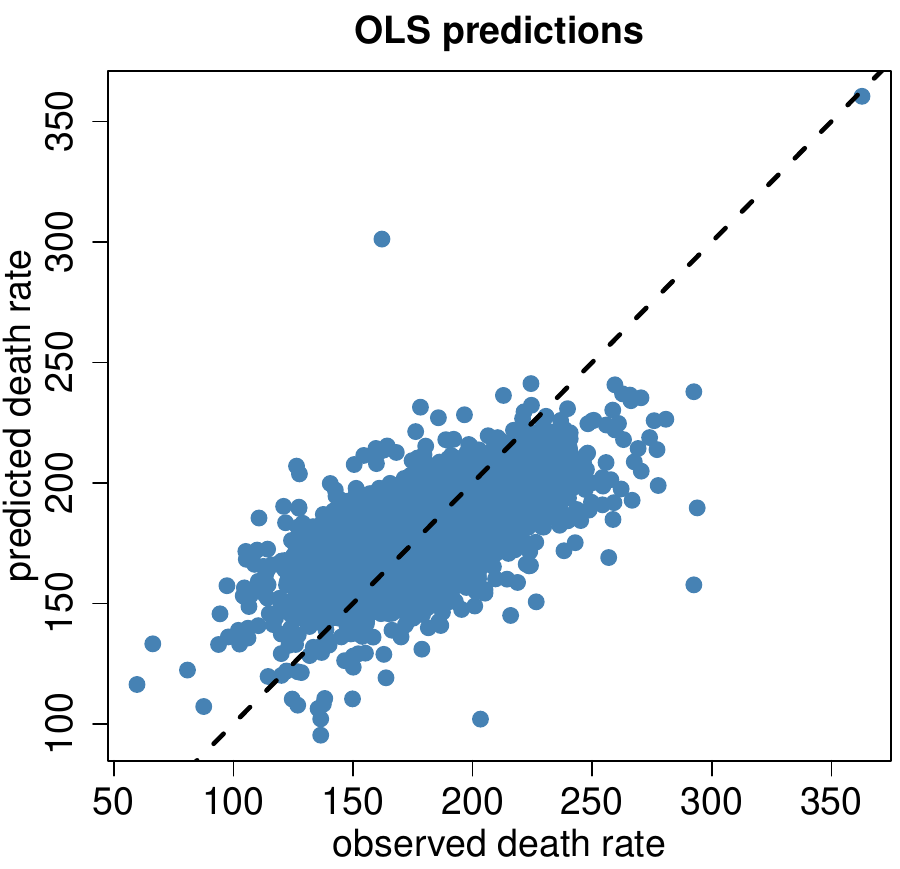}
\hspace{3mm}
\includegraphics[width=0.47\linewidth]
{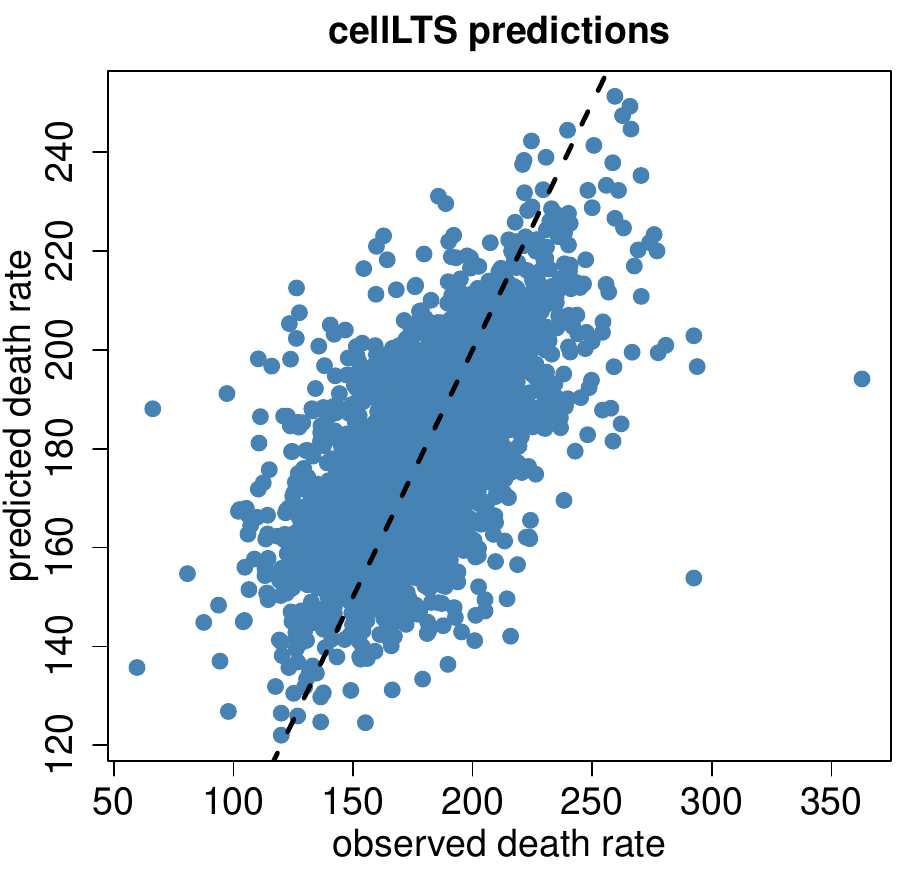}
\caption{Predicted vs. observed death rate for 
  OLS (left) and cellLTS (right).}
\label{fig:USCancersmall_predictedvsobserved}
\end{figure}

Figure~\ref{fig:USCancersmall_predictedvsobserved}
plots the predictions versus the observed response 
for OLS and cellLTS. We see that cellLTS provides 
a better fit overall.

\begin{figure}[!ht]
\centering
\includegraphics[width=0.65\linewidth]
{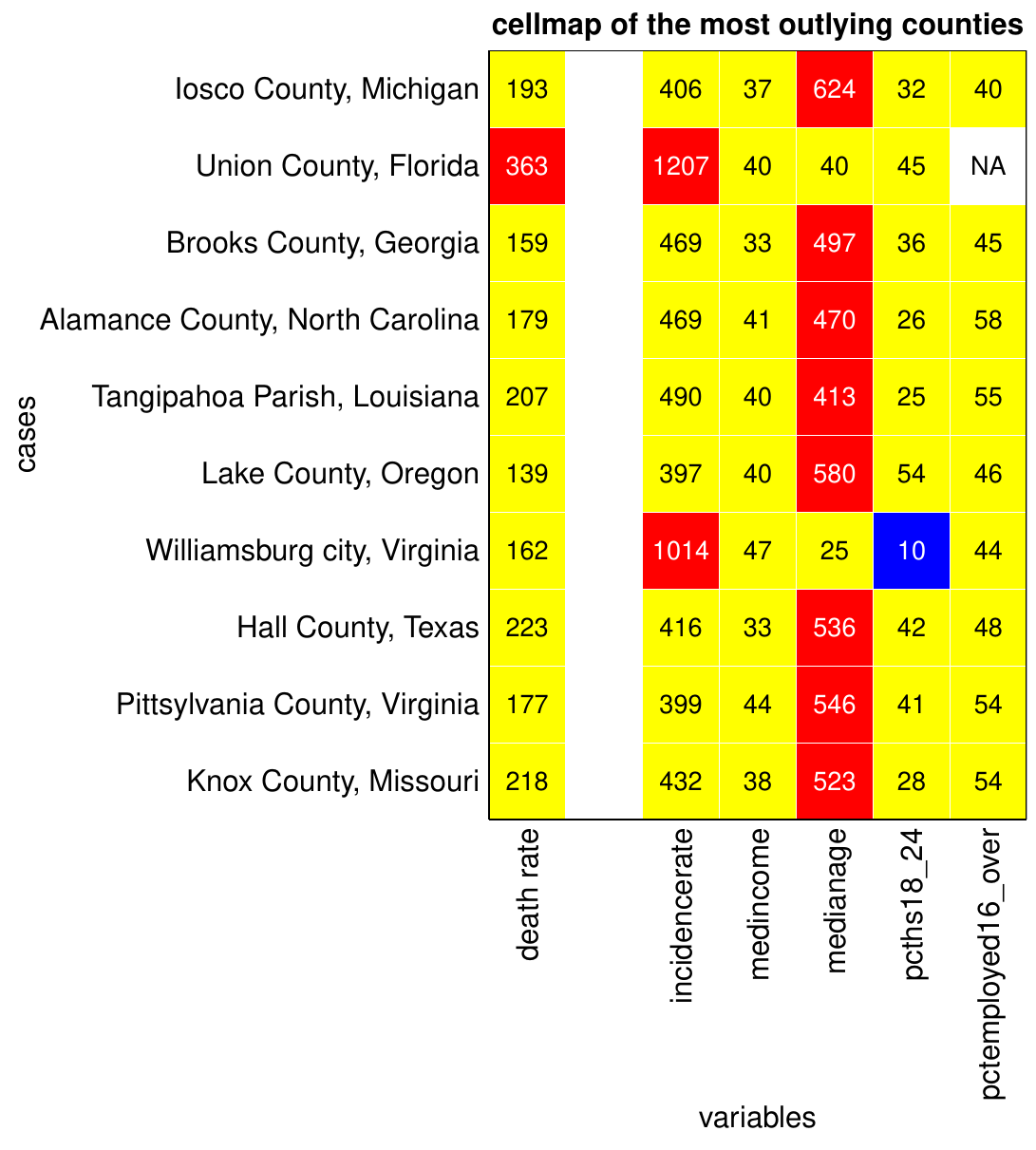}
\caption{Cellmap of the 10 counties with the highest 
sum of absolute standardized cellwise residuals.}
\label{fig:USCancersmall_cellmaplargestresiduals}
\end{figure}

In addition to its fit, cellLTS also provides 
more detailed information. Its cellmap visualizes
the cells that were flagged as suspicious in 
either the response or the regressors. 
Section~ref{supp:uscancer} 
of the Supplementary
Material contains a large cellmap grouped by 
state. To zoom in on some interesting cases, 
Figure~\ref{fig:USCancersmall_cellmaplargestresiduals}
shows the cellmap of the 10 counties with the 
largest sum of absolute standardized cellwise 
residuals. The leftmost column is the response
variable, followed by a blank column to separate
it visually from the columns of the regressors.
Cells with large positive residuals are shown 
in red, whereas those with large negative
residuals are blue. Cells with unremarkable
residuals are yellow, and missing cells are
white.

The cellmap reveals several unusual patterns. 
Most notably, we see median ages of 400 years
and more, that are definitely errors. We saw 
that these influence the OLS coefficient of 
the \texttt{medianage} regressor, whereas 
the cellLTS coefficient of this feature is 
not affected. 
We also see two massive outliers in the cancer
incidence rate, Union County in Florida and 
Williamsburg City in Virginia. Union County 
indeed turns out to be the county with the 
highest cancer incidence rate in the US. 
It also has the highest mortality rate by far. 
The exact reasons are unclear, but appear to 
be related to high smoking rates and 
unhealthy dietary behavior.

The outlier in 
the incidence rate in Williamsburg City, 
however, appears to be an error, as the 
National Cancer Institute gives an incidence 
rate of around 500 for this city. That 
is more in line with its relatively low death 
rate of 162. CellLTS has implicitly imputed 
this cell at an incidence rate of about 450, 
which is much more realistic. 
The value 10 for \texttt{pcths18\_24} 
indicates that only 10\% of the city's 
18-to-24-year-old do not obtain a diploma 
after finishing high school. This is 
exceptionally low, but can be explained by 
the fact that Williamsburg City is a college 
town, home to The College of William \& Mary, 
the second-oldest institution of higher 
learning in the US (with Harvard being the 
oldest).

\begin{figure}[!ht]
\centering
\includegraphics[width=0.75\linewidth]
 {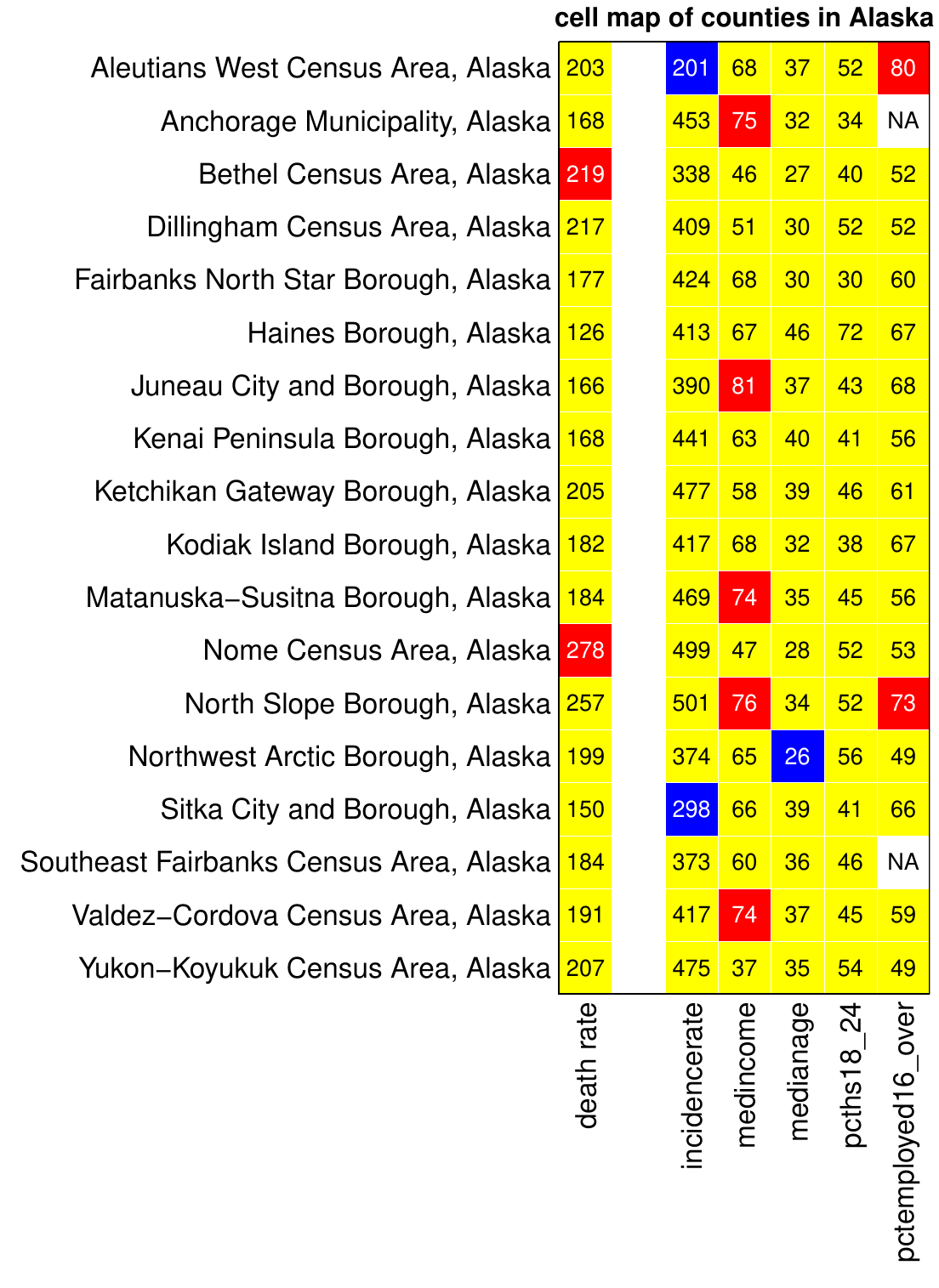} 
\caption{Cellmap of the US Cancer data in Alaska.}
\label{fig:USCancer_cellmap_Alaska}
\end{figure}

We also look at the cellmap of Alaska in 
Figure~\ref{fig:USCancer_cellmap_Alaska}. 
Alaska is an unusual state. It is by far
the largest state in terms of area, 
but with under 1 million inhabitants 
the least densely populated. The cellmap 
reveals some atypical data. For
instance, the Aleutians West Census Area
has a suspiciously low incidence rate and 
a very high employment rate. The first one 
is likely a measurement error, since it is 
lower than the reported death rate due to 
cancer. The employment rate for this area 
is very high, as it is for the North 
Slope Borough. Both of these areas are 
very remote, and mainly attract workers 
in the oil \& gas industry and the 
fishing industry.

We further see some boroughs with 
relatively high median incomes. In fact, 
many Alaskan regions are high-income. 
This compensates for the fairly high cost 
of living due to the remoteness, and the
unemployment rate is generally low. 
Juneau city has the highest income, it 
is the capital and has a lot of 
government services.
Bethel Census Area has a higher death rate 
than would be expected based on its 
explanatory features. Its incidence rate 
is fairly normal, which would predict a 
lower death rate. 
An explanation might be that this area has 
about 85\% of native population, who may 
not find their way to medical attention 
as quickly as the non-native inhabitants. 
A similar demographic is found in the 
Nome Census Area, that also has a flagged 
death rate. 
Finally, Sitka City and Borough has a 
lower-than-expected incidence rate. It 
is not immediately clear why this is the 
case. According to the National Cancer 
Institute  its incidence rate was 355.2 
in 2017-2021, so the value of 298 might 
be due to noise given the 
small population of about 8000 people.

\section{Conclusion}\label{sec:conclusion}

In this paper we have proposed the cellLTS
regression method for handling data that may
contain casewise outliers, cellwise outliers, 
and missing values. It is a two-step procedure
that first applies a cellwise robust estimator 
of multivariate location and covariance to the
regressor data, allowing it to clean cellwise 
outliers and to impute missing values. In the 
second step it regresses the response on the 
cleaned regressors by a ridge version of the 
casewise least trimmed squares method. In both 
steps a symmetrization is applied, to help the 
method deal with skewed data.

An additional feature is that cellLTS contains
an algorithm for robust out-of-sample 
prediction. When presented with a new set of
regressor values, it does not automatically
assume that all of its entries are 
well-behaved. Instead it investigates whether
some of its cells should be flagged and 
imputed, and it also imputes missing values.

Simulations indicate that cellLTS outperforms
the three existing methods for cellwise
robust regression, both in terms of 
accuracy of the regression coefficients and
in terms of prediction performance as
measured by mean squared error. Finally, the
method is illustrated on a substantial real
data set. In the main text the data is 
analyzed with only five regressors, and in 
Section~\ref{supp:uscancer} 
of the Supplementary Material an analysis is 
performed that takes all 27 regressors 
into account.

\vspace{5mm}
\noindent \textbf{Software availability.}
The \textsf{R} code and an example script 
reproducing the data analysis are at 
\url{https://wis.kuleuven.be/statdatascience/code/cellreg_code_script.zip}\,.

\small
\spacingset{1}
\setlength{\bibsep}{5pt plus 0.2ex}



\clearpage
\pagenumbering{arabic}
\appendix

\spacingset{1.2} 
\begin{center}

\Large{Supplementary Material to:\\ Least Trimmed
       Squares Regression\\ with Missing Values and 
       Cellwise Outliers}\\

\vspace{3mm}

\end{center}

\setcounter{equation}{0} 
\renewcommand{\theequation}
  {A.\arabic{equation}} 

\section{Proofs} \label{supp:proofs}

\noindent{\bf Proof of the breakdown value.} 
The breakdown value satisfies
$\eps^*_n \leqslant (n-h+1)/n$ because
by contaminating at least $n-h+1$ cells 
in the columns of $\bX$ we can break
down the cellMCD estimate in the first 
part of the algorithm. Indeed, we can 
replace $n-h+1$ cells in each 
$\bX_{\bcdot j}$ by 
$\hmu_j+s, \hmu_j+2s, \hmu_j+3s,\ldots$ 
for a large $s$ that we can let go to
$+\infty$, which creates at least
$(n-h+1)(n-h) +2(n-h+1)(h-1)$ far
cellwise outliers in 
$\Sym(\bX_{\bcdot j})$. This
number is larger than 
$\tn - \tth = n(n-1) - h(h-1)$, so every
$\tth$-subset of $\Sym(\bX_{\bcdot j})$
contains at least one of these cells,
so $\bhSigma(\Sym(\bX))$ explodes.

It remains to show that the breakdown 
value of cellLTS is at least $(n-h+1)/n$.
If we constrain the parameter vector to 
$\bbeta = \bzero$, on the clean data the 
maximal value of the objective function 
over all $\tth$-subsets $\ttH$ with
$\tth = h(h-1)$ is
\begin{equation}\label{eq:ridgemax2}
  M \coloneqq \max_{\ttH} 
  \sum_{\ell \in \ttH}
  r^2(\ty_{\ell}, \btx_{\ell},
  \bbeta = \bzero) + \lambda||\bzero||_2^2
  = \max_{\ttH} \sum_{\ell \in \ttH}
	\ty_{\ell}^2 < \infty
\end{equation}
where the maximum is over all 
$\tth$-subsets $\tH$ of 
$\{1,\ldots,\tn\}$.

Next we contaminate $\bX$ to $\bX^\m$ by 
replacing at most $n-h$ cells in any 
column. It follows that each column 
$\Sym(\bX^{\m}_{\bcdot j})$ of the 
symmetrized data contains at least
$\tth = h(h-1)$ values from the
uncontaminated $\Sym(\bX_{\bcdot j})$,
so the cellMCD estimates 
$\bhmu^{\m}_{\bX}$ and 
$\bhSigma^{\m}_{\bX}$ do not break down.
We then compute the imputed $\Ximp^\m$ 
as in~\eqref{eq:Ximp}.
Note that step (e) scales the entries 
in $\Sym(\bX^{\m}_{\bcdot j})$ by
$\hsigma_j^\m \coloneqq 
\sqrt{(\bhSigma_{\bX}^\m)_{jj}}$\,.
Since $\bhSigma_{\bX}^\m$ does not break
down, for any $j=1,\ldots,\p$ there exist
some $0 < a_j < A_j < \infty$ such that 
$a_j \leqslant \hsigma_j^\m \leqslant A_j$ 
where $a_j$ and $A_j$ only depend on
the uncontaminated $\bX$.
In step (e) we also compute $\hsigma_y^\m$.
Analogously there exist some 
$0 < a_y < A_y < \infty$ such that 
$a_y \leqslant \hsigma_y^\m \leqslant A_y$\,,
yielding the scaled $\bty^\m$.

The contaminated response $\bty^\m$ must 
contain at least $\tth$ of the original 
$\ty_i$\,, so after scaling we have that
$\tth$ of the new responses 
$\ty_i^\m$ equal $c_y \ty_i$\,, where
$c_y := \hsigma_y/\hsigma_y^\m$ is a
constant with $c_y \leqslant \hsigma_y/a_y$.
So after symmetrization we have 
$\tth = h(h-1)$ values that are $c_y$
times the original symmetrized response 
values. Let us
denote the corresponding set of indices
by $\ttH' \subset \{1,\ldots,\tn\}$. 
For all $\ell$ in $\ttH'$ we thus have 
$\ty^\m_\ell = c_y \ty_\ell$\,.
We now regress $\bty^\m$ on $\btX^\m$.
Under the constraint $\bbeta = \bzero$, the 
minimal objective value $V(\bzero)$ on the
contaminated data satisfies
\begin{align*}\label{eq:minObj}
  V(\bzero) &= \min_{\ttH} \sum_{\ell \in \ttH}
  r^2(\ty^{\,\m}_{\ell}, \btx_{\ell}^\m, \bzero)
  + \lambda||\bzero||_2^2\;  \leqslant
  \Big[\sum_{\ell \in \ttH} r^2(\ty_{\ell}^{\,\m},
  \btx_{\ell}^\m, \bzero)\Big]_{\ttH = \ttH'}\\
	 &= \sum_{\ell \in \ttH'} (\ty_{\ell}^{\,\m})^2 
  = \sum_{\ell \in \ttH'} c_y^2 \ty_{\ell}^2
  \;\leqslant\; c_y^2 M\;.
\end{align*} 
where the last inequality is due 
to~\eqref{eq:ridgemax2}.

We now want to minimize the objective without 
the constraint $\bbeta = \bzero$. For any
candidate $\bbeta$ with 
$||\bbeta||_2 > c_y \sqrt{M/\lambda}$ it
holds that 
\begin{align*}
V(\bbeta) = \min_{\ttH} \sum_{\ell \in \ttH}
  r^2(\ty^{\,\m}_{\ell}, \btx_{\ell}^\m, 
  \bzero) + ||\bbeta||_2^2 
  \geqslant ||\bbeta||_2^2 > c_y^2 M
\end{align*} 
which exceeds $V(\bzero)$, so $\bbeta$ 
cannot be a minimizer. Therefore we 
can restrict the search space to the
closed ball $B(\bzero,c_y \sqrt{M/\lambda})$
around the origin. Since this set is compact
and $V(\bcdot)$ is continuous, there exists
a minimizer that we denote as $\bhbeta^*$,
with \mbox{$||\bhbeta^*||_2 \leqslant 
c_y\sqrt{M/\lambda}$\,.} 
Next we need to transform the linear regression
(prediction) formula back to the units of the
original data. Undoing the standardization of 
the response in step (e) multiplies $\ty^{\,\m}$ 
by $\hsigma_y^\m$\,, so it multiplies 
$\bhbeta^*$ by $1/\hsigma_y^\m\leqslant 1/a_y$\,. 
Moreover, transforming back the standardization 
of the $j$-th regressor in step (e) multiplies 
it by $\hsigma_j^{\,\m}$\,, which 
multiplies the slope $\hbeta_j$ by 
$1/\hsigma_j^{\,\m} \leqslant 1/a_j$\,.
Doing this for all slopes multiplies the 
norm of $\bhbeta^*$ by at most $1/a_{\min}$
where $a_{\min} := \min(a_1,\ldots,a_\p) > 0$.
Combining these effects yields
$$||\bhbeta||_2  \leqslant 
 \frac{c_y\sqrt{M/\lambda}}{a_{\min} a_y}
 \leqslant \frac{\hsigma_y\sqrt{M/\lambda}}
  {a_{\min} a_y^2} \,,$$
in which the finite upper bound on the right 
only depends on the uncontaminated data. 
Therefore $\bhbeta$ does not break down,
hence $\eps^*_n \geqslant (n-h+1)/n$.
Since we saw before that 
$\eps^*_n \leqslant (n-h+1)/n$, the breakdown 
value of cellLTS equals $(n-h+1)/n$.
\hfill $\square$\\

\noindent \textbf{Proof of the Corollary.}
This is because the casewise breakdown value
is defined by replacing $m$ entire rows of
the data, which is a special case of 
replacing $m$ cells in each column. So the
casewise breakdown value is always at least 
as high as the cellwise one. The casewise 
breakdown value of cellLTS is also not 
higher than this, because the reasoning in 
the above proof showing that 
$\eps^*_n \leqslant (n-h+1)/n$
still holds when replacing cases.
\hfill $\square$\\

\clearpage
\section{Additional simulation results}
\label{supp:sim}

Section~\ref{sec:simulation} in the main 
text has presented simulation results
for $n = 400$ cases with $\p = 20$
regressors, with a fraction $\eps = 20\%$ 
of cells at random positions in each 
column of the data including the response.
The results for the other combinations of
$n$, $\p$ and $\eps$ are shown here.
We see that they are all qualitatively 
similar to those in 
section~\ref{sec:simulation}.

\subsection{Results for \texorpdfstring{$n=100$,
$\p=10$, and $\eps=10\%$}{n = 100, p = 10, and 
eps = 10\%}}

\begin{figure}[!ht]
\centering
\vspace{-4mm}
\includegraphics[width = 0.37\columnwidth]
{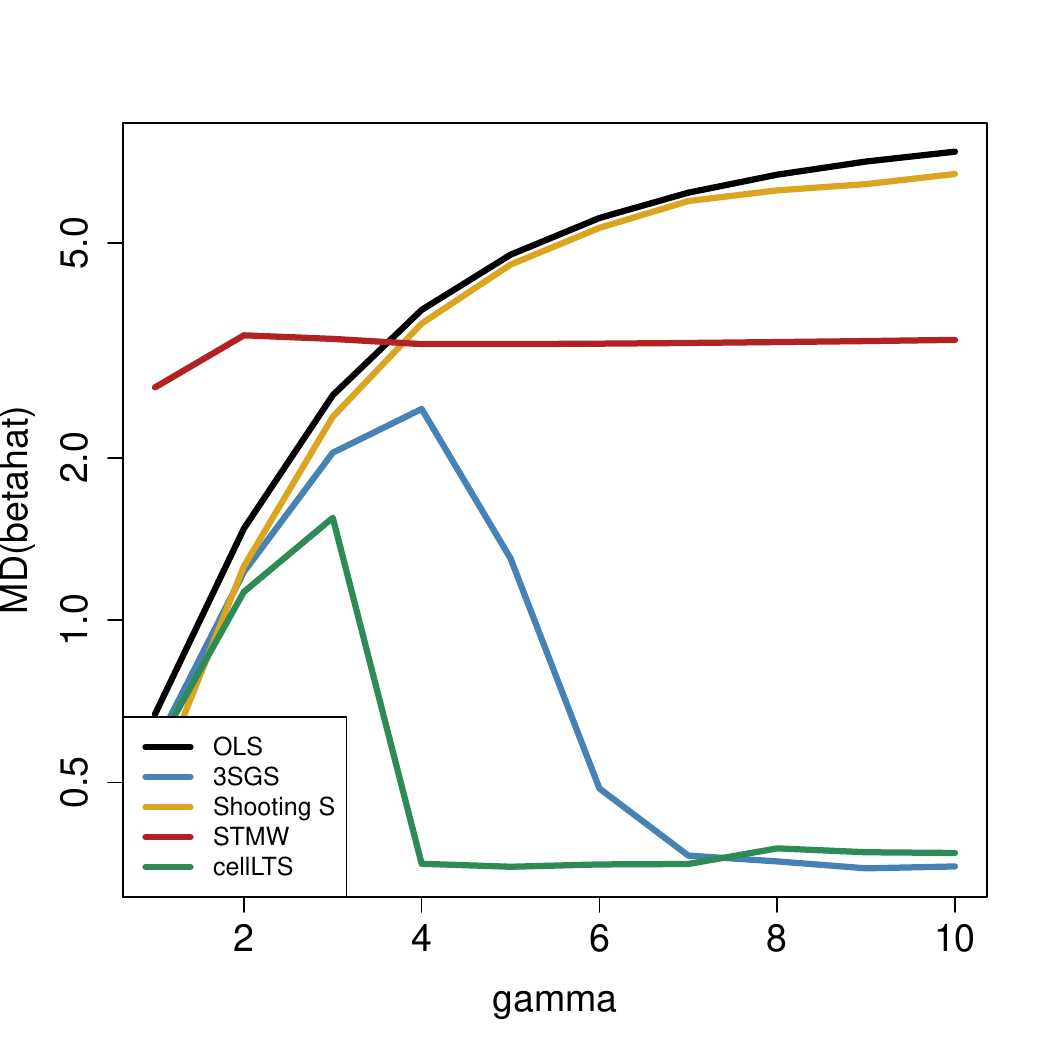}
\includegraphics[width = 0.37\columnwidth]
{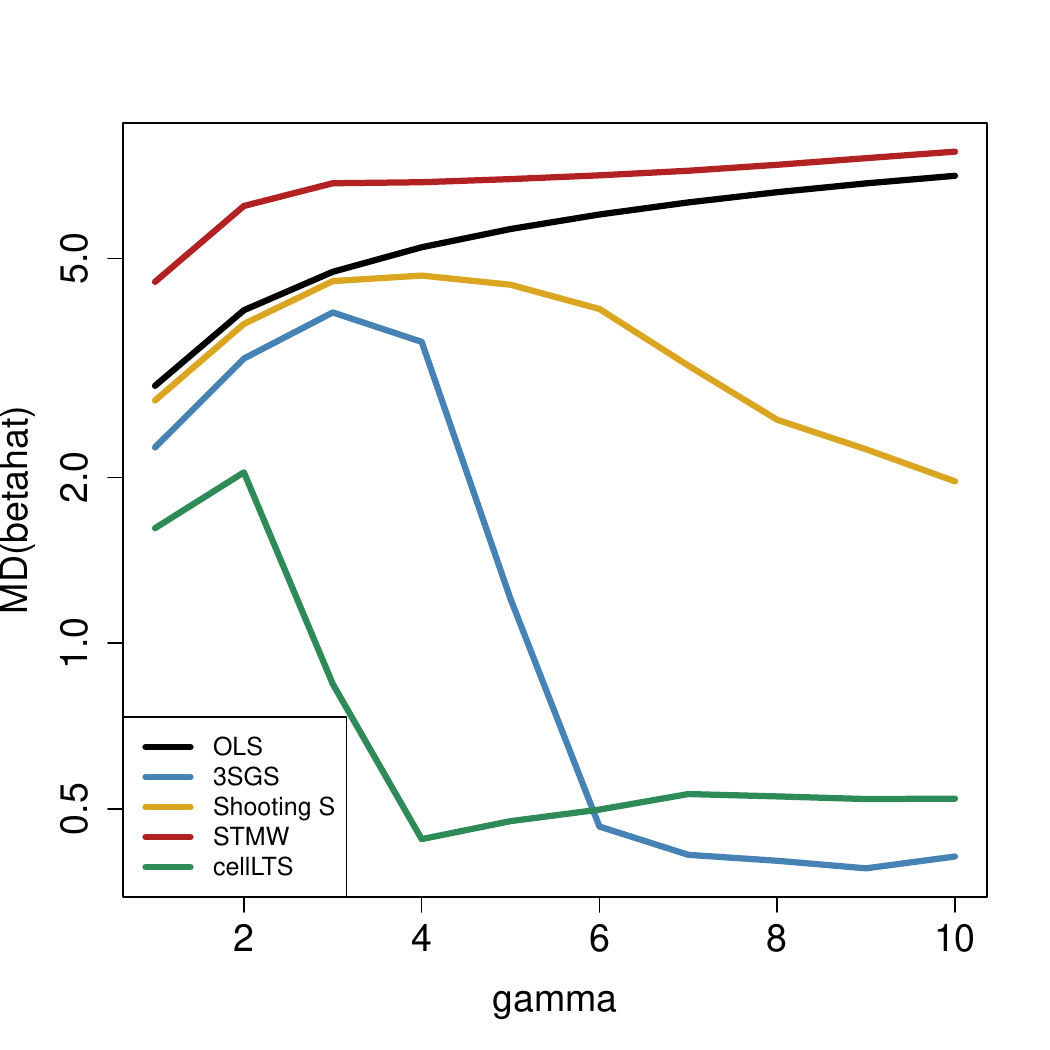}\\
\vspace{-6mm}
\includegraphics[width = 0.37\columnwidth]
{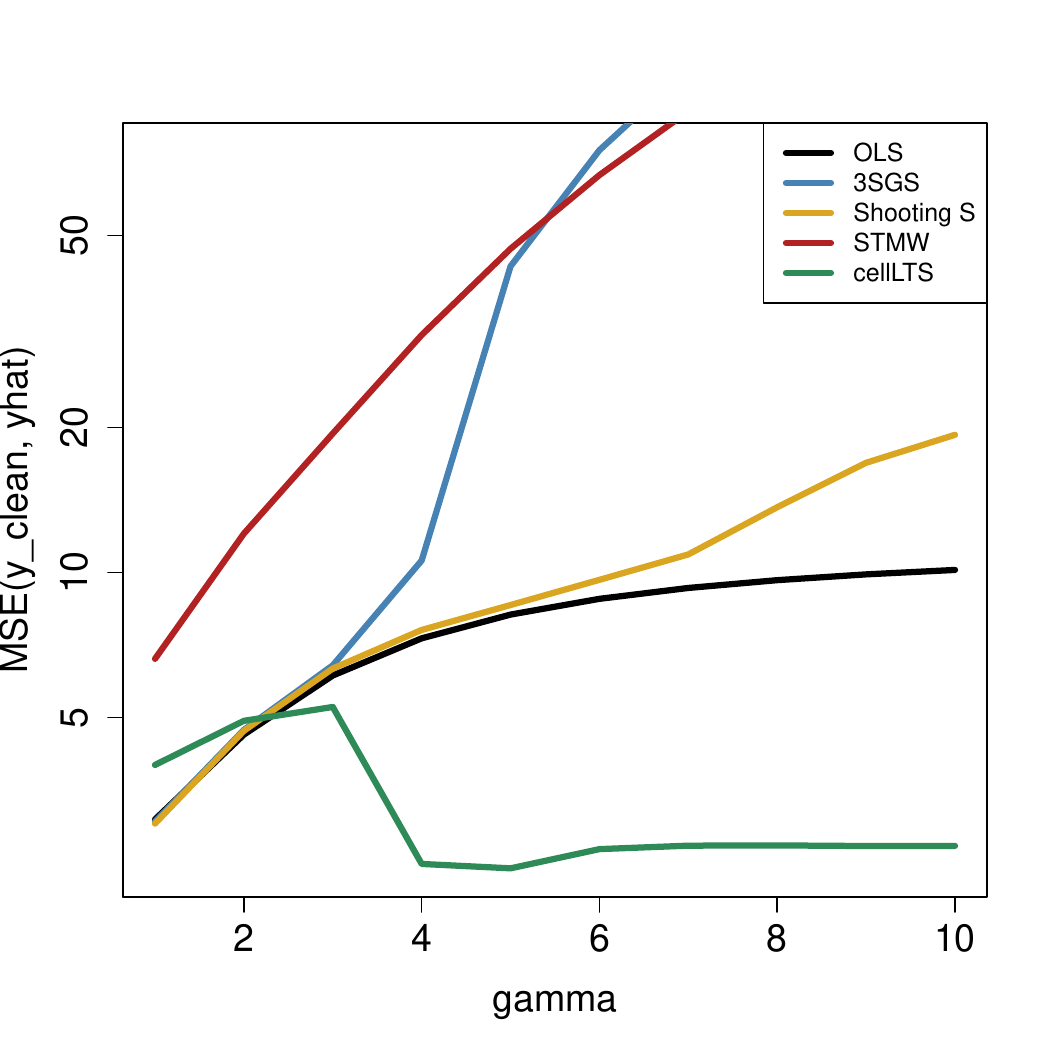}
\includegraphics[width = 0.37\columnwidth]
{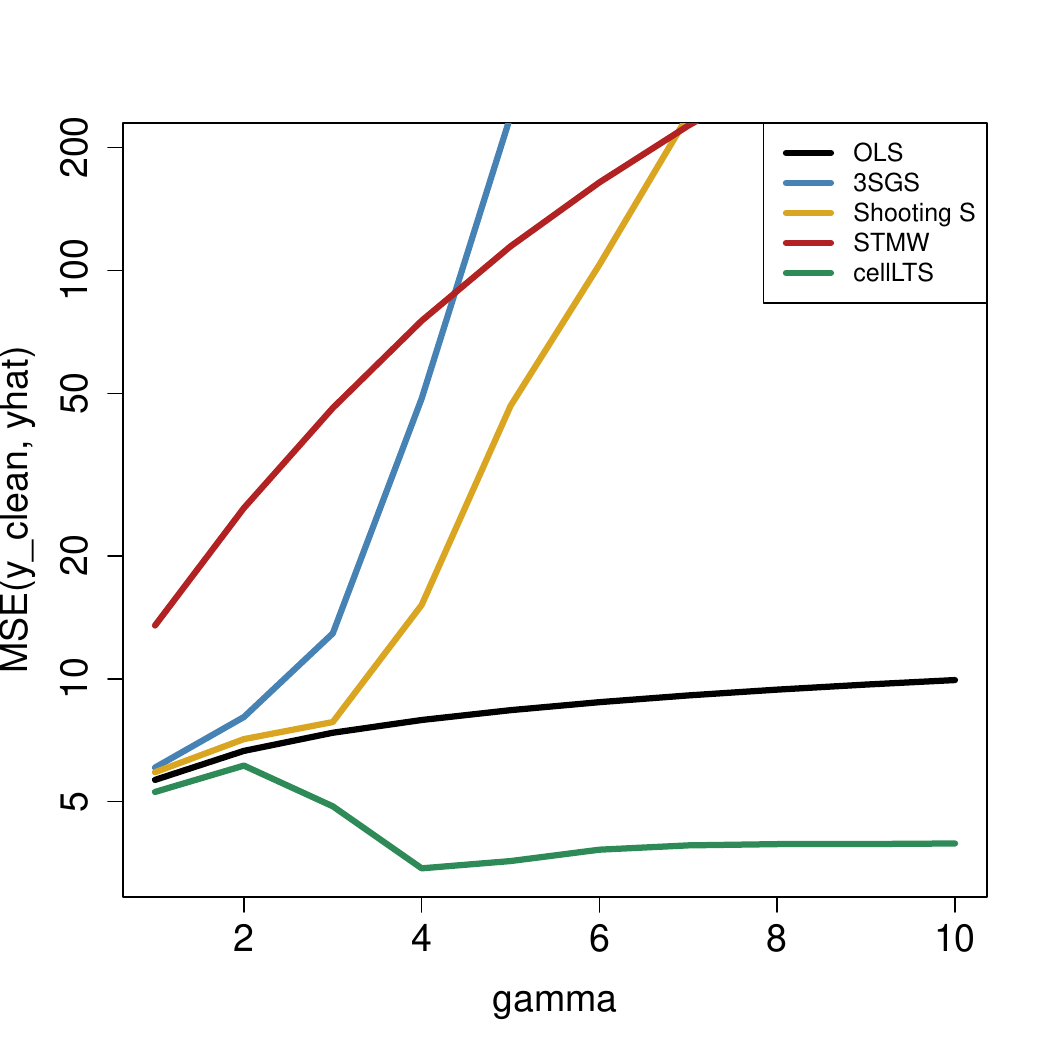}\\
\caption{Top: average MD (on log scale) of the 
estimated coefficients for $n = 100$, dimension
$\p = 10$, 
$\eps = 10\%$ of cellwise outliers, and 
$\bSigma = \bSigma_{\ALYZ}$ (left) or 
$\bSigma = \bSigma_{\AN}$ (right), for normal
predictors.
Bottom: corresponding MSE, also on log scale.}
\label{fig:MD_MSE_d10_e10_normal}
\end{figure}

\begin{figure}[!ht]
\centering
\vspace{-4mm}
\includegraphics[width = 0.45\columnwidth]
{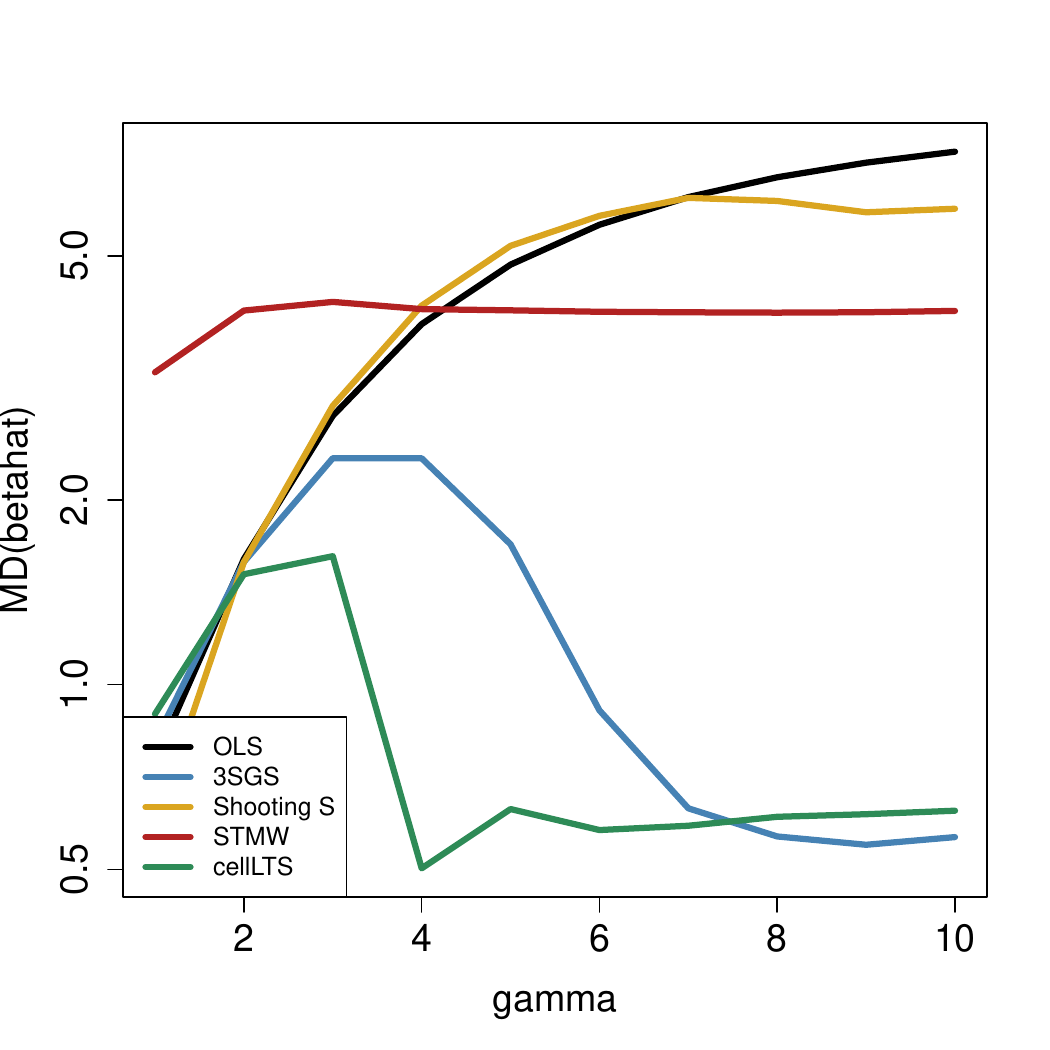}
\includegraphics[width = 0.45\columnwidth]
{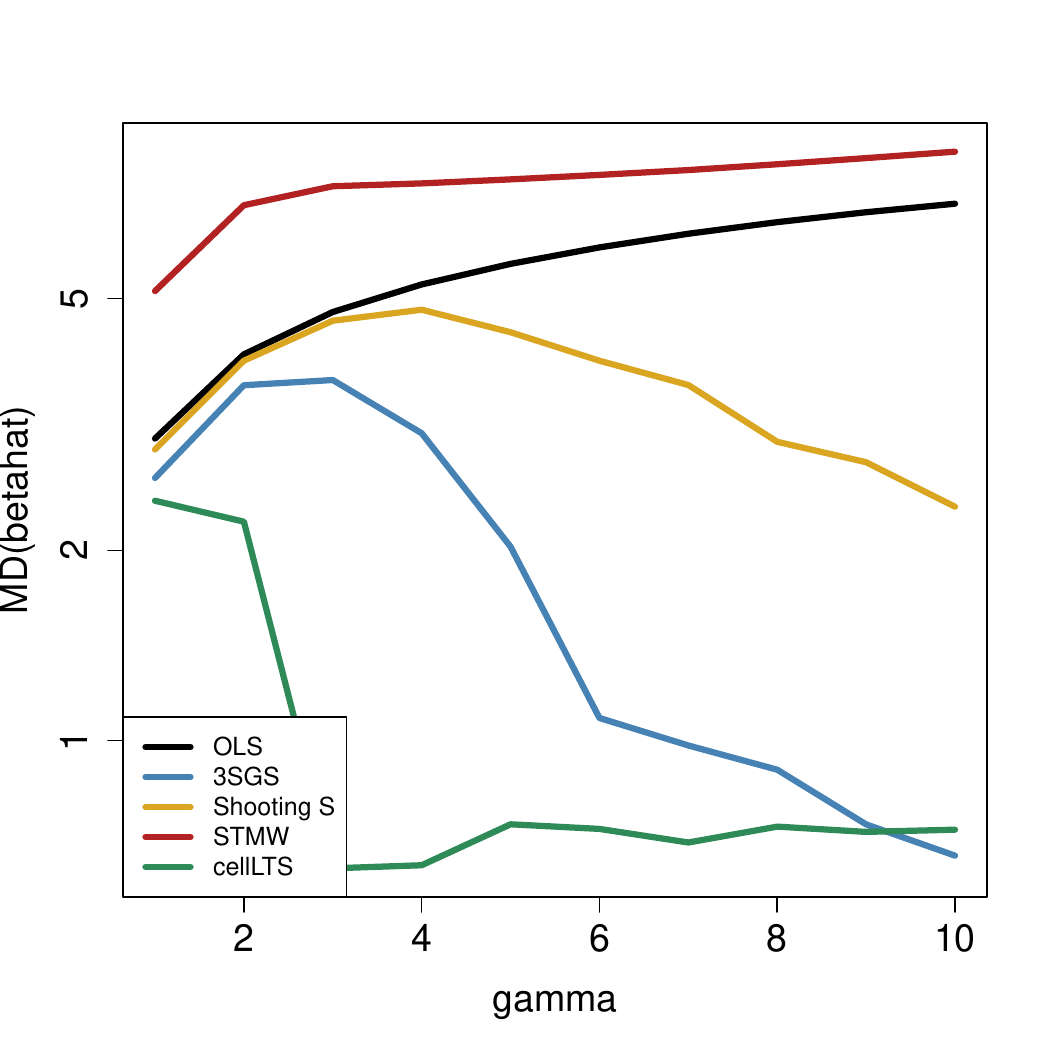}\\
\vspace{-6mm}
\includegraphics[width = 0.45\columnwidth]
{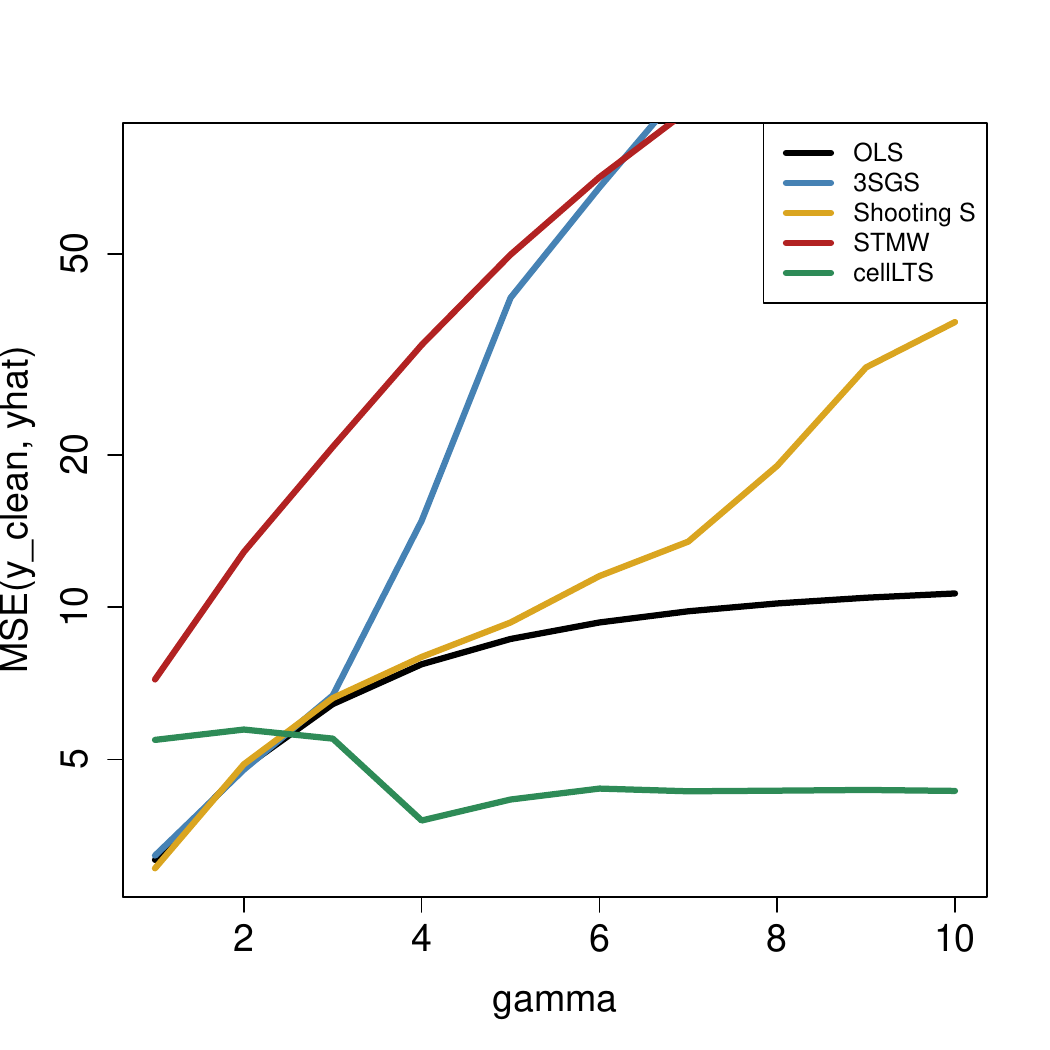}
\includegraphics[width = 0.45\columnwidth]
{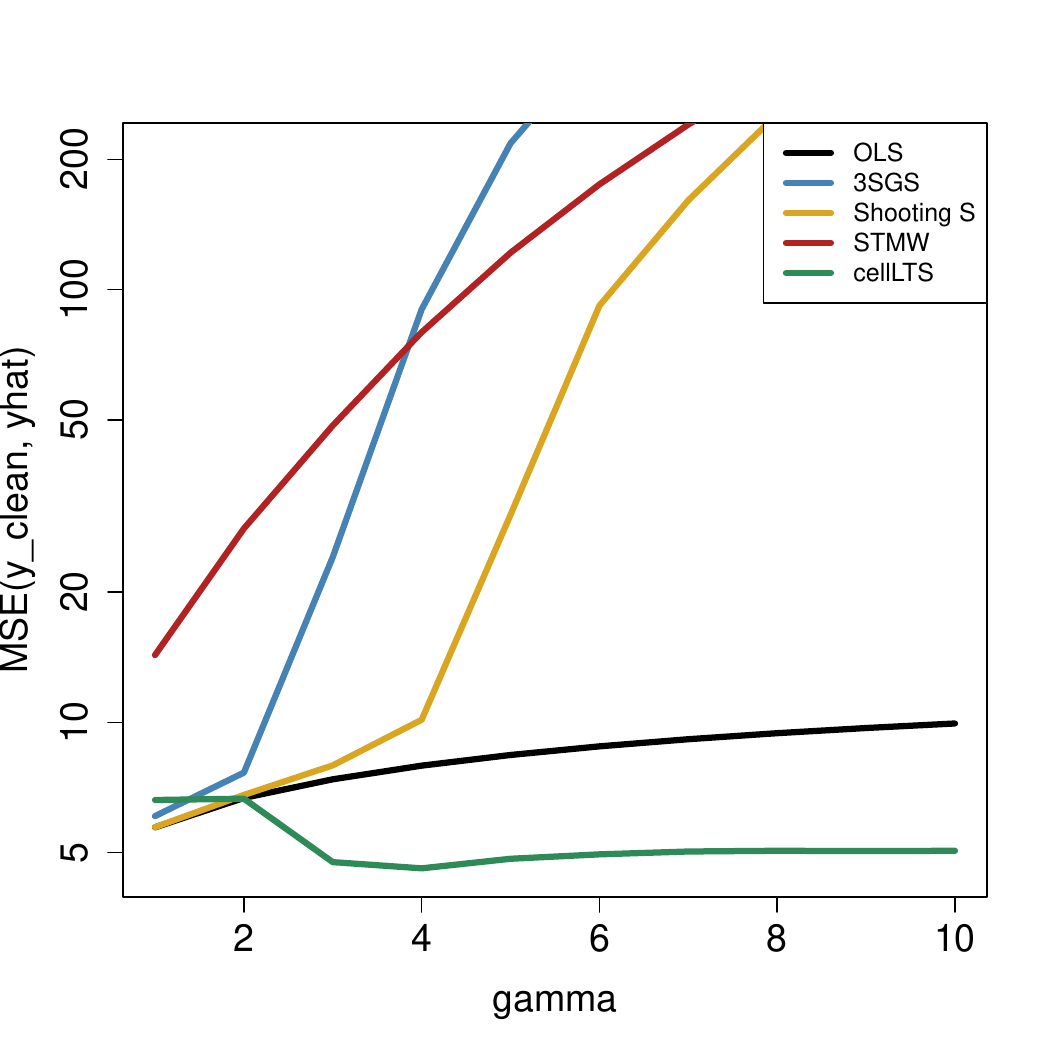}\\
\caption{Like Figure~\ref{fig:MD_MSE_d10_e10_normal}, 
but for exponential predictors.}
\end{figure}

\begin{figure}[!ht]
\centering
\vspace{-4mm}
\includegraphics[width = 0.45\columnwidth]
{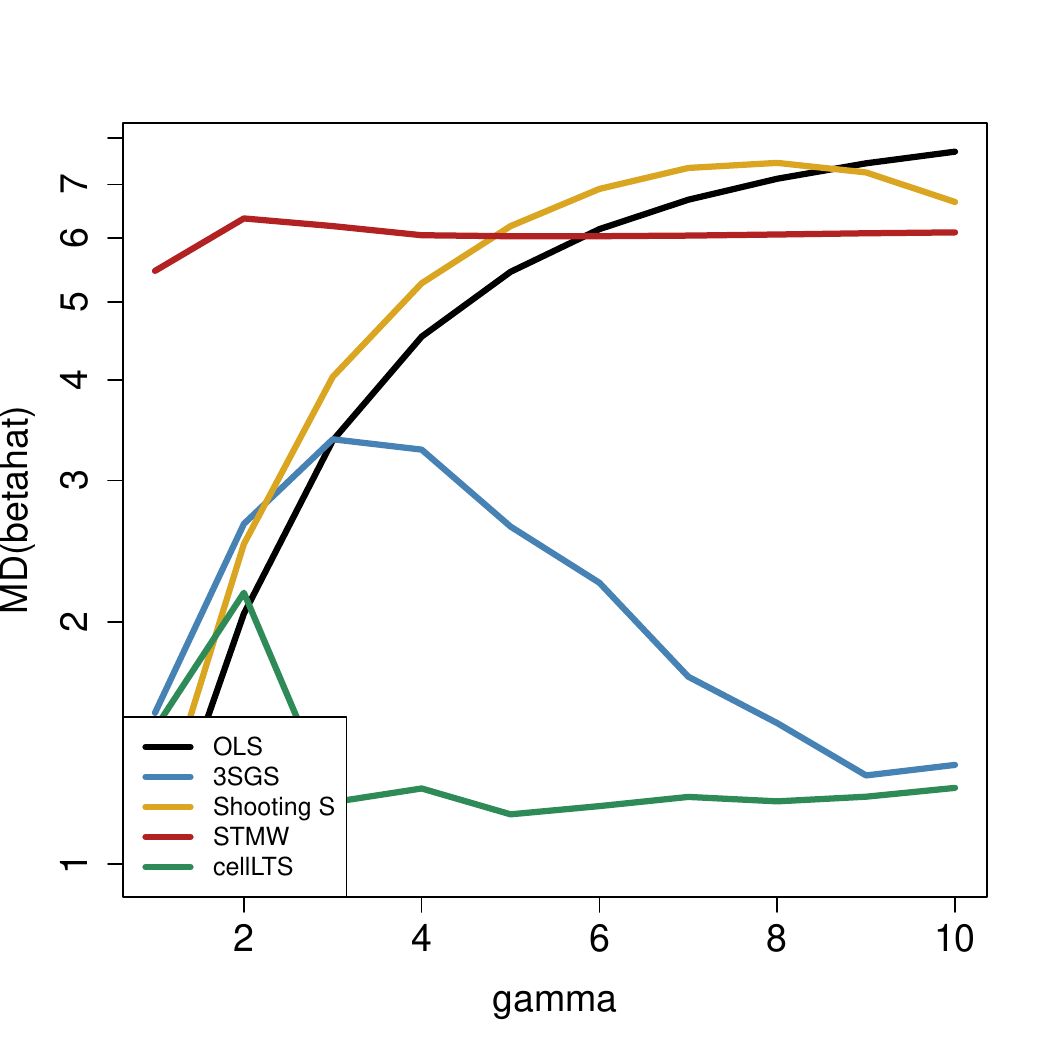}
\includegraphics[width = 0.45\columnwidth]
{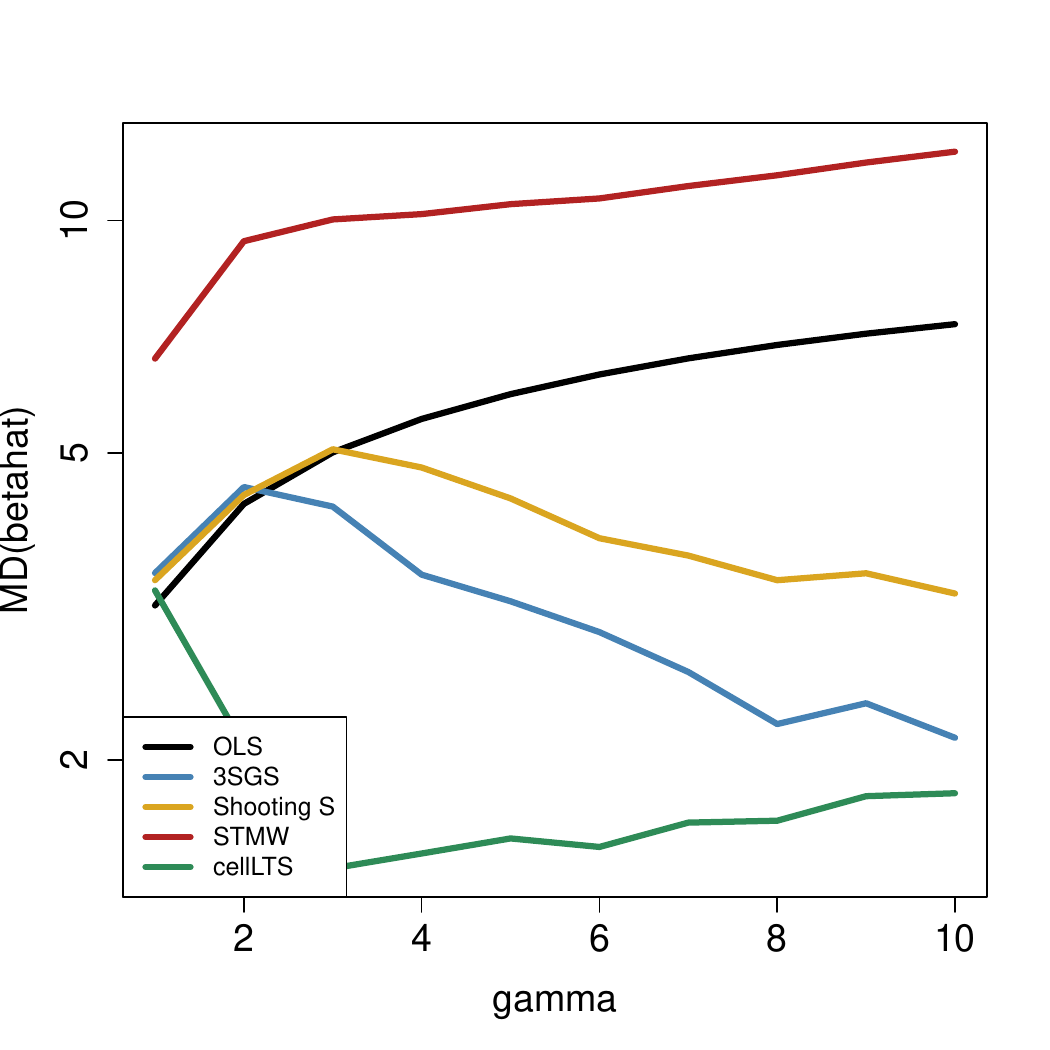}\\
\vspace{-6mm}
\includegraphics[width = 0.45\columnwidth]
{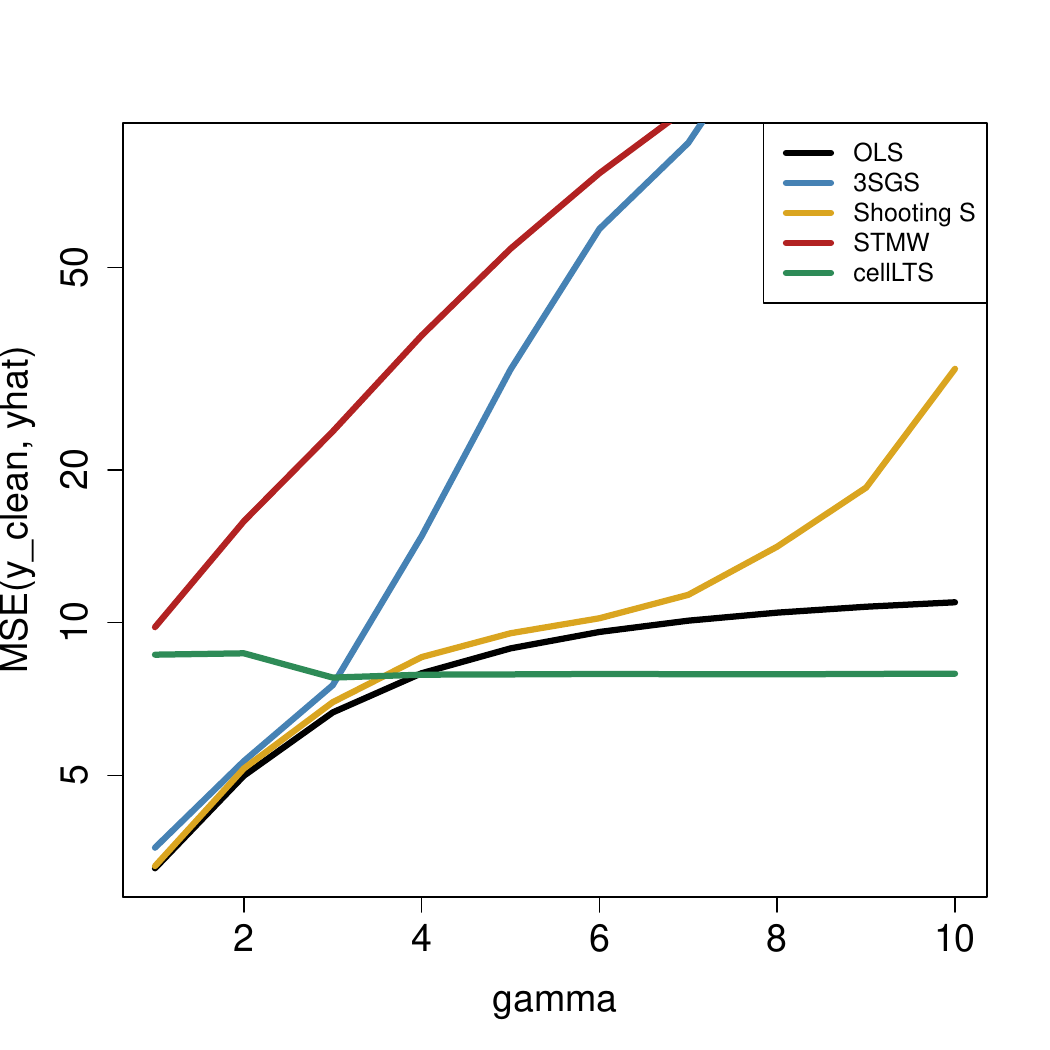}
\includegraphics[width = 0.45\columnwidth]
{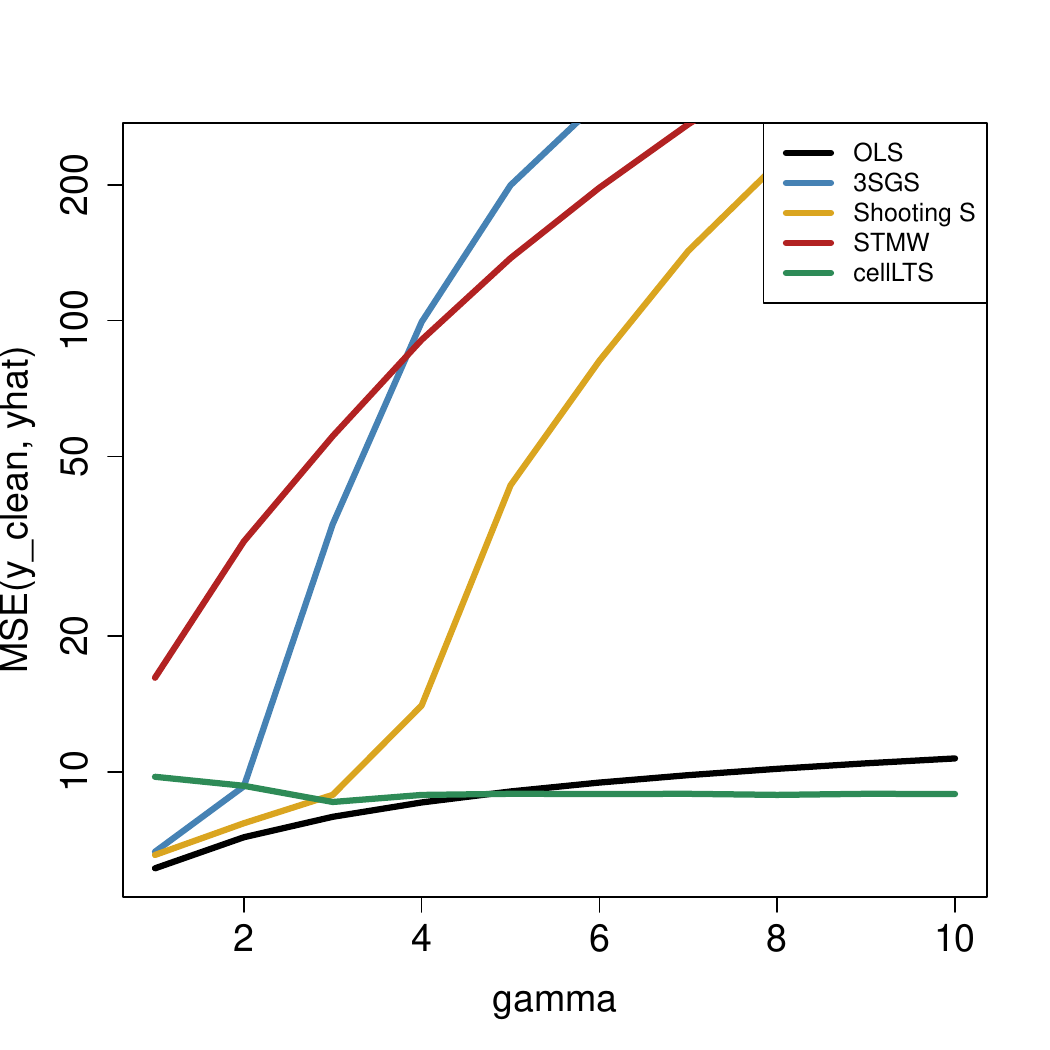}\\
\caption{Like Figure~\ref{fig:MD_MSE_d10_e10_normal}, 
but for lognormal predictors.}
\end{figure}

\clearpage

\begin{figure}[!ht]
\centering
\vspace{-4mm}
\includegraphics[width = 0.45\columnwidth]
{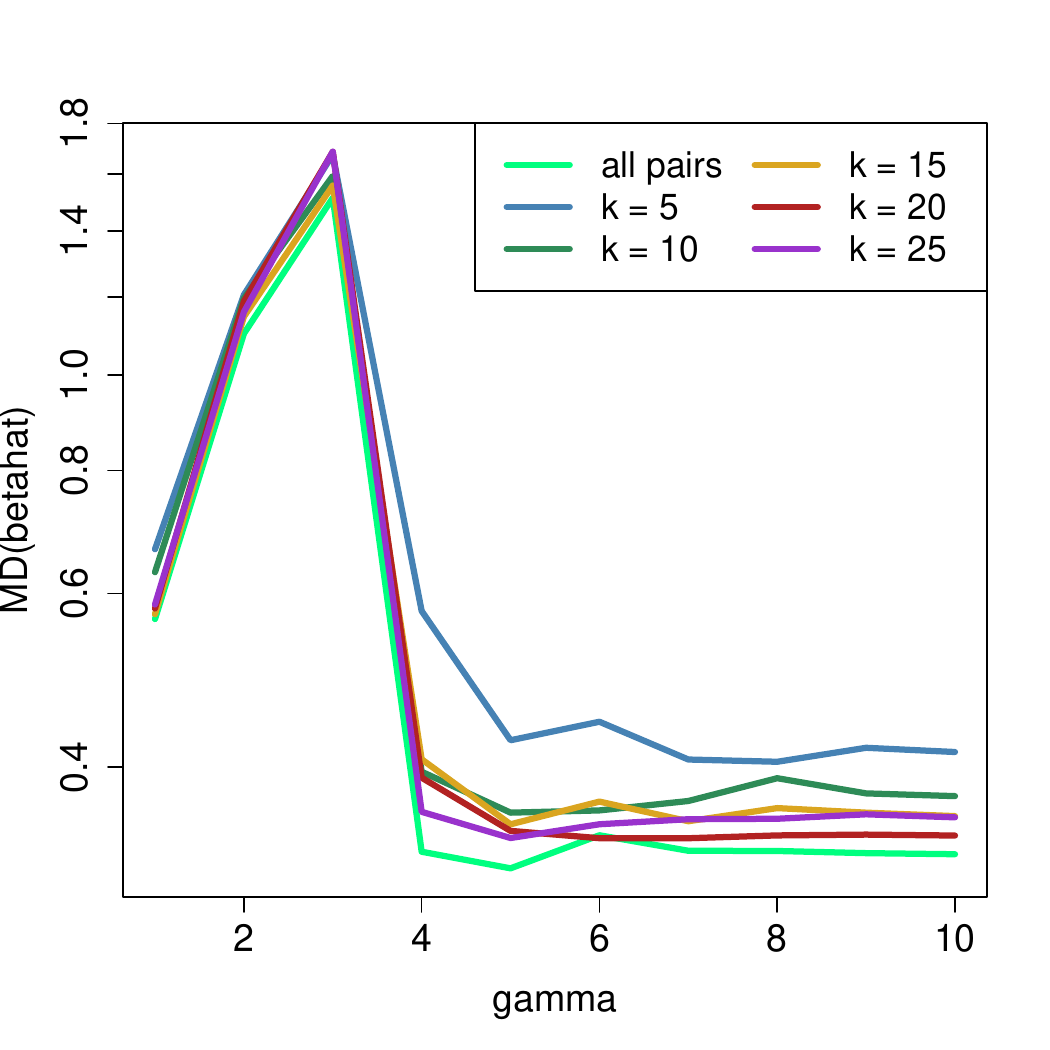}
\includegraphics[width = 0.45\columnwidth]
{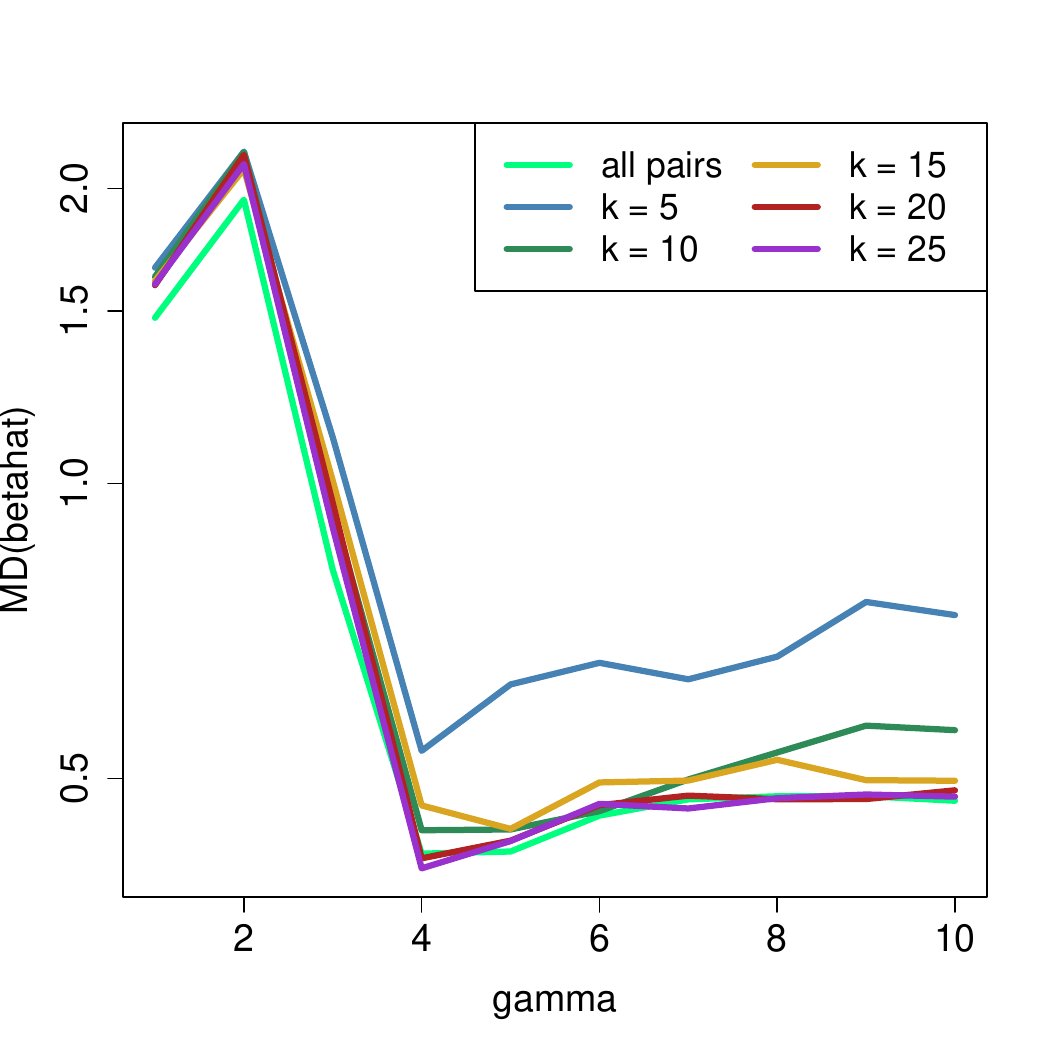}\\
\vspace{-6mm}
\includegraphics[width = 0.45\columnwidth]
{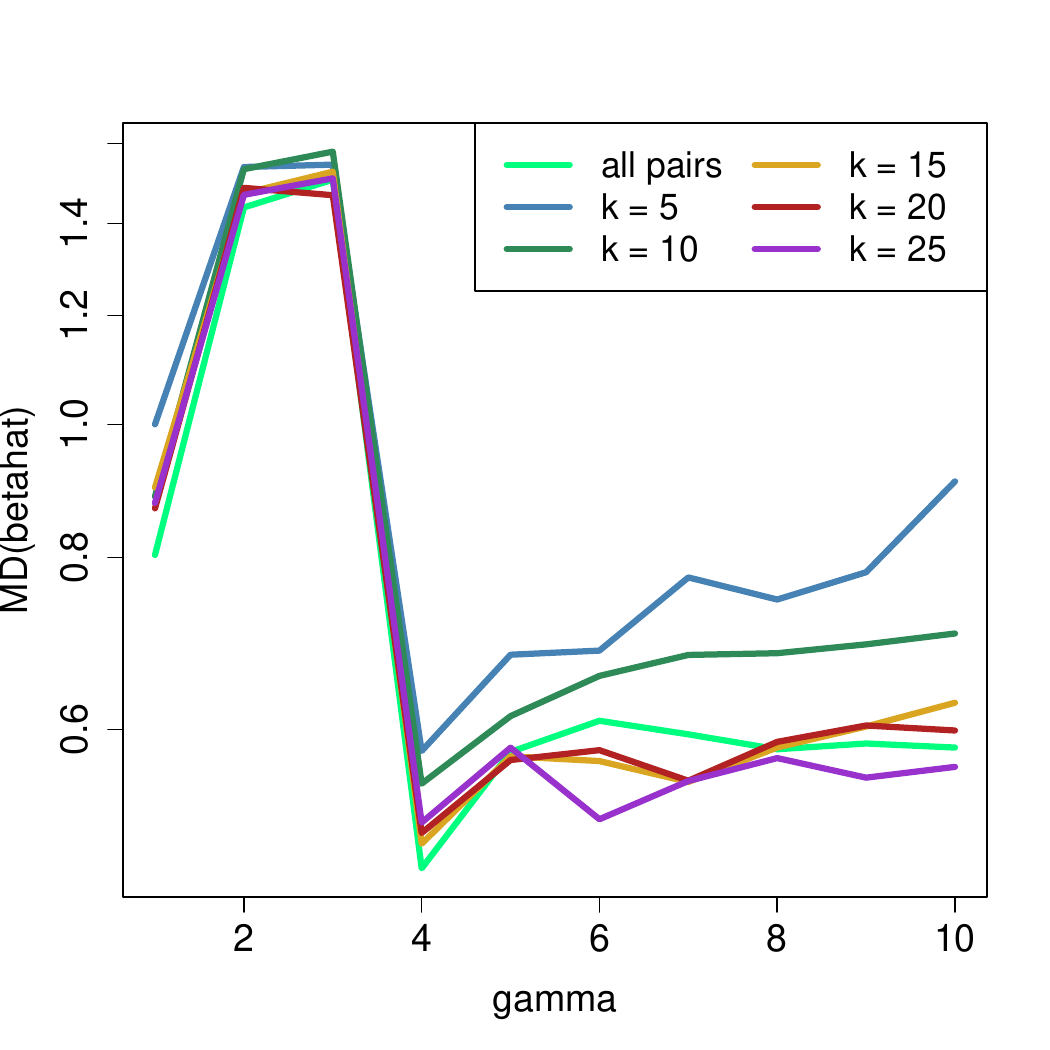}
\includegraphics[width = 0.45\columnwidth]
{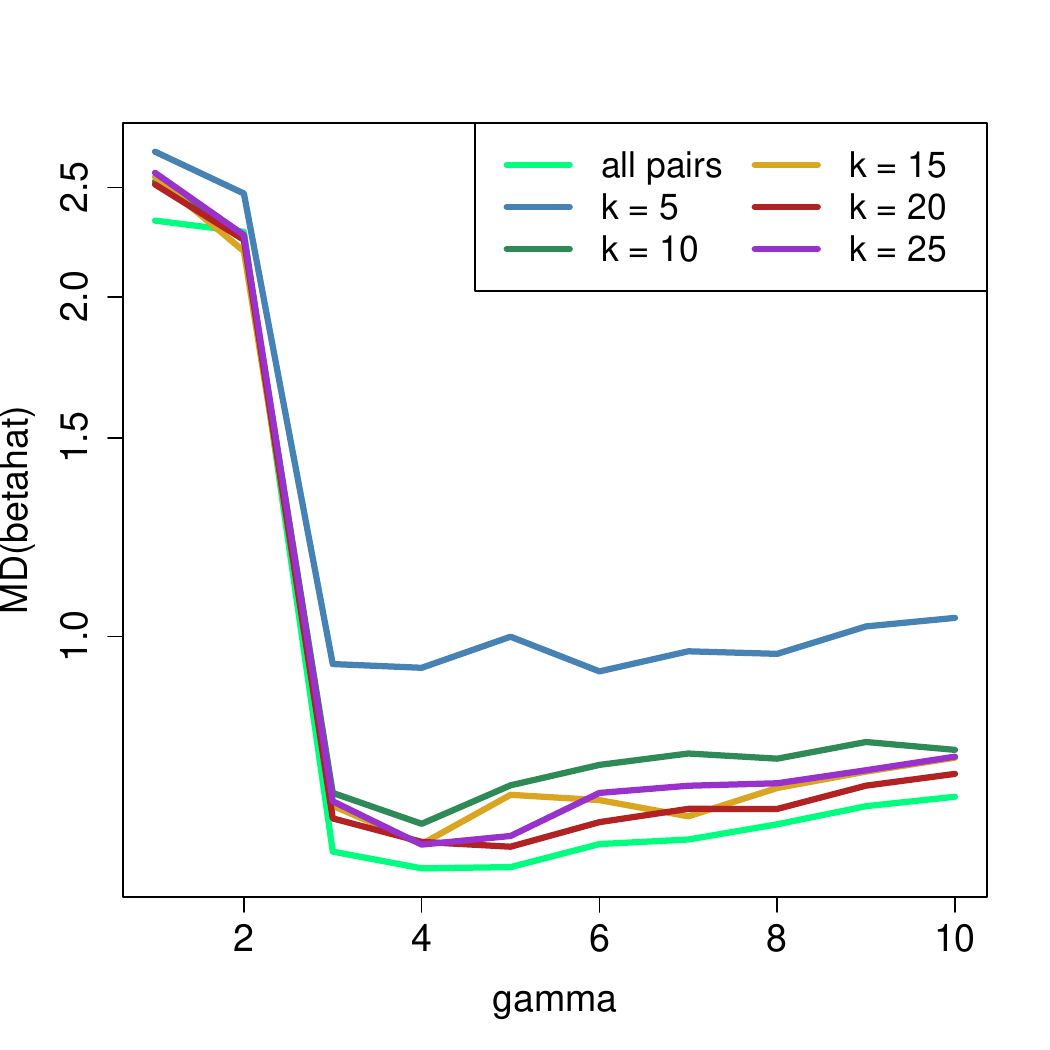}\\
\vspace{-6mm}
\includegraphics[width = 0.45\columnwidth]
{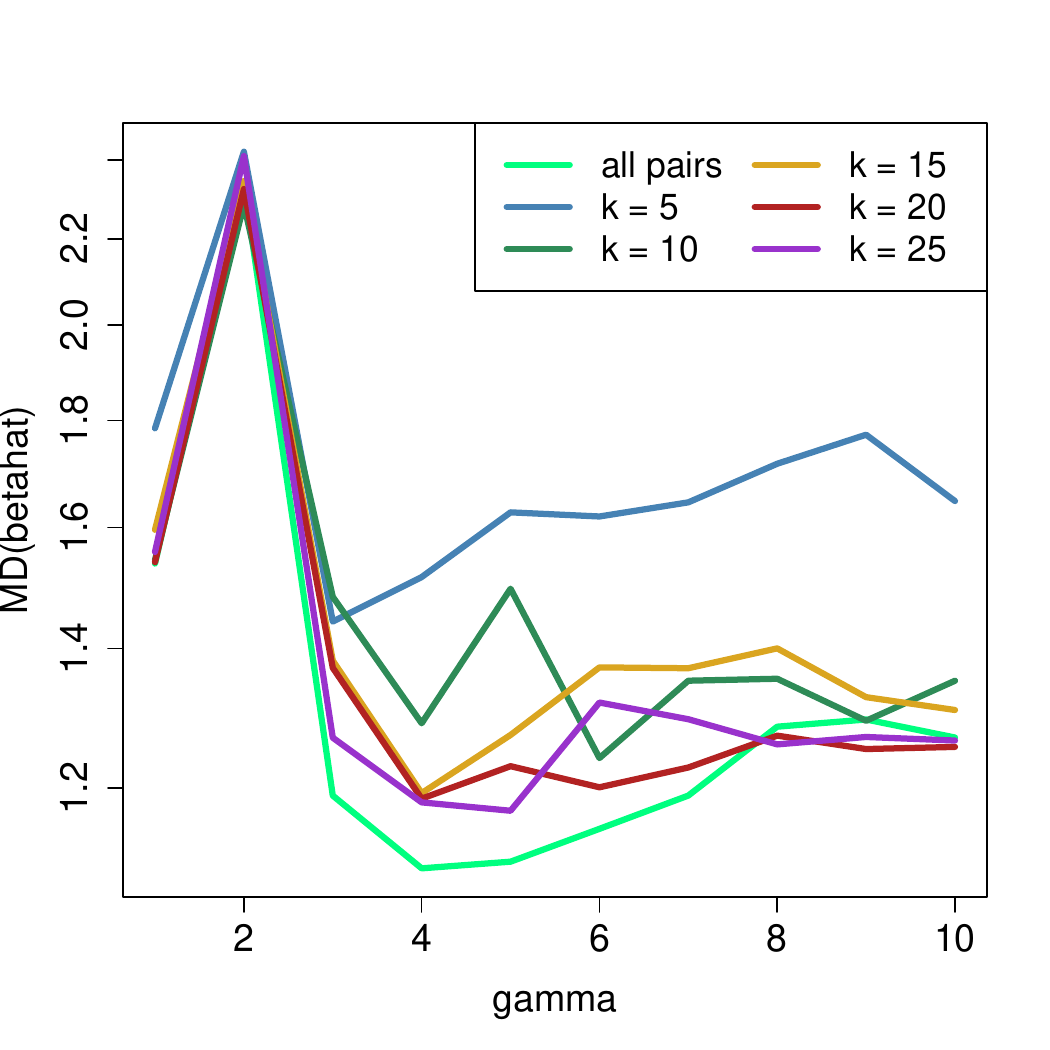}
\includegraphics[width = 0.45\columnwidth]
{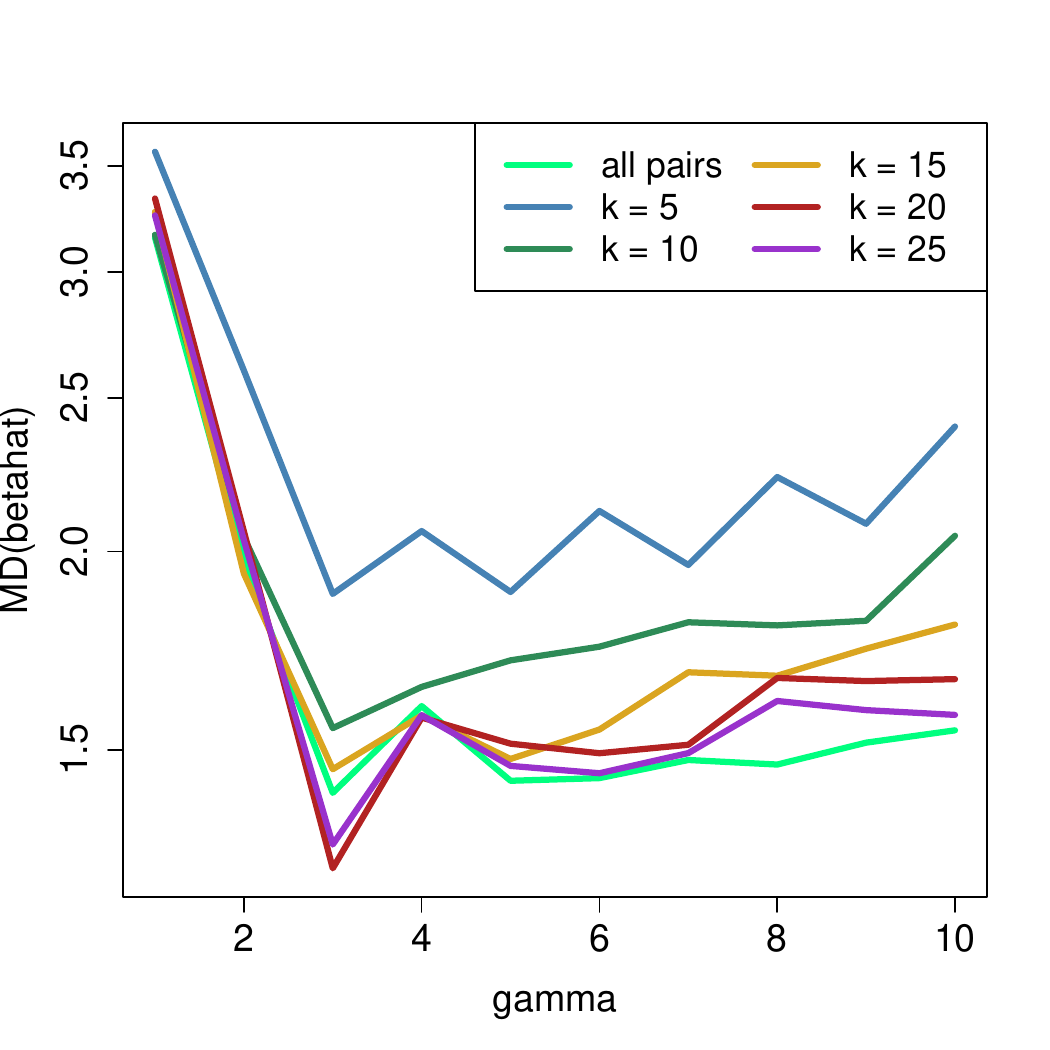}\\
\caption{Top row: average MD (on log scale) of the 
estimated coefficients for different symmetrization 
strategies and normal predictors. 
The data has $n=100$, $\p = 10$, 
$\eps = 10\%$ of cellwise outliers, and 
$\bSigma = \bSigma_{\ALYZ}$ (left) or 
$\bSigma = \bSigma_{\AN}$ (right).
Middle row: same for exponential predictors.
Bottom row: same for lognormal predictors.}
\end{figure}

\clearpage
\subsection{Results for \texorpdfstring{$n=100$,
$\p=10$, and $\eps=20\%$}{n = 100, p = 10, and 
eps = 20\%}}

\begin{figure}[!ht]
\centering
\vspace{-4mm}
\includegraphics[width = 0.45\columnwidth]
{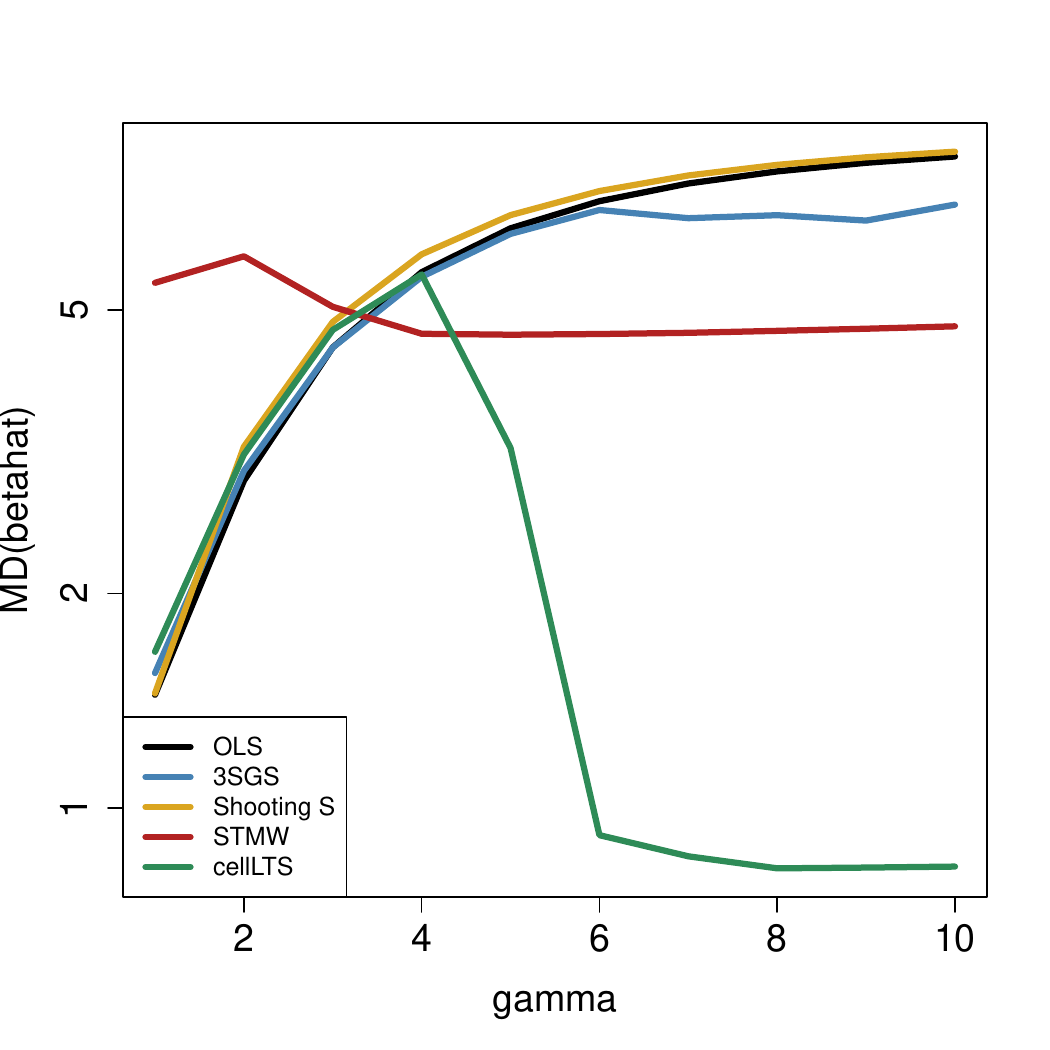}
\includegraphics[width = 0.45\columnwidth]
{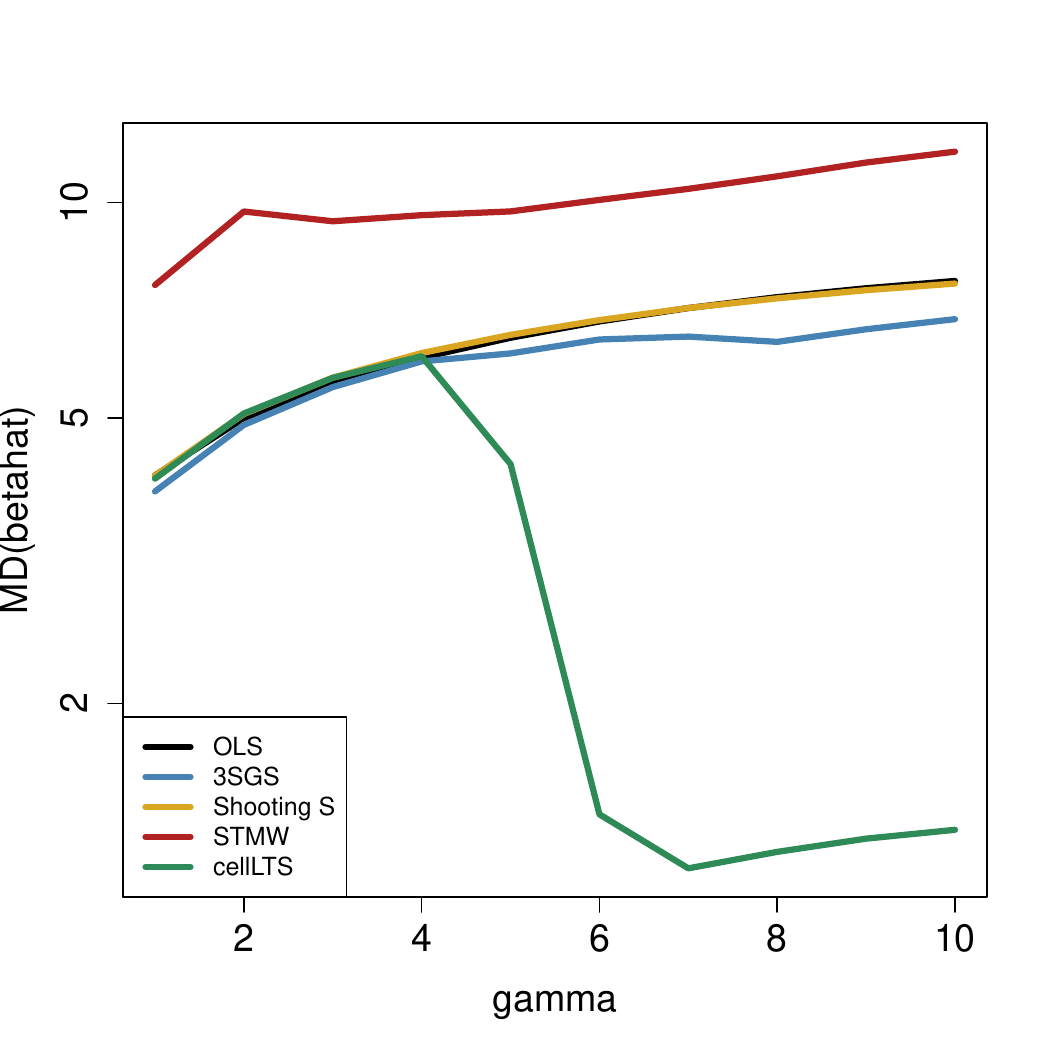}\\
\vspace{-6mm}
\includegraphics[width = 0.45\columnwidth]
{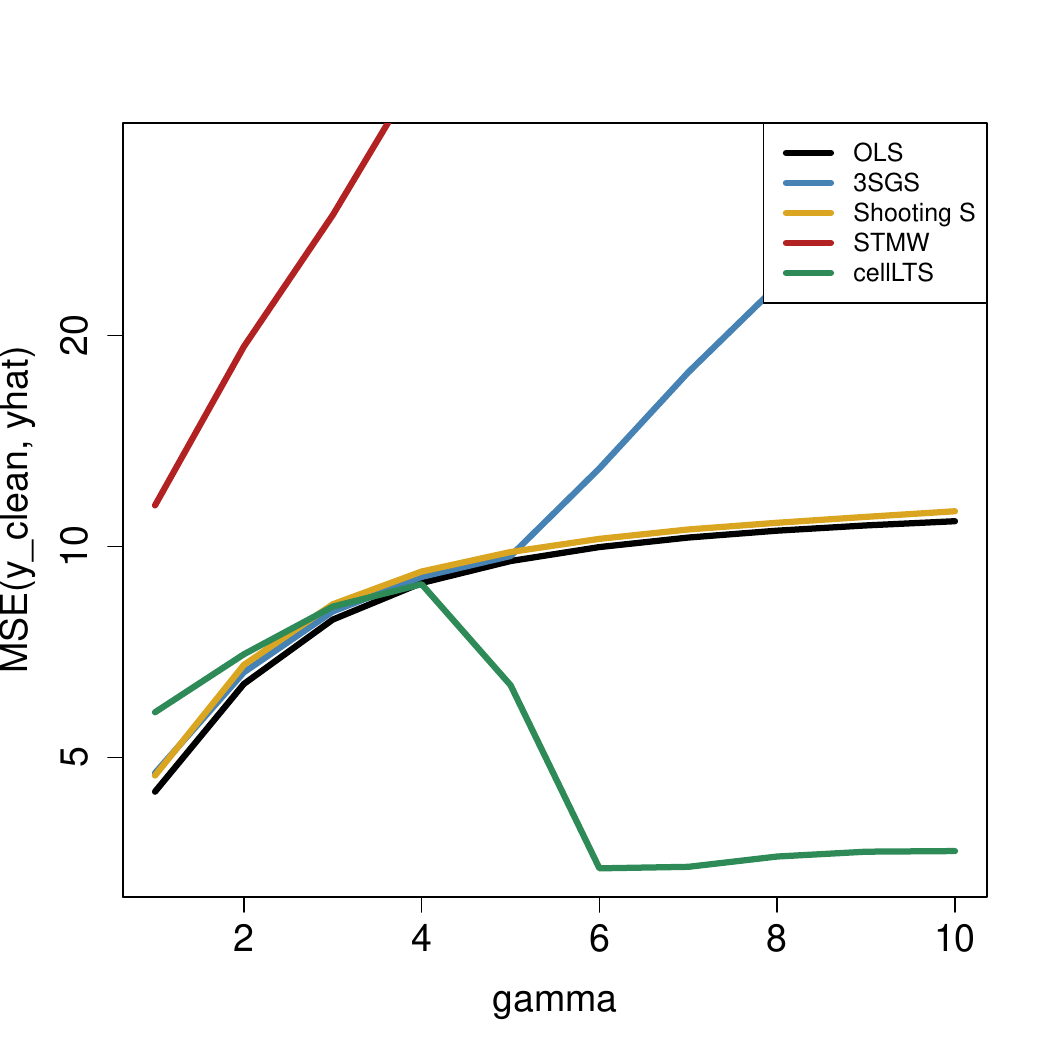}
\includegraphics[width = 0.45\columnwidth]
{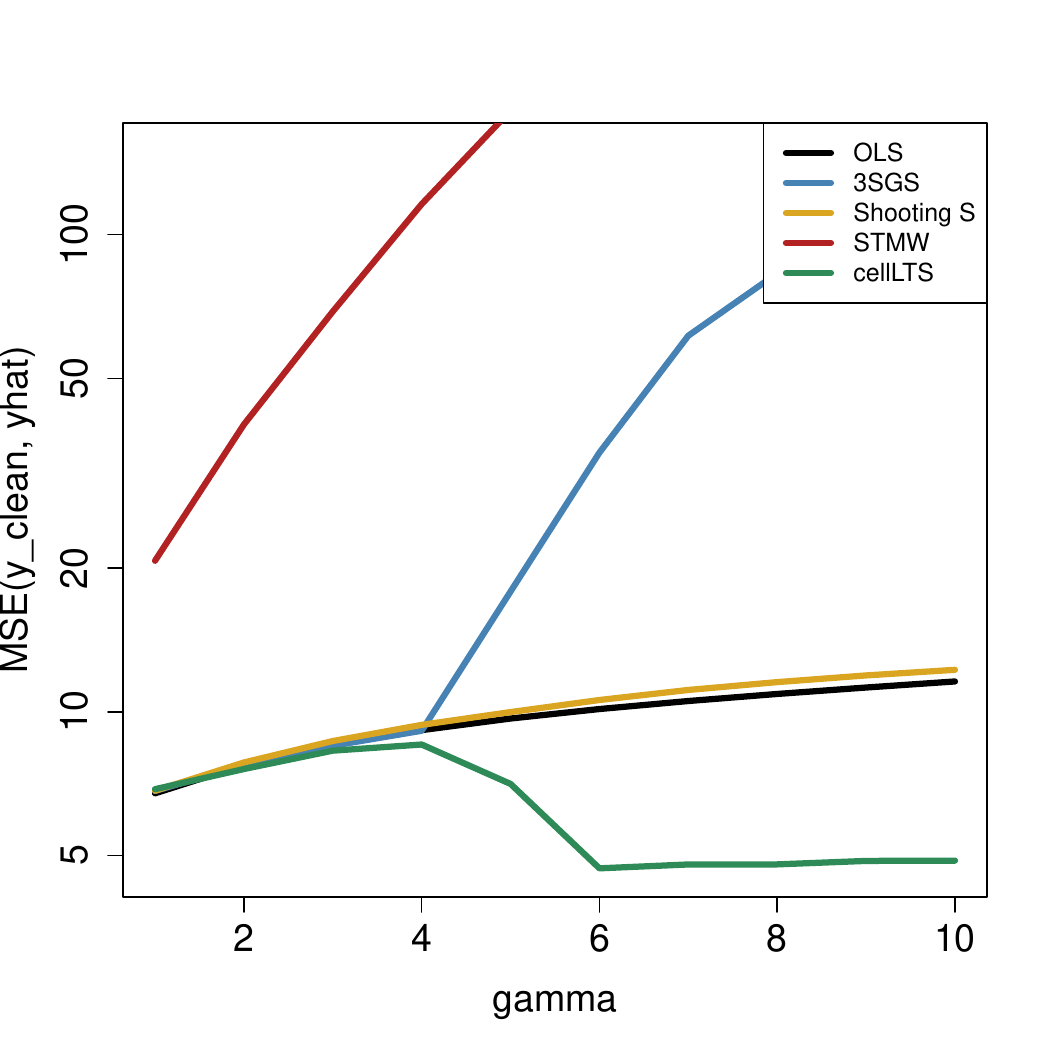}\\
\caption{Top: average MD (on log scale) of the 
estimated coefficients for $n = 100$, $\p = 10$, 
$\eps = 20\%$ of cellwise outliers, and 
$\bSigma = \bSigma_{\ALYZ}$ (left) or 
$\bSigma = \bSigma_{\AN}$ (right), for normal
predictors.
Bottom: corresponding MSE, also on log scale.}
\label{fig:MD_MSE_d10_e20_normal}
\end{figure}

\begin{figure}[!ht]
\centering
\vspace{-4mm}
\includegraphics[width = 0.45\columnwidth]
{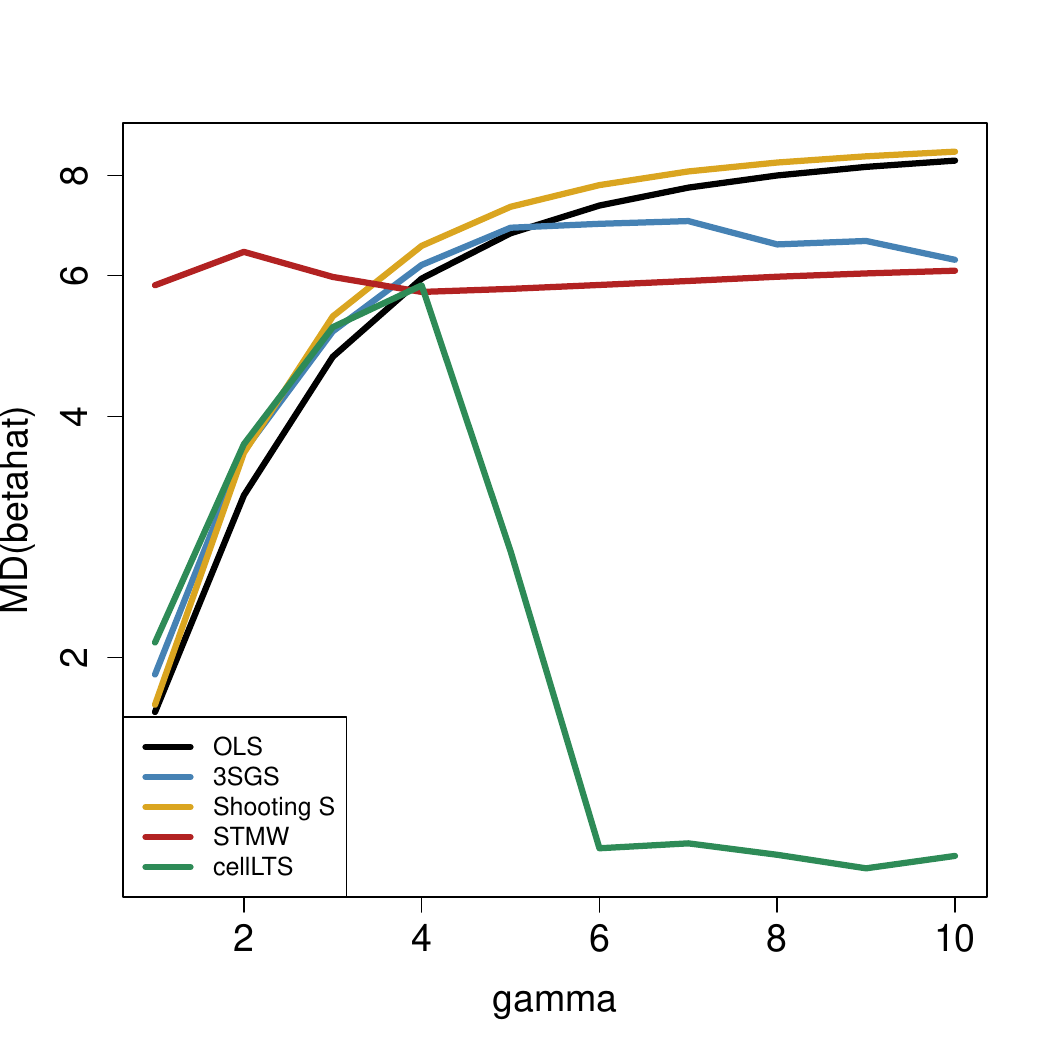}
\includegraphics[width = 0.45\columnwidth]
{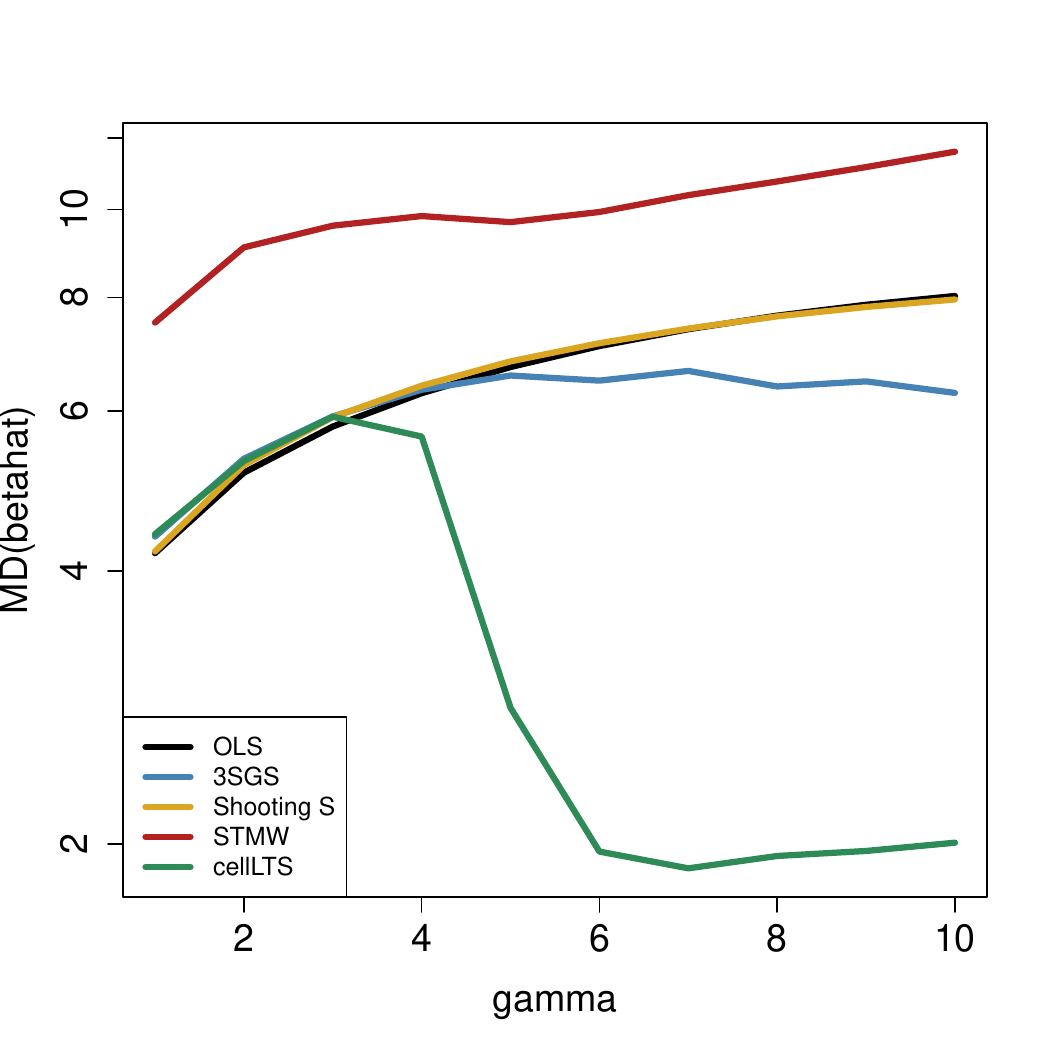}\\
\vspace{-6mm}
\includegraphics[width = 0.45\columnwidth]
{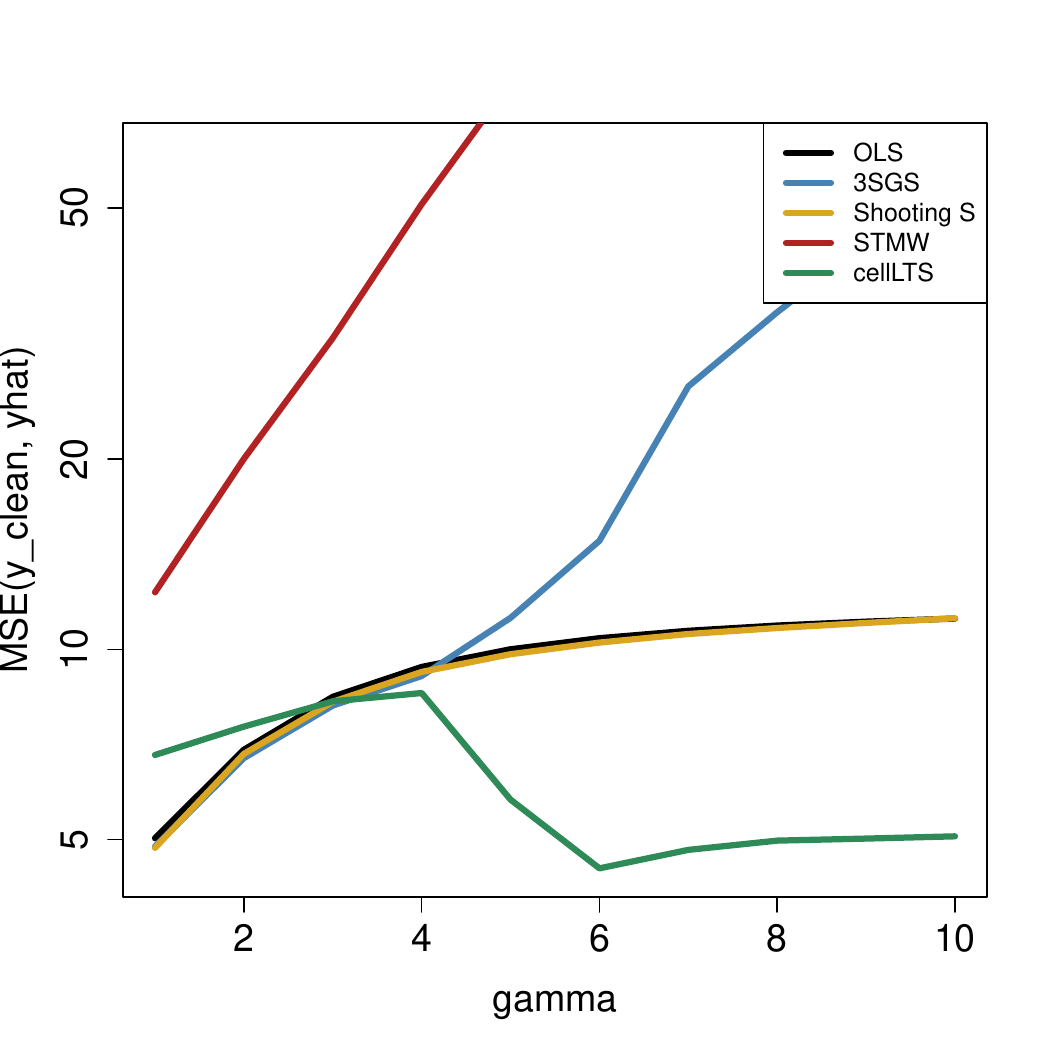}
\includegraphics[width = 0.45\columnwidth]
{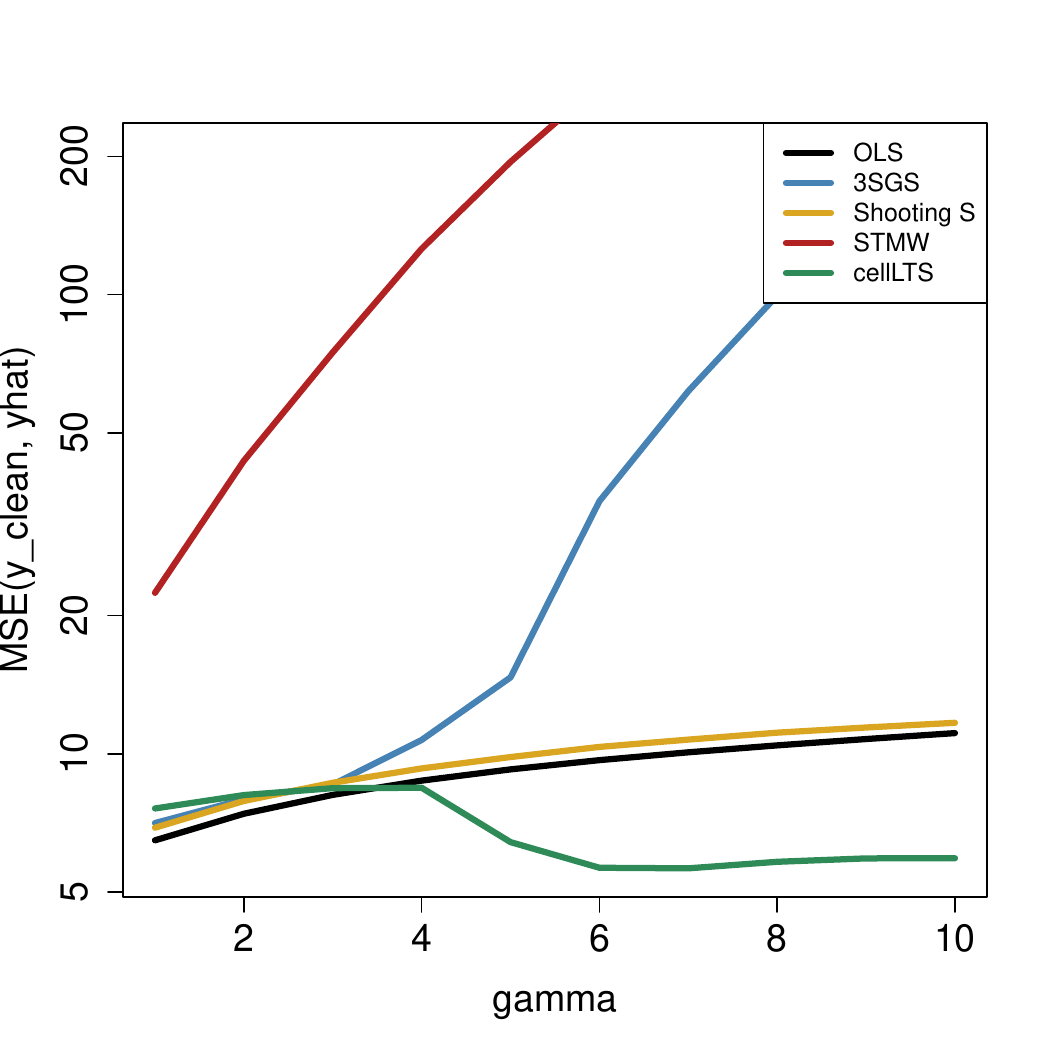}\\
\caption{Like Figure~\ref{fig:MD_MSE_d10_e20_normal}, 
but for exponential predictors.}
\end{figure}

\begin{figure}[!ht]
\centering
\vspace{-4mm}
\includegraphics[width = 0.45\columnwidth]
{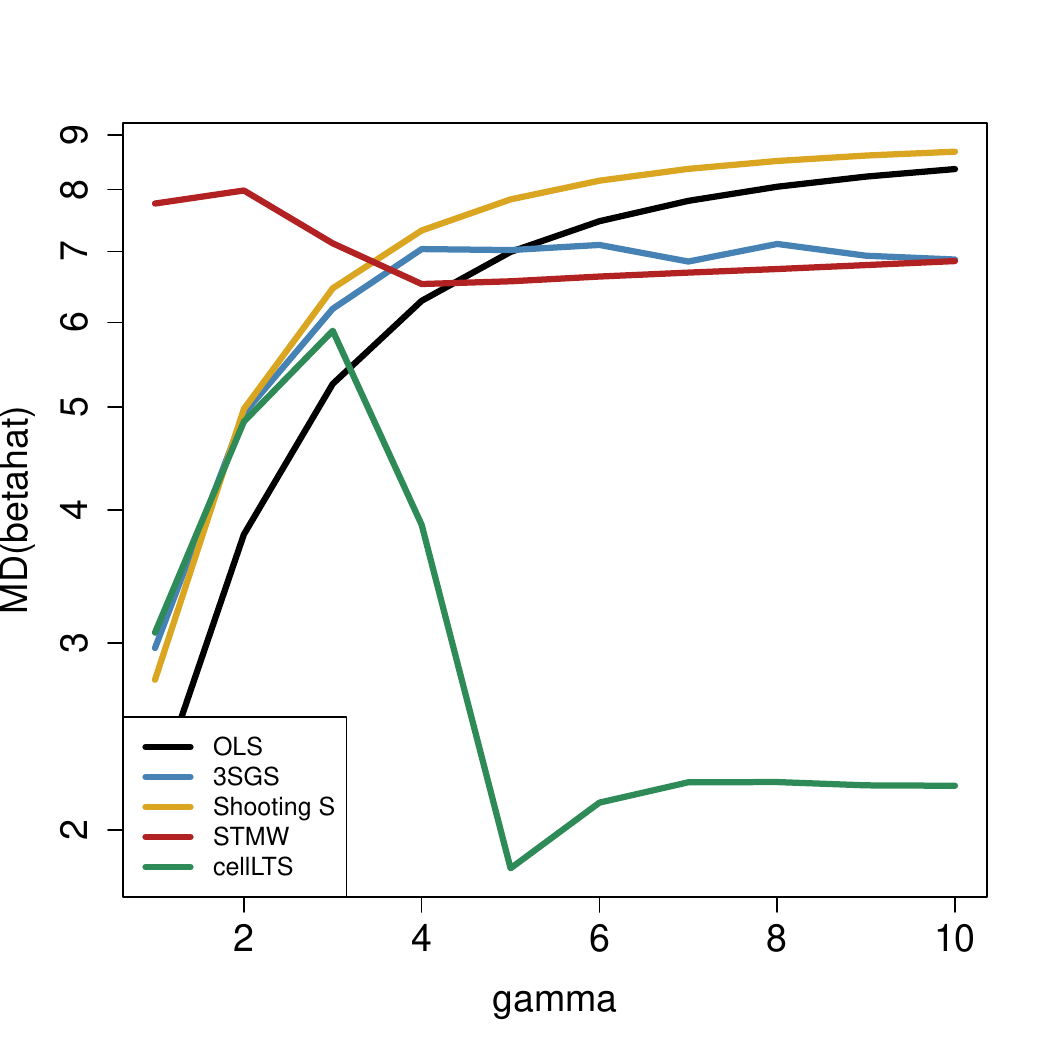}
\includegraphics[width = 0.45\columnwidth]
{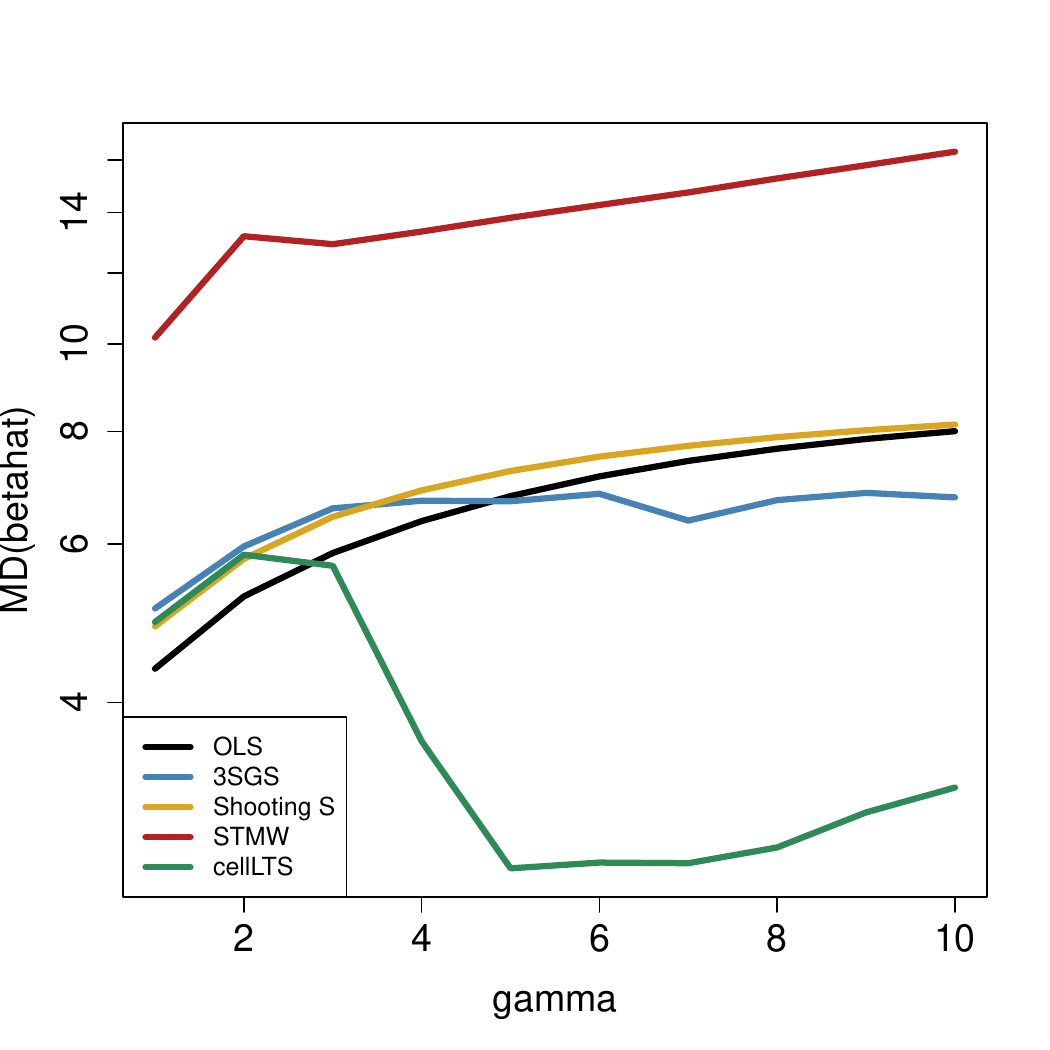}\\
\vspace{-6mm}
\includegraphics[width = 0.45\columnwidth]
{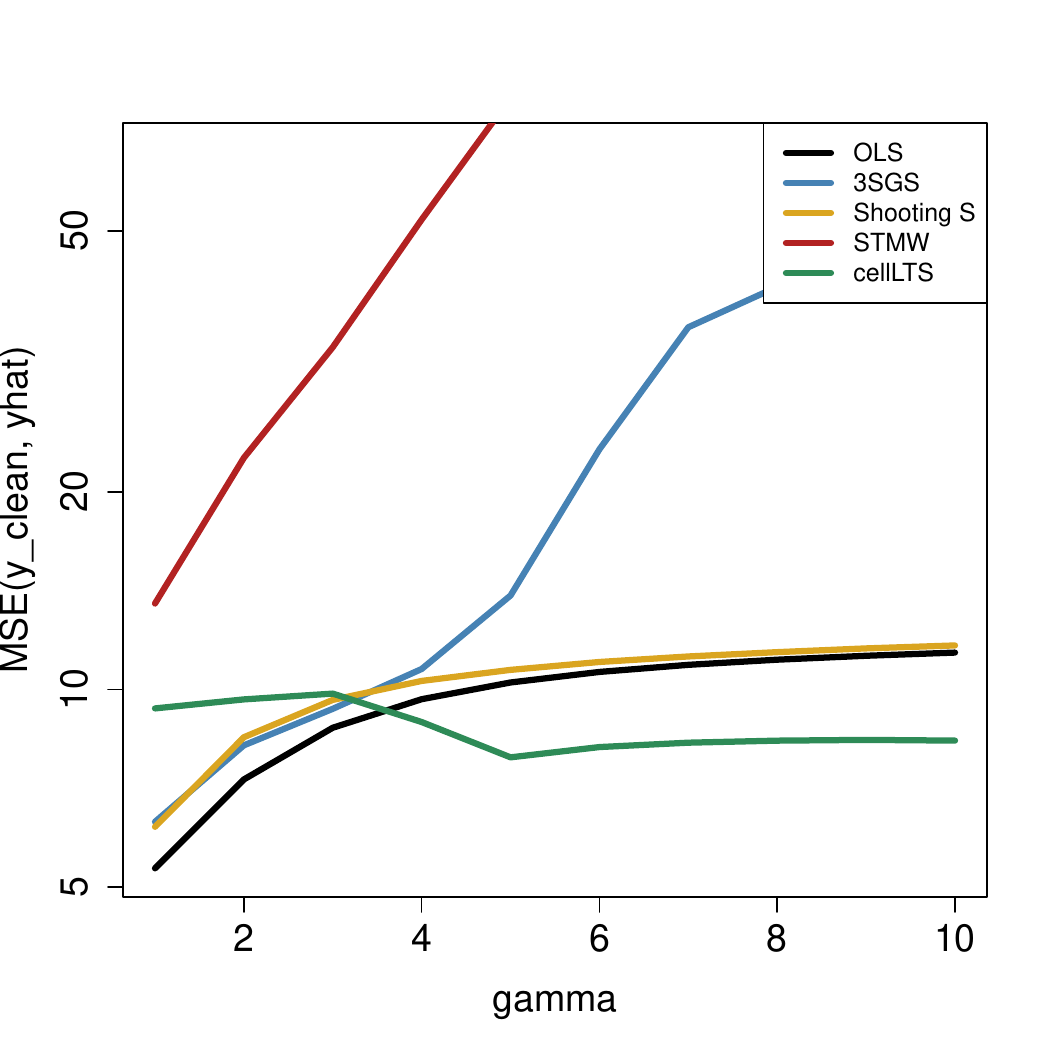}
\includegraphics[width = 0.45\columnwidth]
{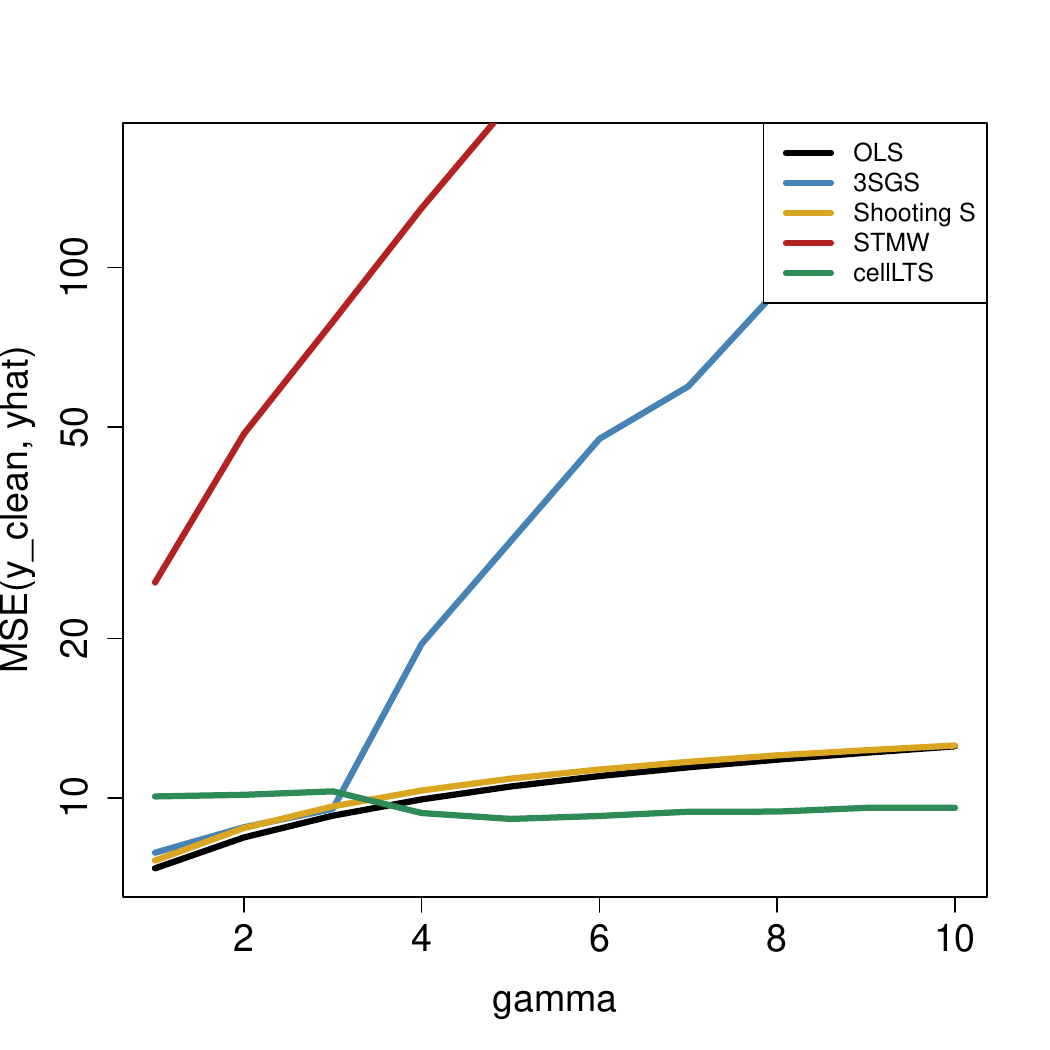}\\
\caption{Like Figure~\ref{fig:MD_MSE_d10_e20_normal}, 
but for lognormal predictors.}
\end{figure}

\clearpage

\begin{figure}[!ht]
\centering
\vspace{-4mm}
\includegraphics[width = 0.45\columnwidth]
{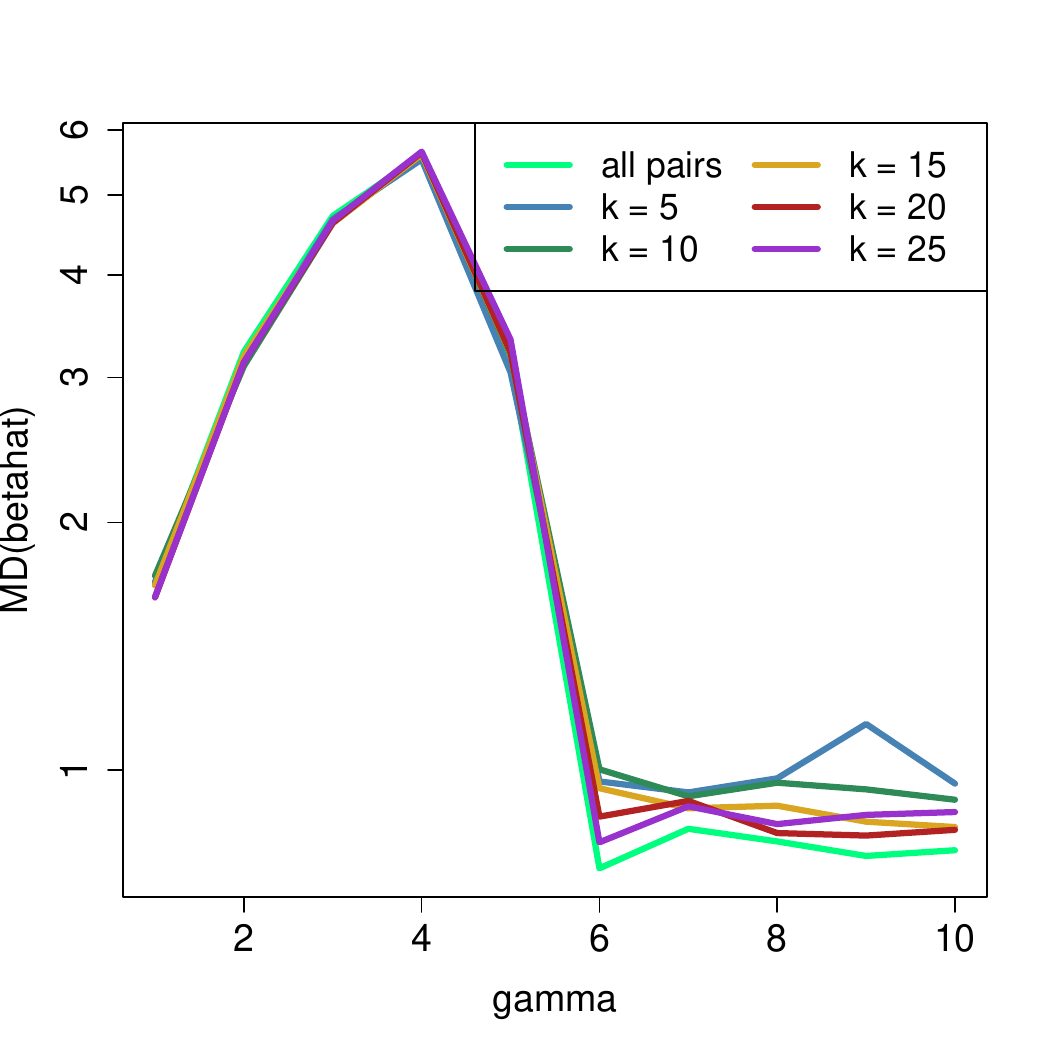}
\includegraphics[width = 0.45\columnwidth]
{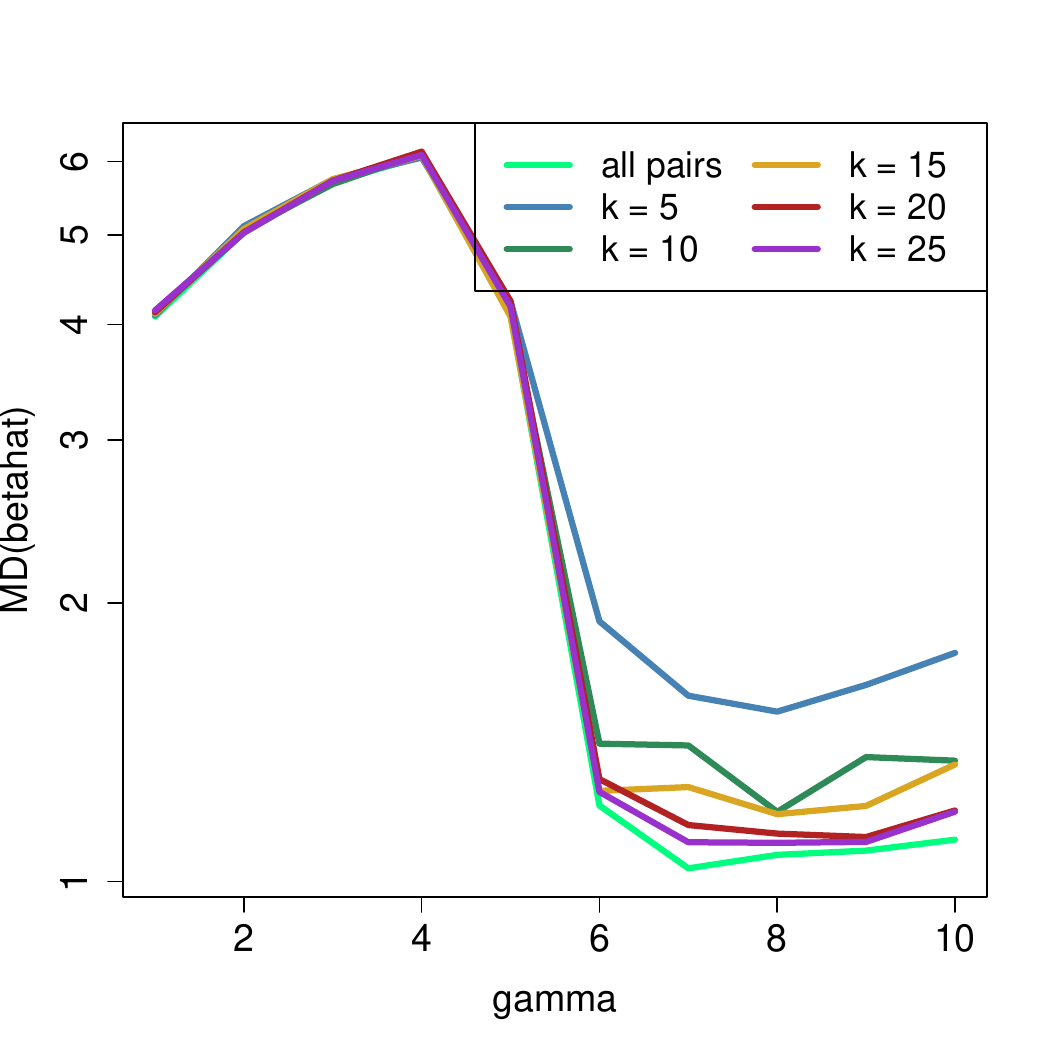}\\
\vspace{-6mm}
\includegraphics[width = 0.45\columnwidth]
{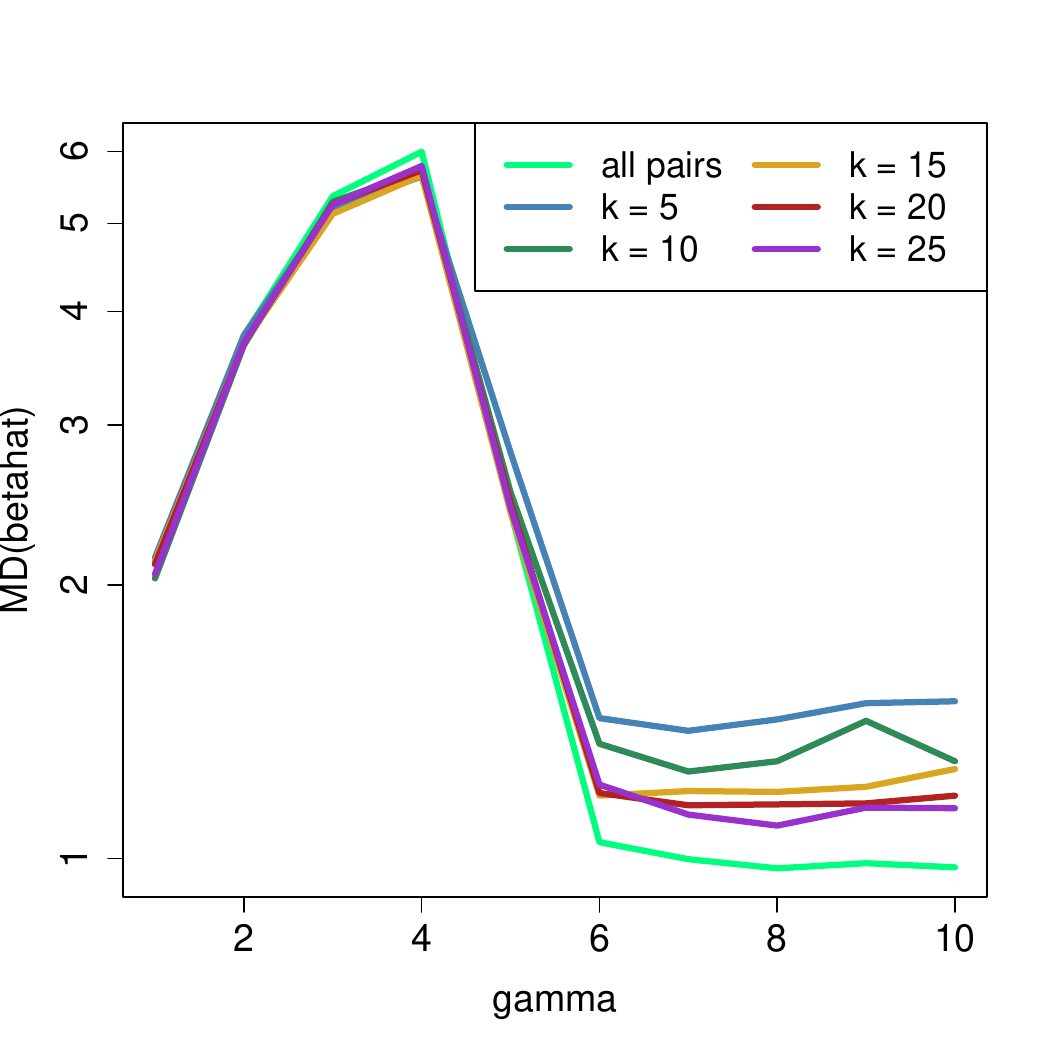}
\includegraphics[width = 0.45\columnwidth]
{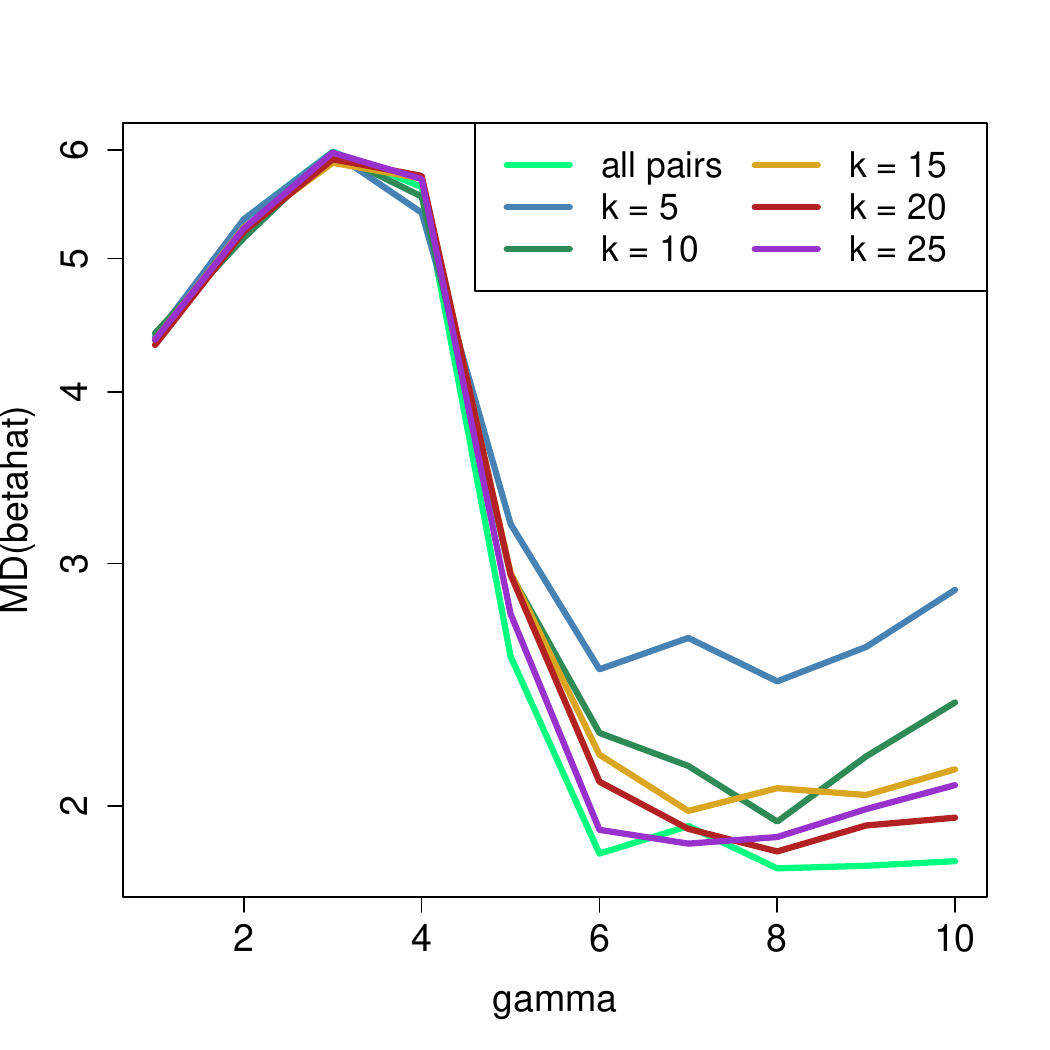}\\
\vspace{-6mm}
\includegraphics[width = 0.45\columnwidth]
{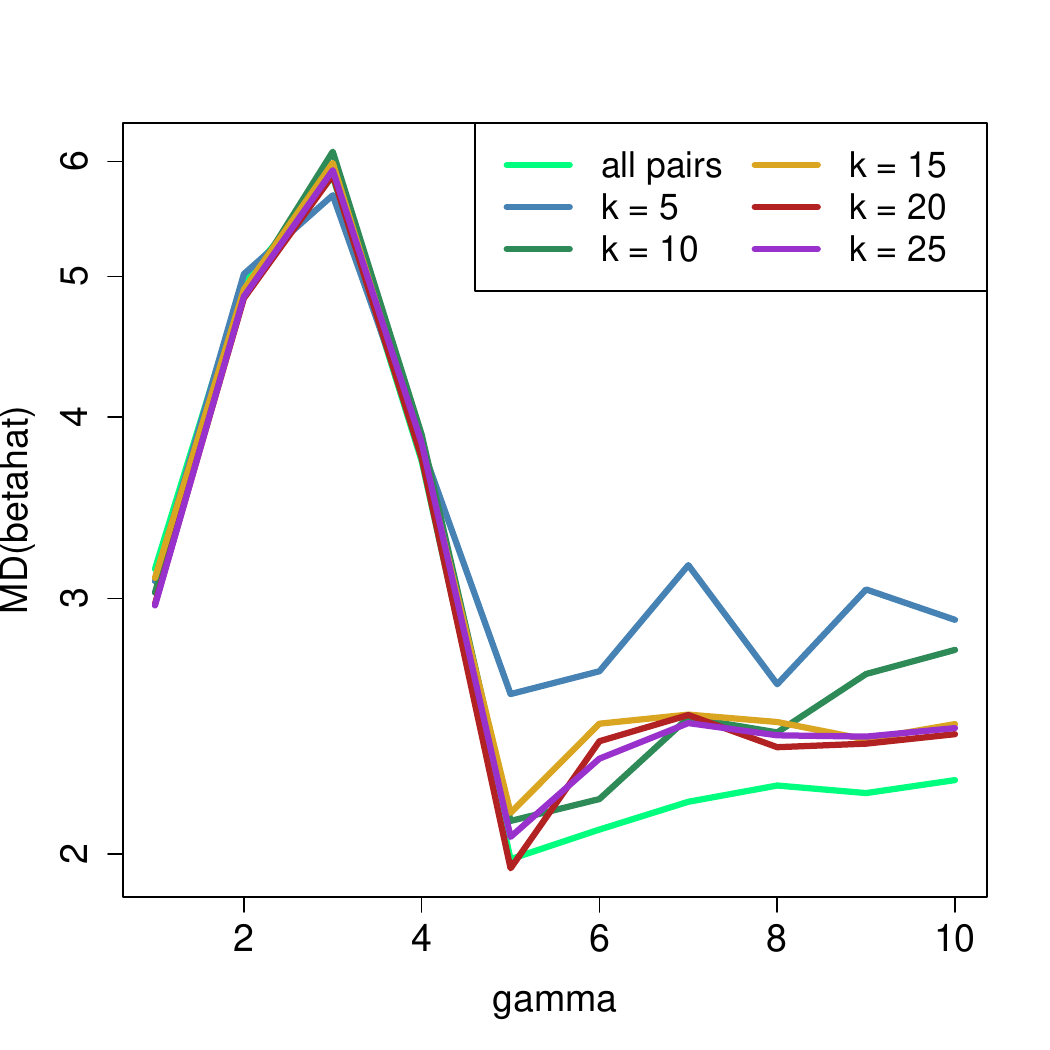}
\includegraphics[width = 0.45\columnwidth]
{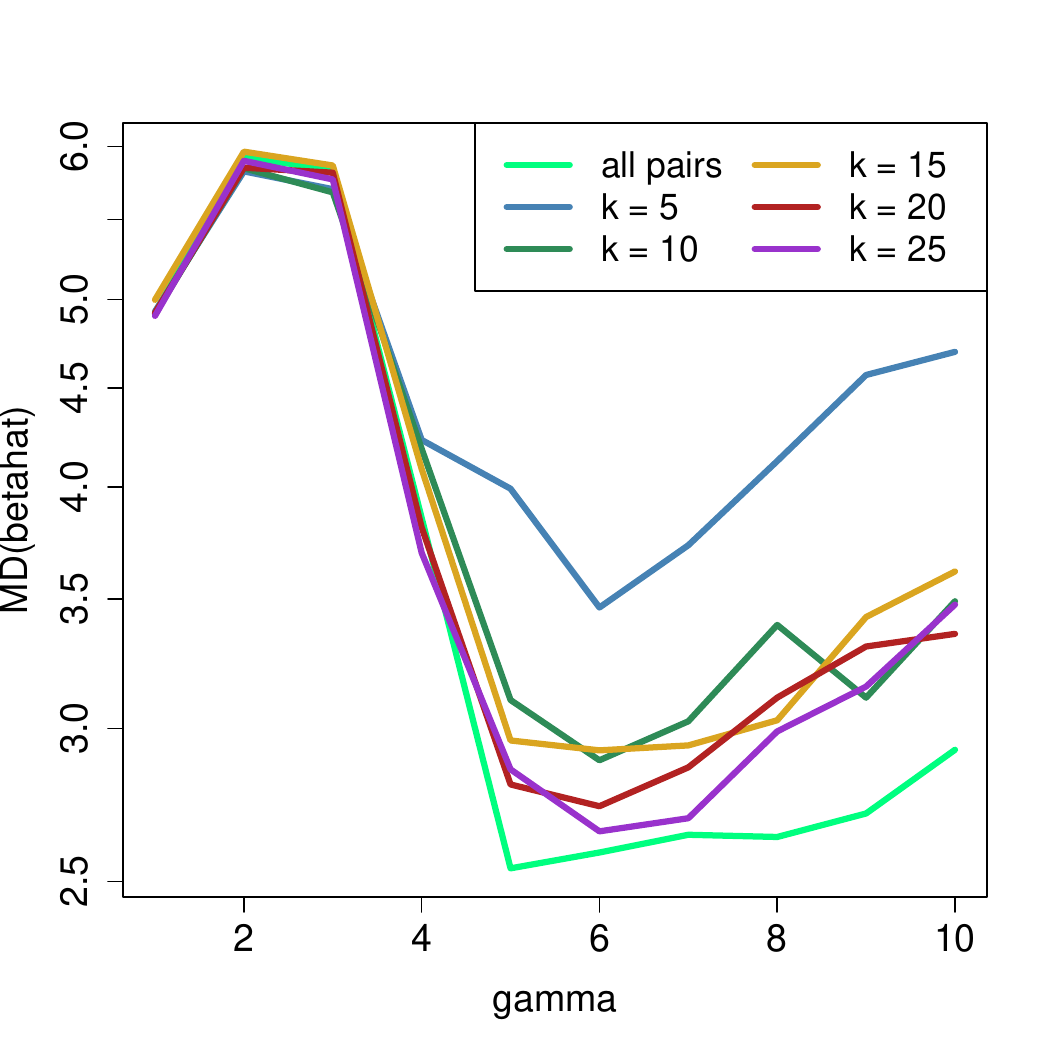}\\
\caption{Top row: average MD (on log scale) of the 
estimated coefficients for different symmetrization 
strategies and normal predictors. 
The data has $n=100$, $\p = 10$, 
$\eps = 20\%$ of cellwise outliers, and 
$\bSigma = \bSigma_{\ALYZ}$ (left) or 
$\bSigma = \bSigma_{\AN}$ (right).
Middle row: same for exponential predictors.
Bottom row: same for lognormal predictors.}
\end{figure}

\clearpage
\subsection{Results for \texorpdfstring{$n=400$, 
$\p=20$, and $\eps=10\%$}{n = 400, p = 20, and
eps = 10\%}}

\begin{figure}[!ht]
\centering
\vspace{-4mm}
\includegraphics[width = 0.45\columnwidth]
{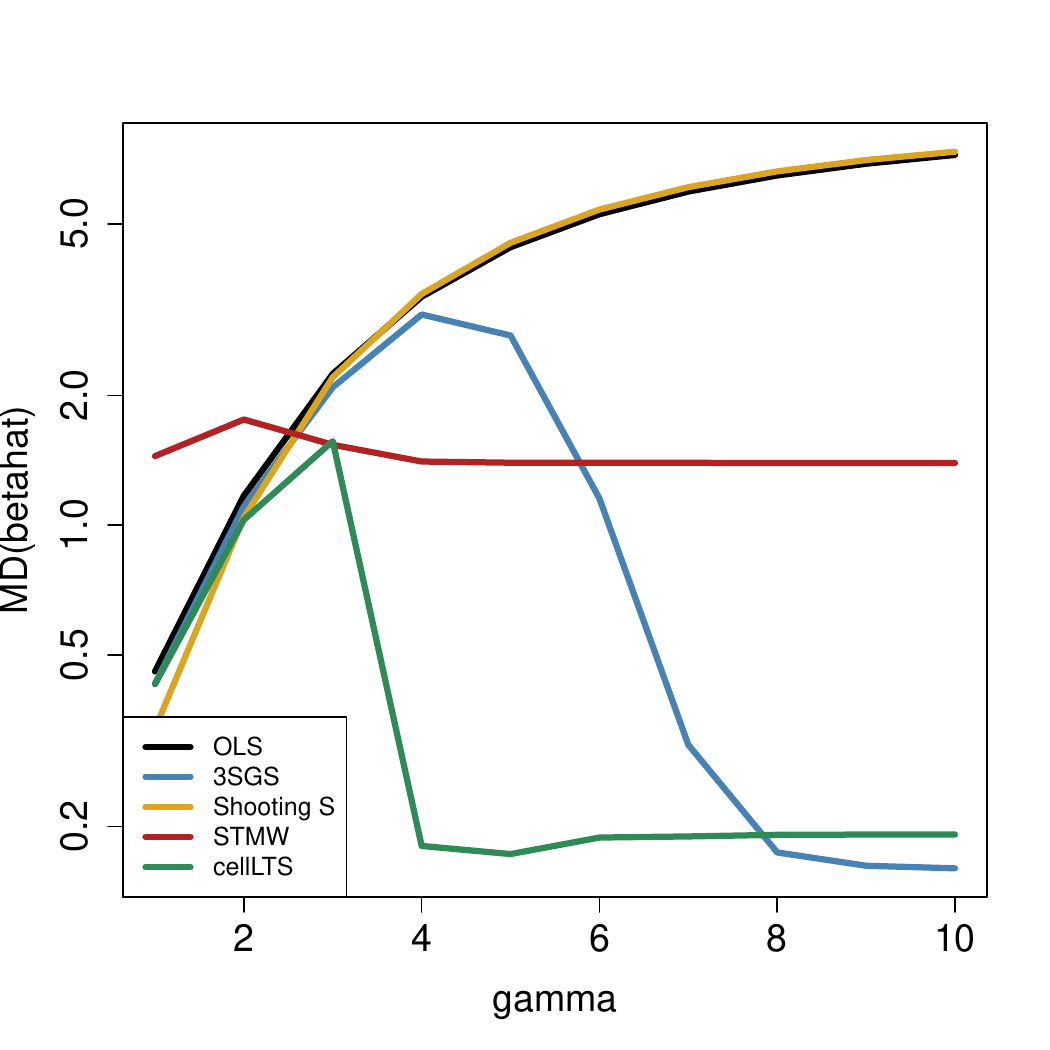}
\includegraphics[width = 0.45\columnwidth]
{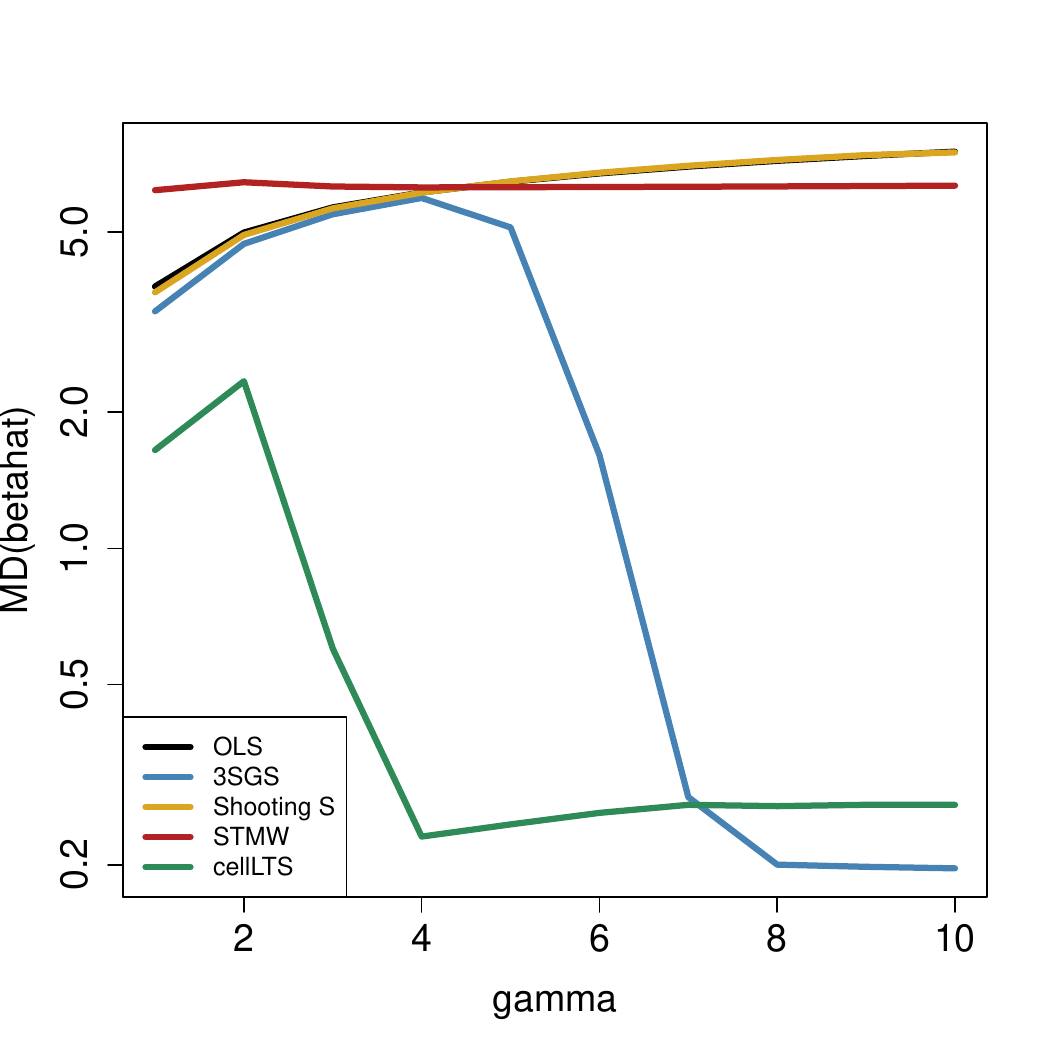}\\
\vspace{-6mm}
\includegraphics[width = 0.45\columnwidth]
{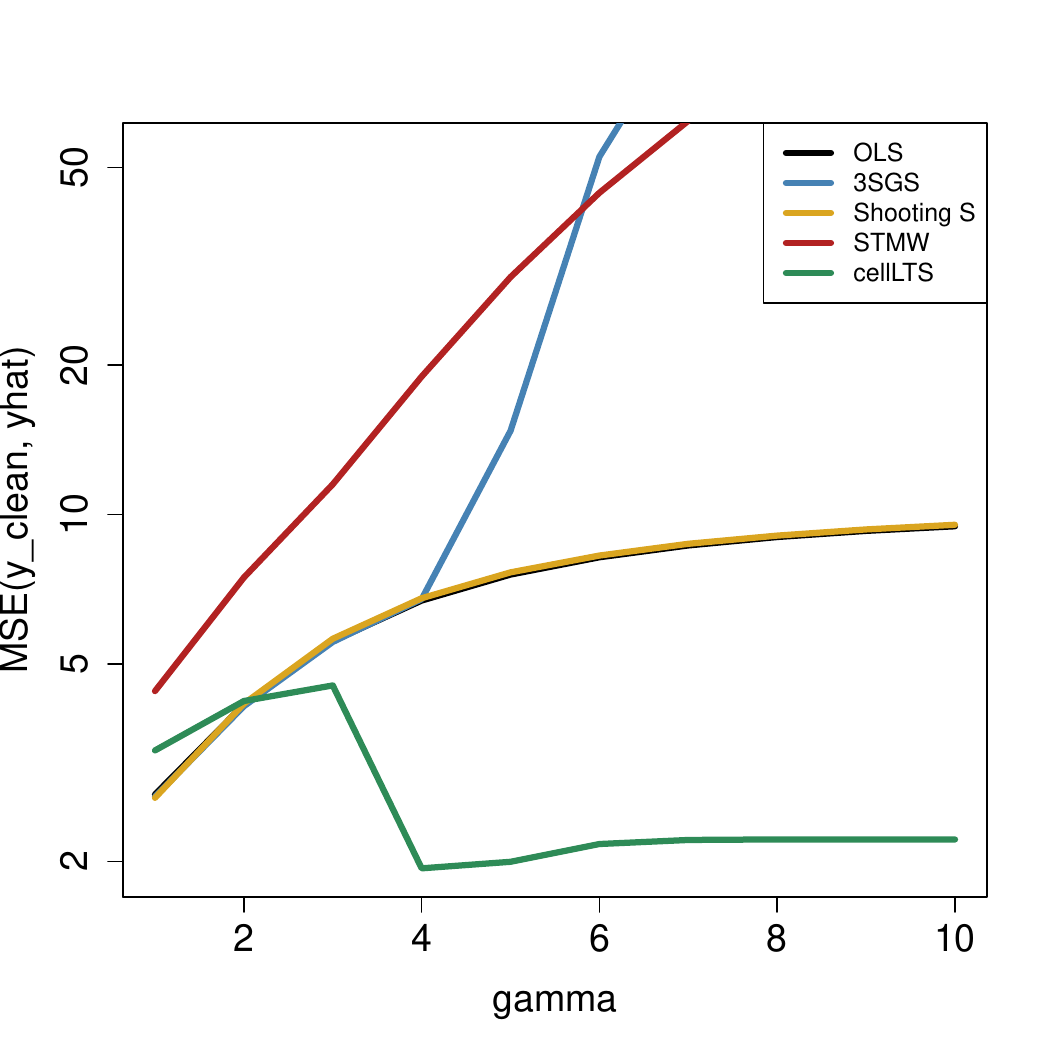}
\includegraphics[width = 0.45\columnwidth]
{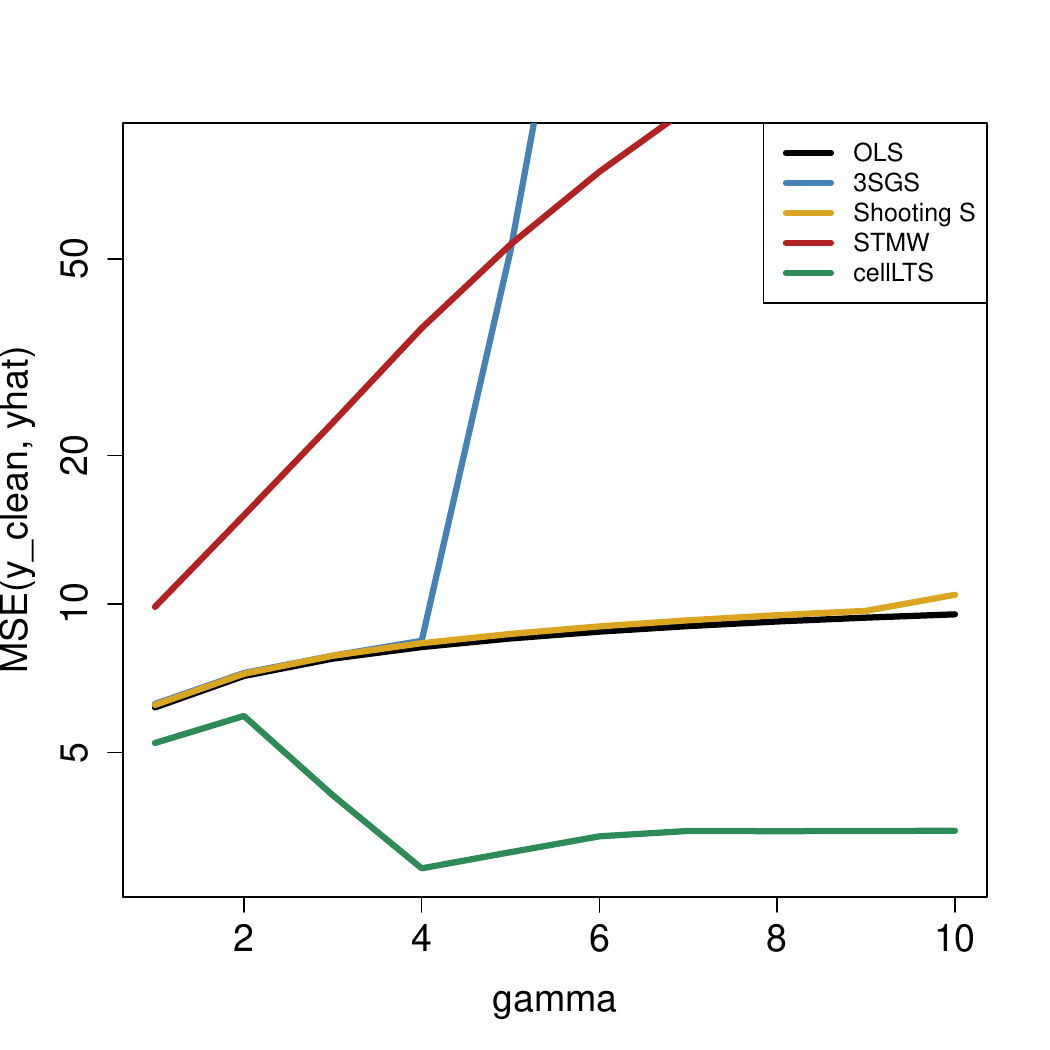}\\
\caption{Top: average MD (on log scale) of the 
estimated coefficients for $n = 400$, $\p = 20$, 
$\eps = 10\%$ of cellwise outliers, and 
$\bSigma = \bSigma_{\ALYZ}$ (left) or 
$\bSigma = \bSigma_{\AN}$ (right), for normal
predictors.
Bottom: corresponding MSE, also on log scale.}
\label{fig:MD_MSE_d20_e10_normal}
\end{figure}

\begin{figure}[!ht]
\centering
\vspace{-4mm}
\includegraphics[width = 0.45\columnwidth]
{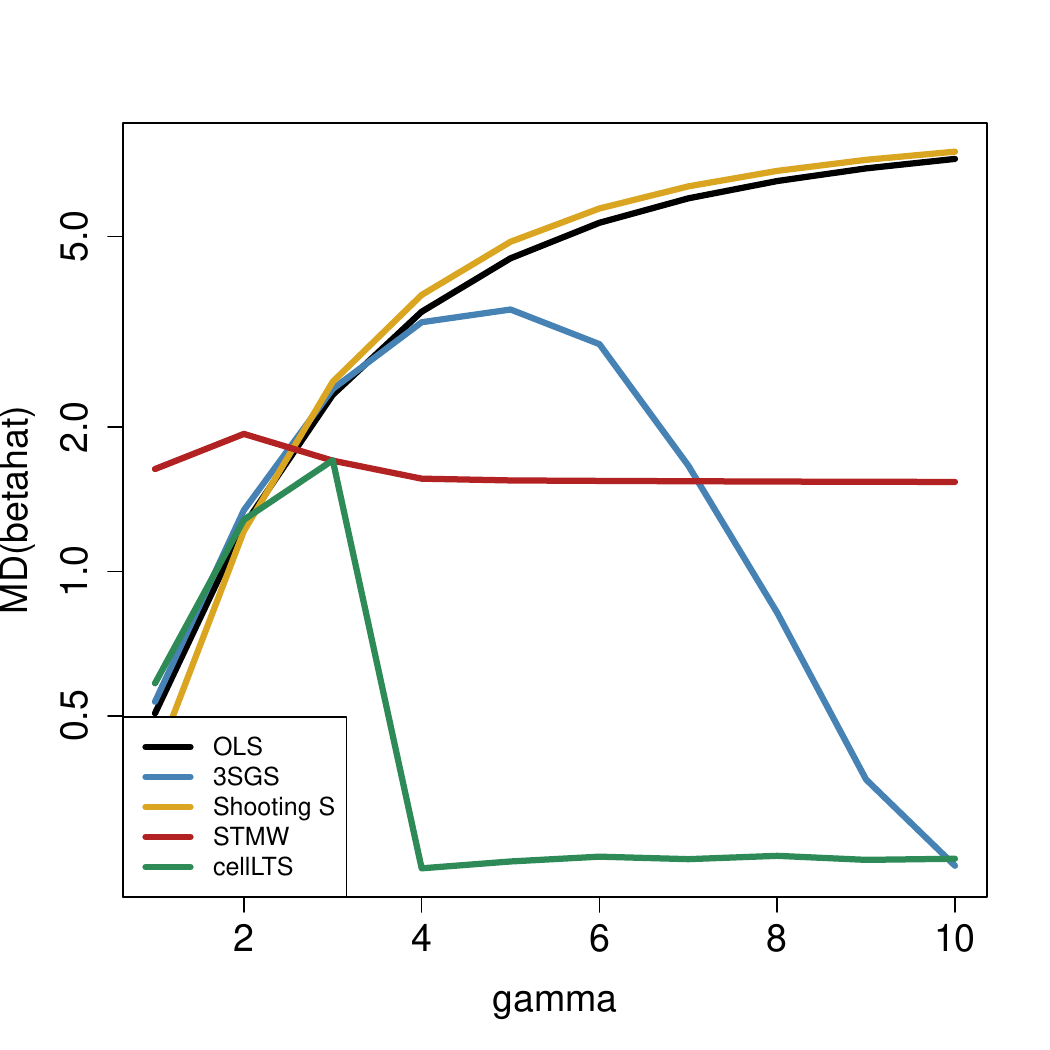}
\includegraphics[width = 0.45\columnwidth]
{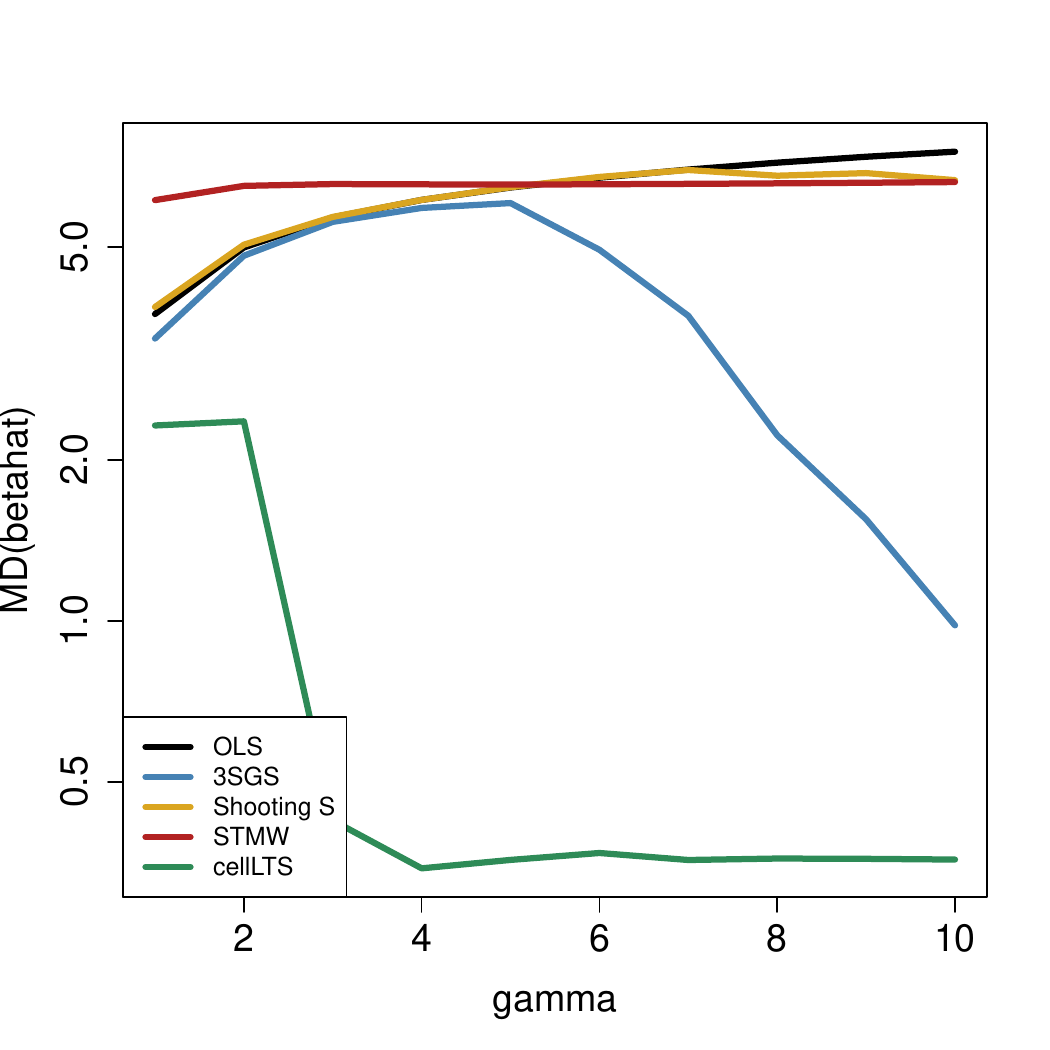}\\
\vspace{-6mm}
\includegraphics[width = 0.45\columnwidth]
{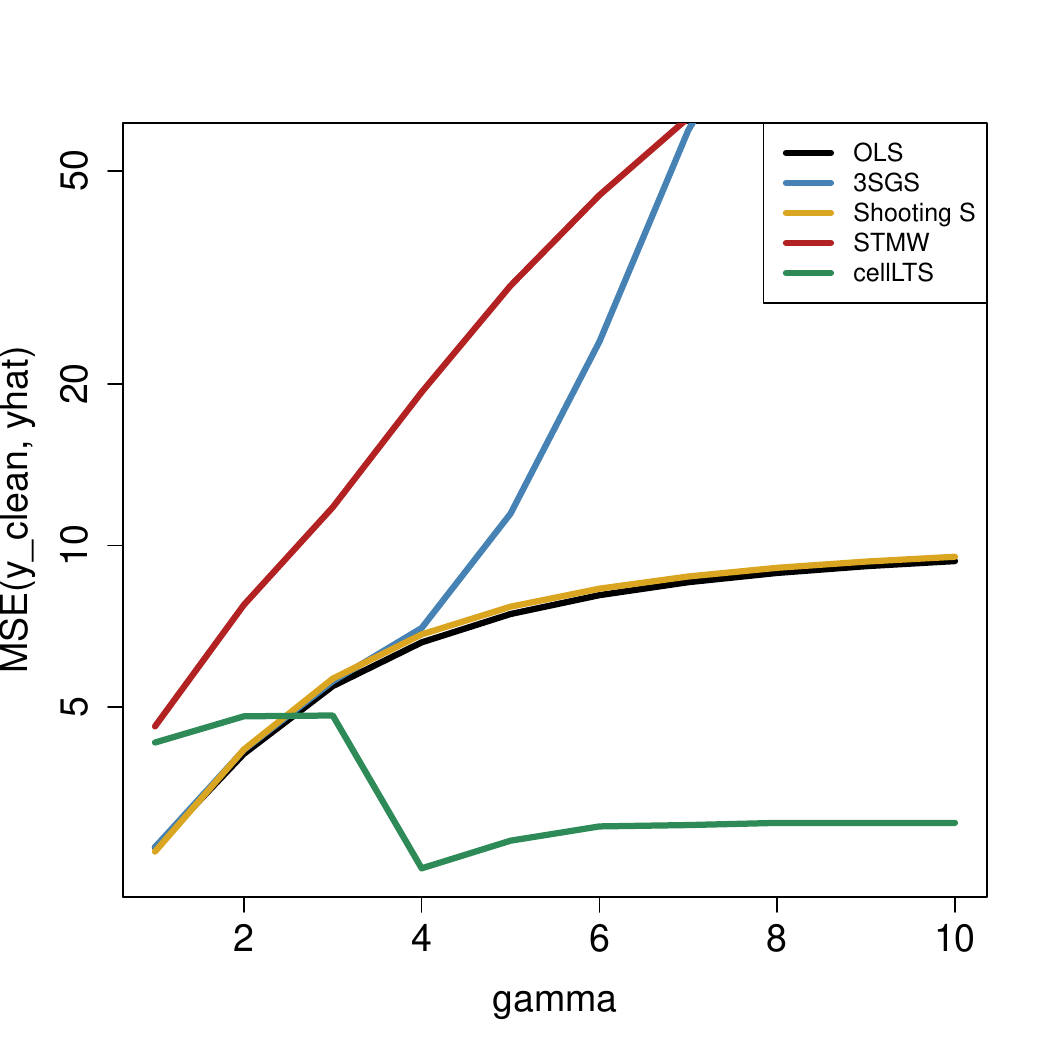}
\includegraphics[width = 0.45\columnwidth]
{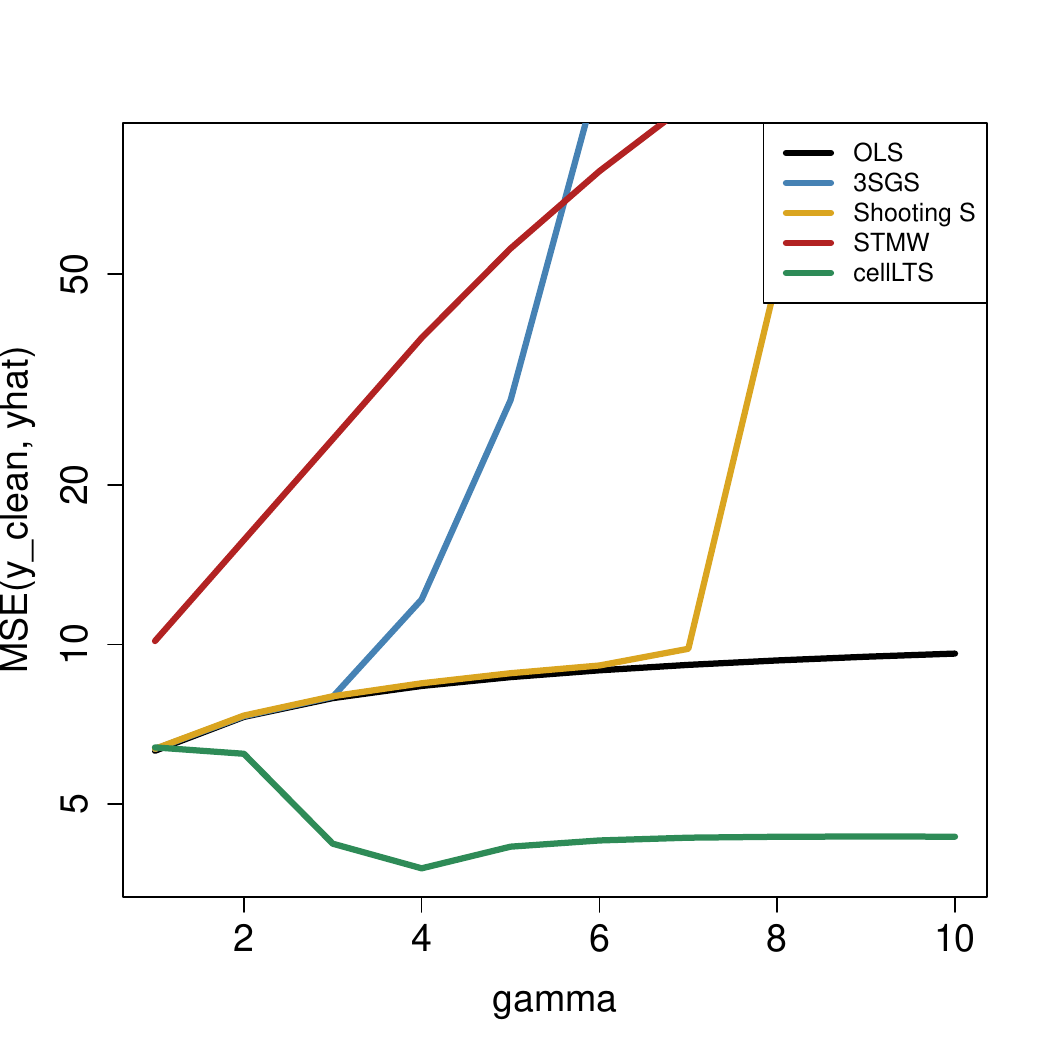}\\
\caption{Like Figure~\ref{fig:MD_MSE_d20_e10_normal}, 
but for exponential predictors.}
\end{figure}

\begin{figure}[!ht]
\centering
\vspace{-4mm}
\includegraphics[width = 0.45\columnwidth]
{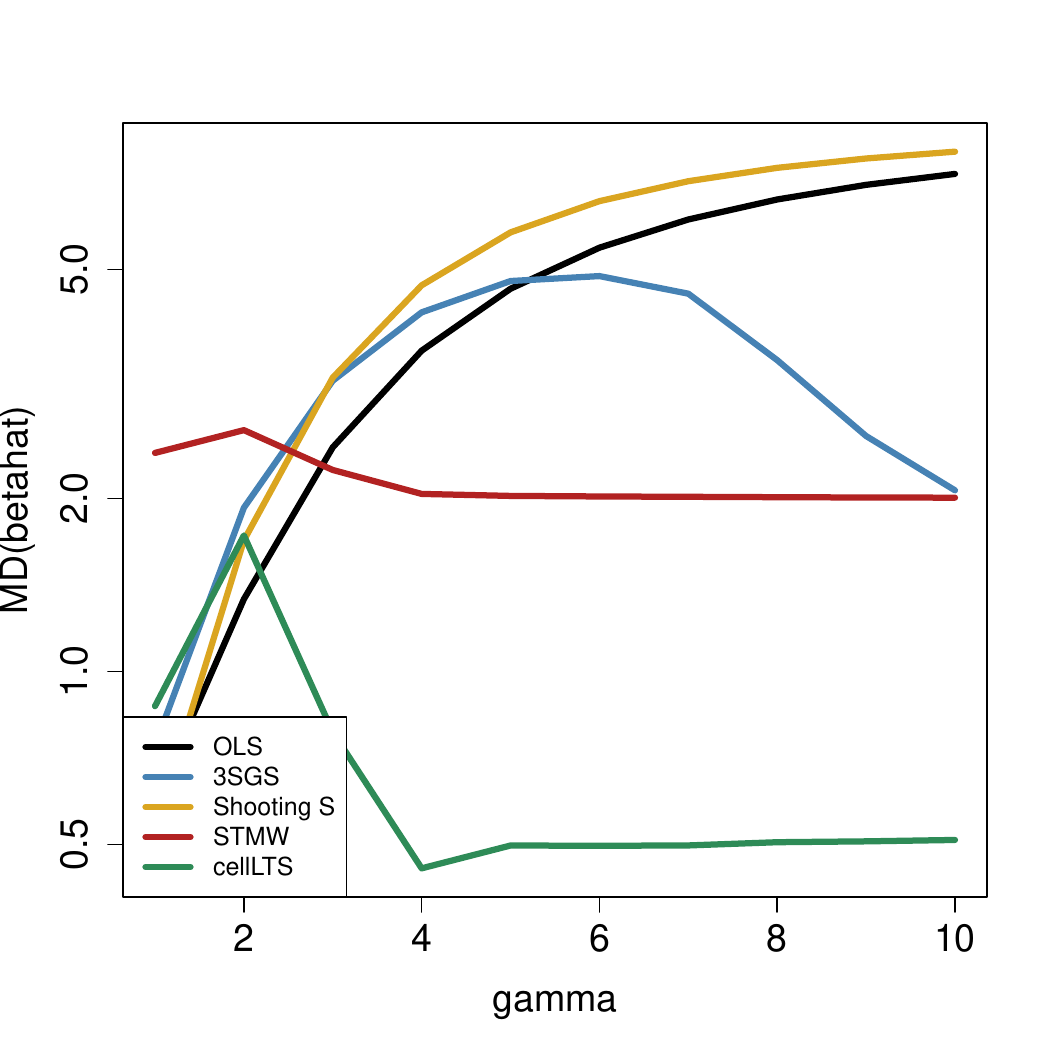}
\includegraphics[width = 0.45\columnwidth]
{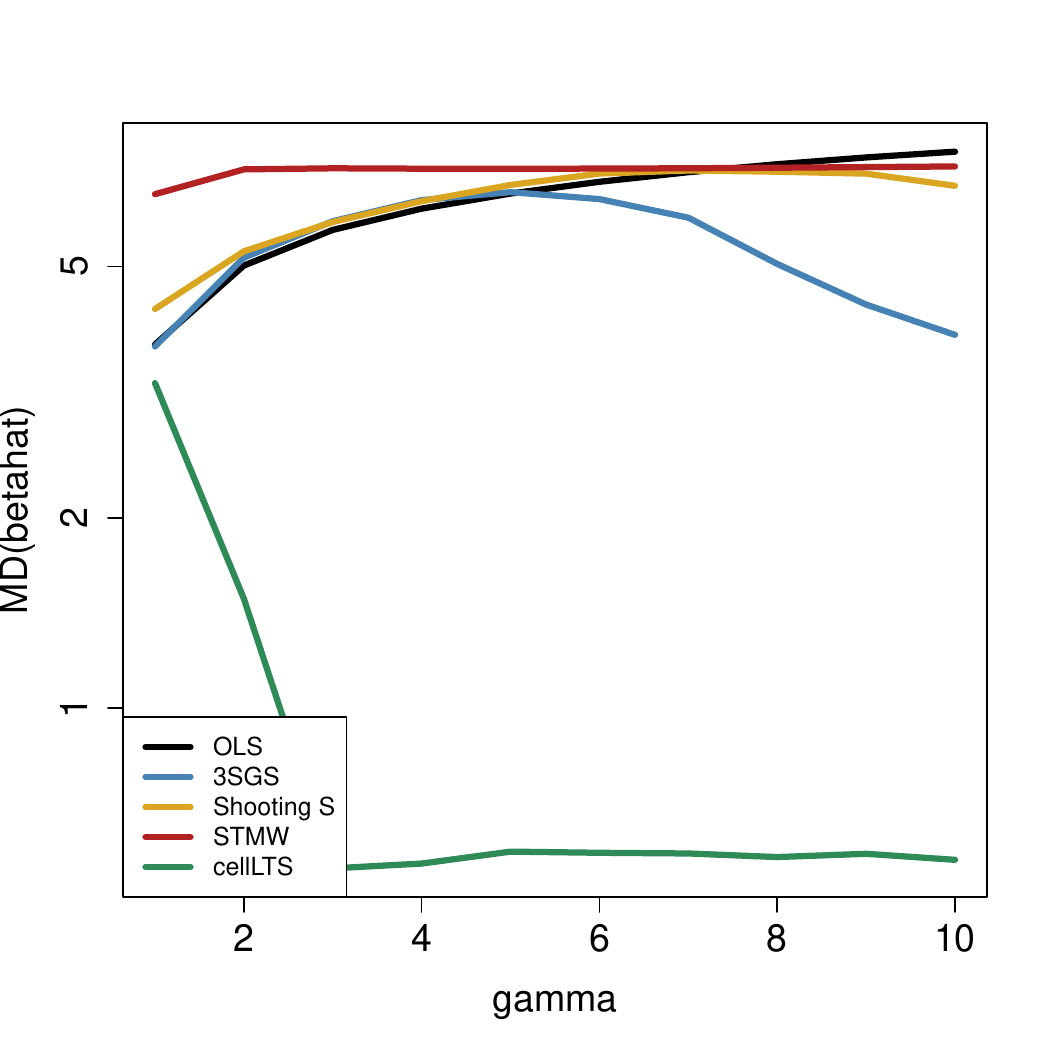}\\
\vspace{-6mm}
\includegraphics[width = 0.45\columnwidth]
{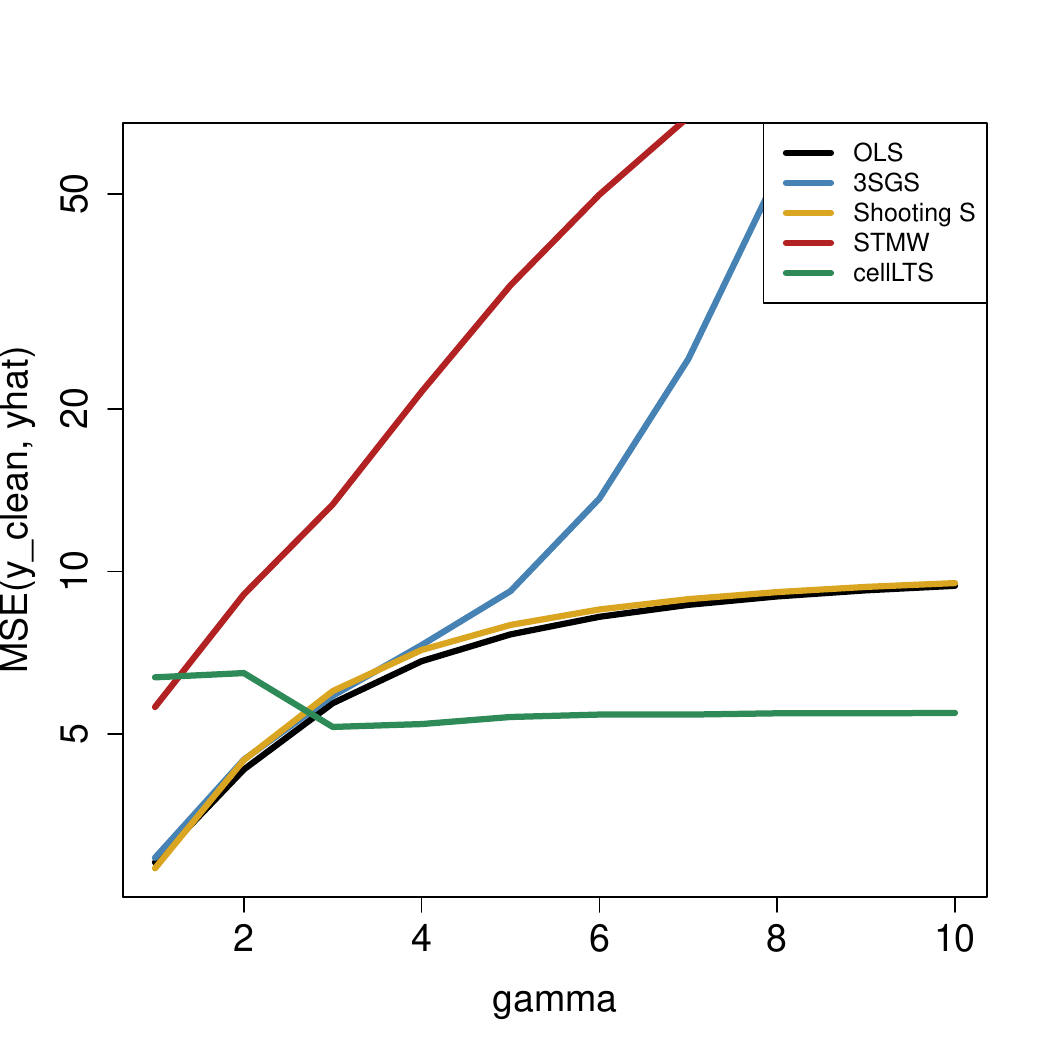}
\includegraphics[width = 0.45\columnwidth]
{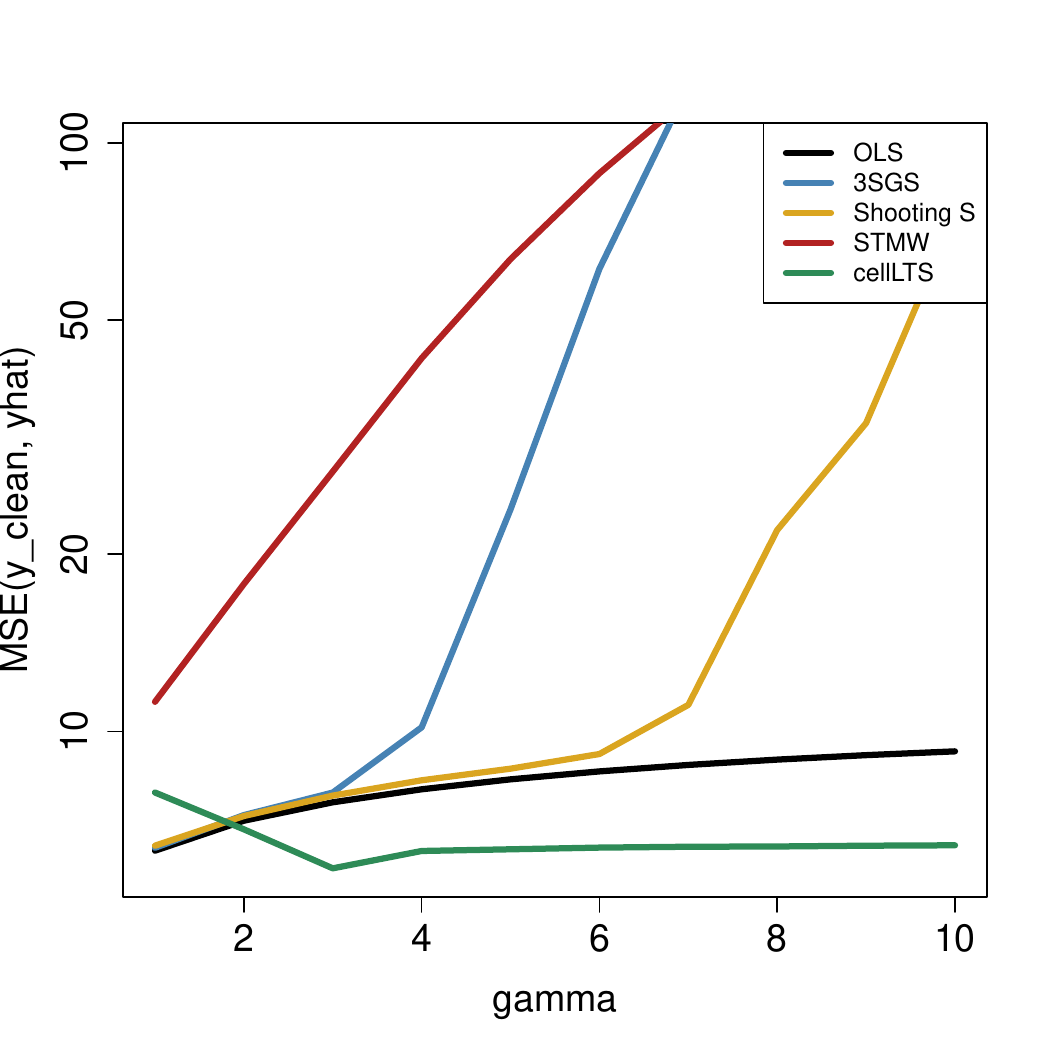}\\
\caption{Like Figure~\ref{fig:MD_MSE_d20_e10_normal}, 
but for lognormal predictors.}
\end{figure}

\clearpage

\begin{figure}[!ht]
\centering
\vspace{-4mm}
\includegraphics[width = 0.45\columnwidth]
{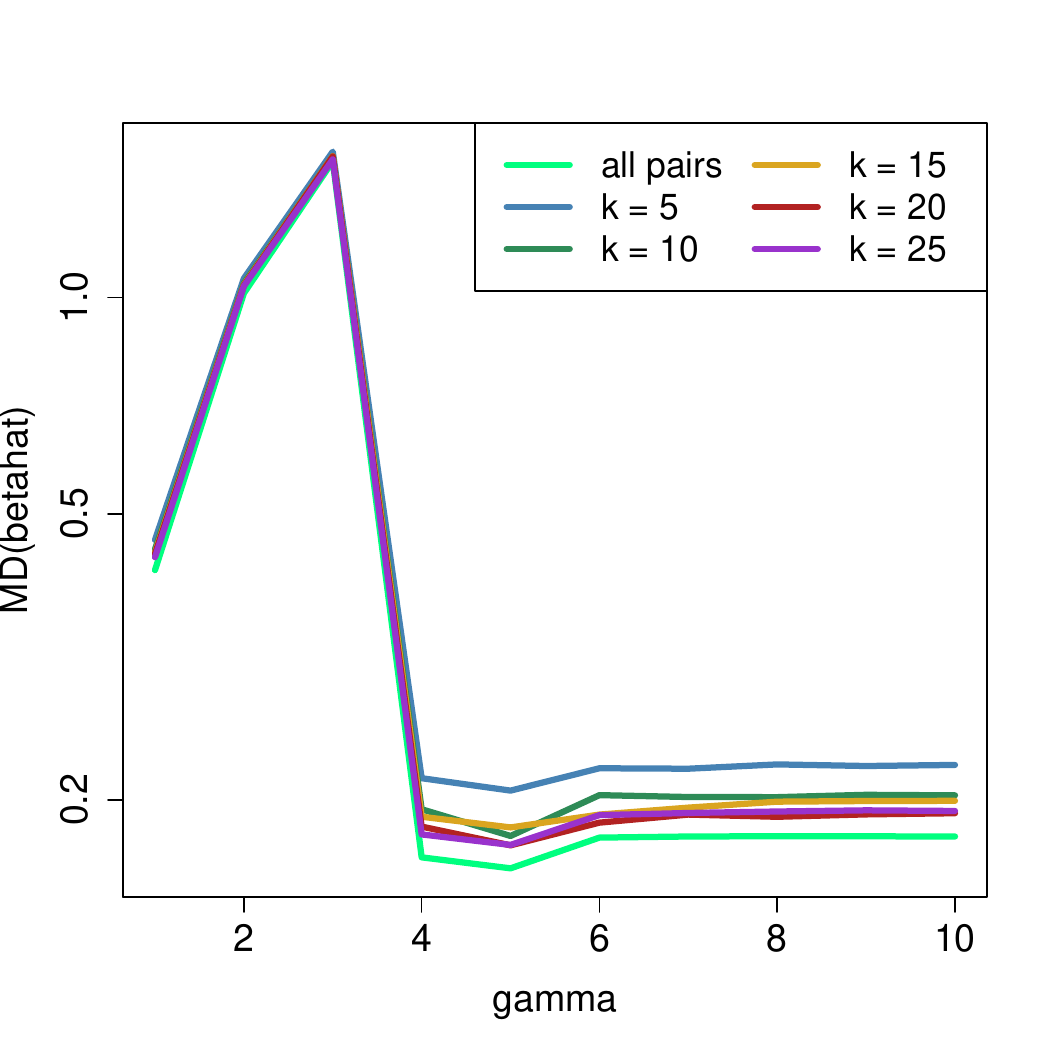}
\includegraphics[width = 0.45\columnwidth]
{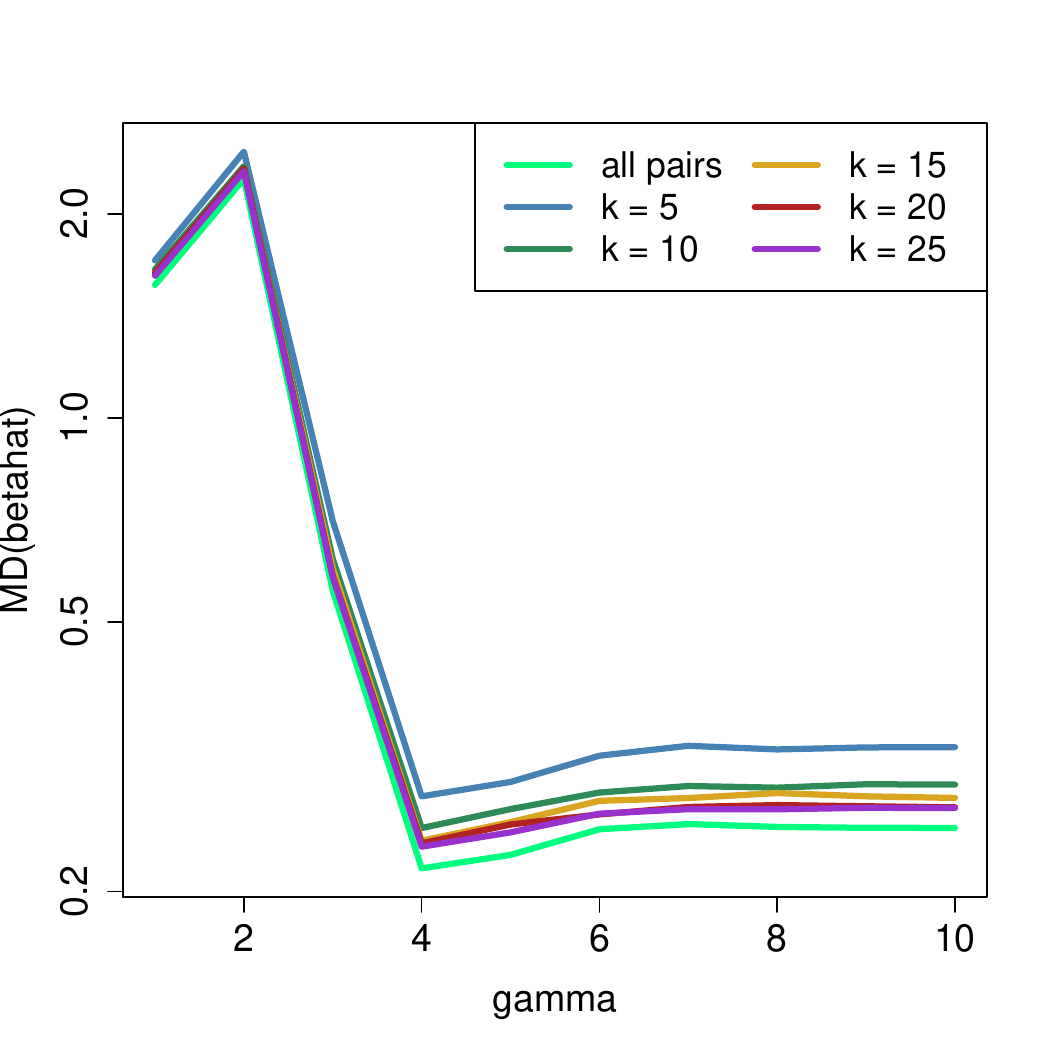}\\
\vspace{-6mm}
\includegraphics[width = 0.45\columnwidth]
{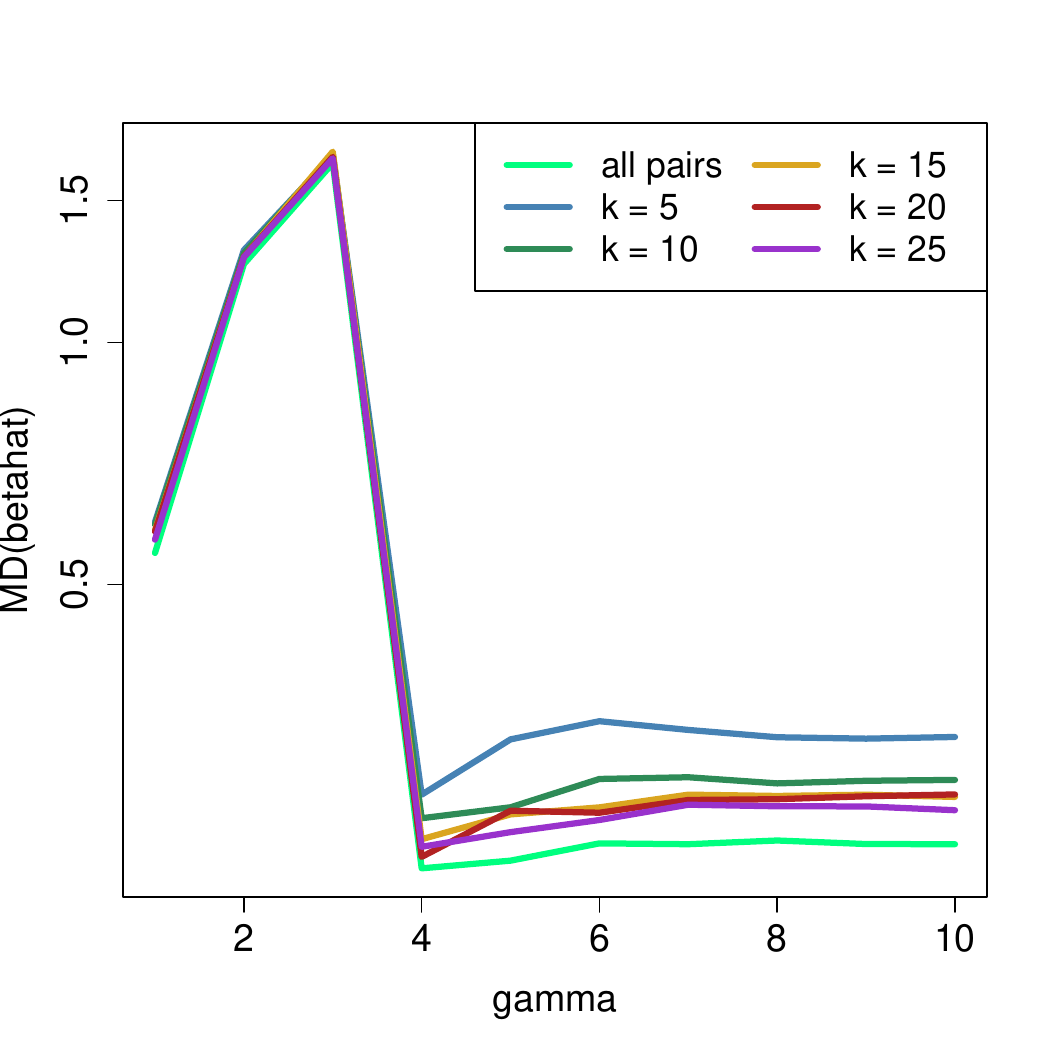}
\includegraphics[width = 0.45\columnwidth]
{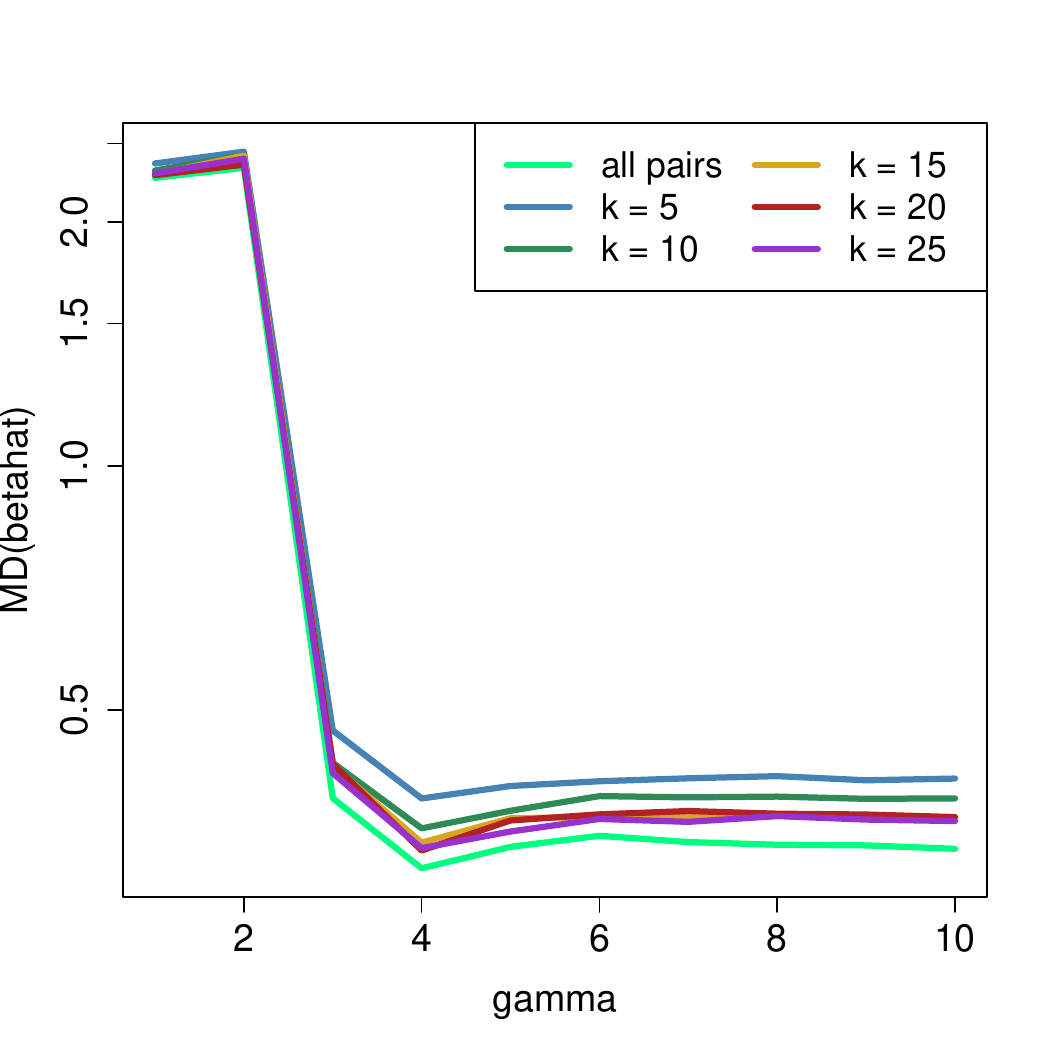}\\
\vspace{-6mm}
\includegraphics[width = 0.45\columnwidth]
{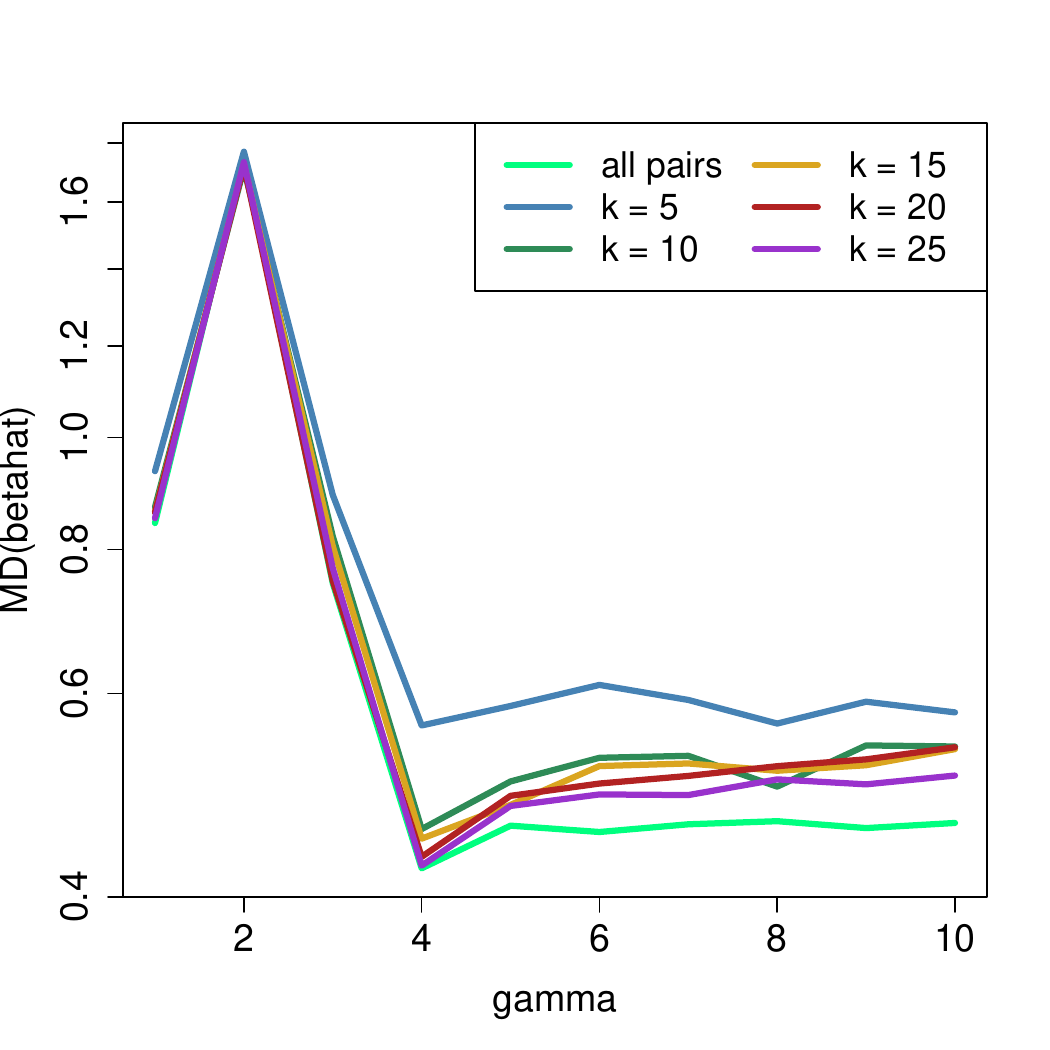}
\includegraphics[width = 0.45\columnwidth]
{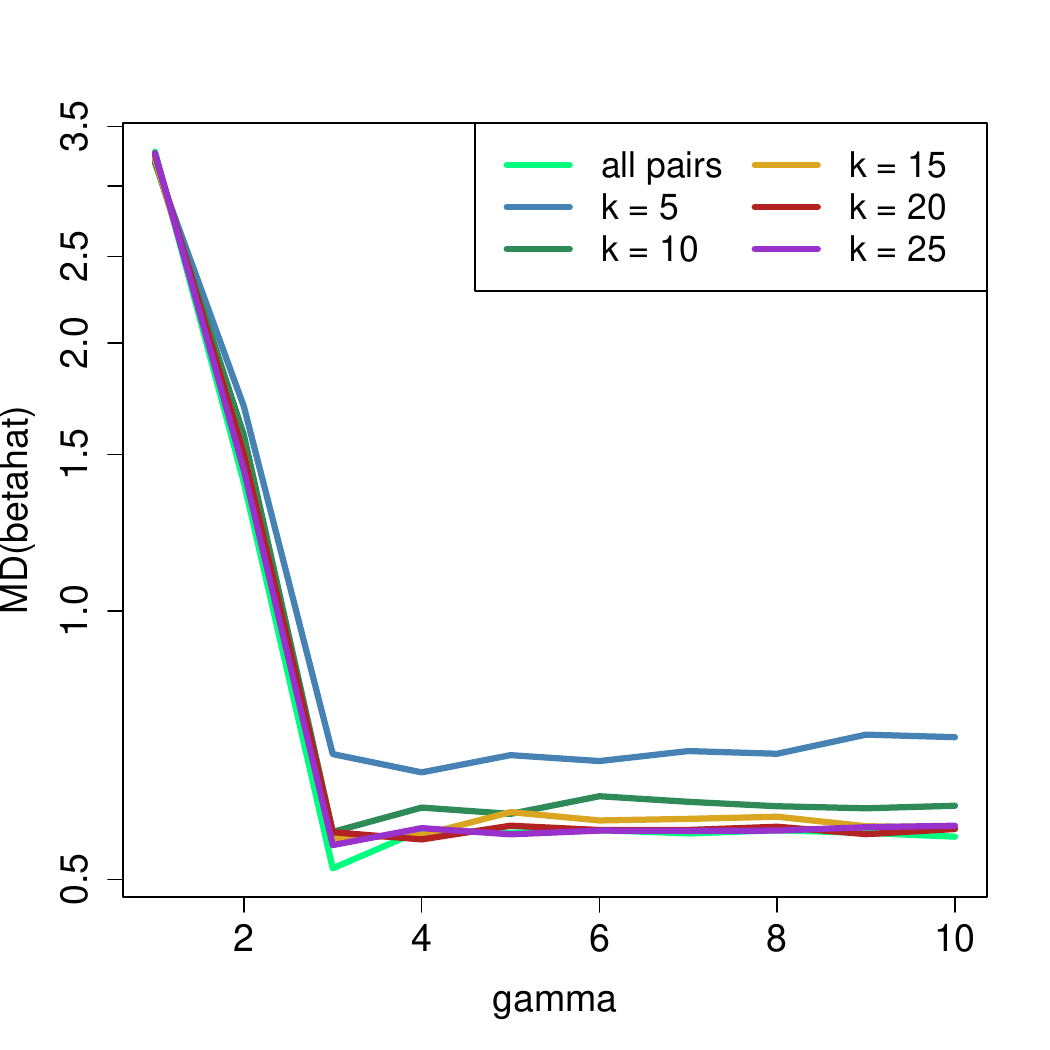}\\
\caption{Top row: average MD (on log scale) of the 
estimated coefficients for different symmetrization 
strategies and normal predictors. 
The data has $n = 400$, $\p = 20$, 
$\eps = 10\%$ of cellwise outliers, and 
$\bSigma = \bSigma_{\ALYZ}$ (left) or 
$\bSigma = \bSigma_{\AN}$ (right).
Middle row: same for exponential predictors.
Bottom row: same for lognormal predictors.}
\end{figure}

\section{Additional analyses of the US Cancer data}
\label{supp:uscancer}

\subsection{Cellmap grouped per state}

\begin{figure}[H]
\centering
\includegraphics[width=0.29\linewidth]
{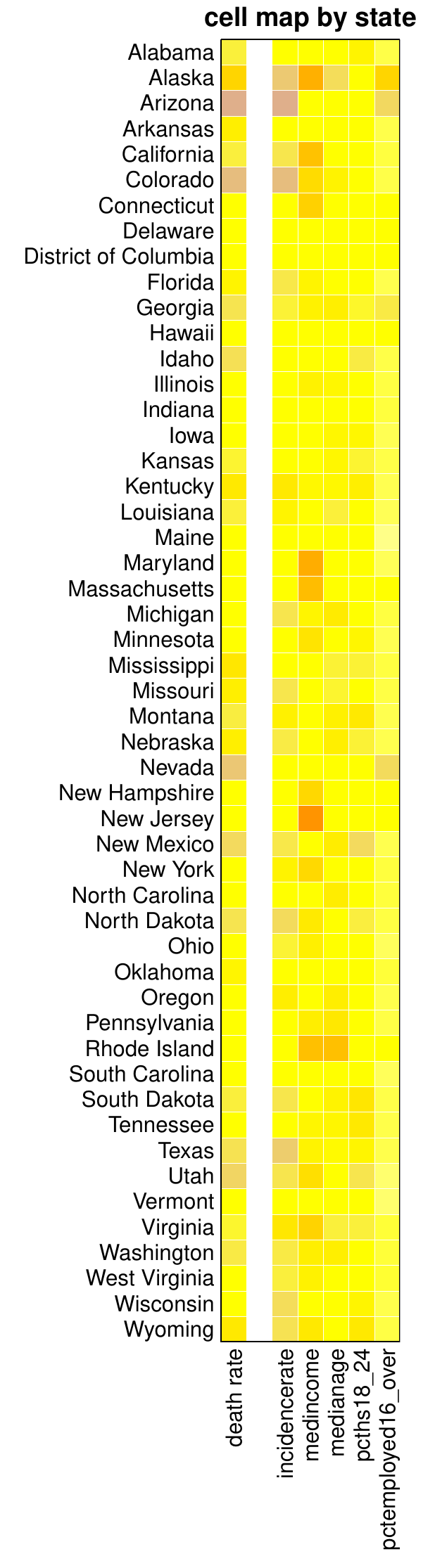}
\caption{Cellmap of the US Cancer data per state.}
\label{fig:USCancersmall_cellmapstates}
\end{figure}

The analysis in Section~\ref{sec:realdata}
modeled cancer mortality by five regressors.
Figure~\ref{fig:USCancersmall_cellmapstates}
shows the cellmap where the counties are
grouped by state. 
We notice unusually low cancer death rates 
in Arizona and Colorado, in tandem with low
incidence rates.
The median income is higher than expected
from the other regressors in Alaska,
California, Maryland, Massachusetts,
New Jersey, and Rhode Island.

\subsection{Analysis using all variables
of the US Cancer data}

To further illustrate the workings of the 
cellLTS method, we now investigate the 
complete set of variables in the US Cancer 
dataset. It contains 3047 counties and 33 
variables including the response, which is
the death rate due to cancer, labeled as 
\texttt{target\_deathrate}. The remaining 
features are listed in 
Table~\ref{tab:USCancerlarge_features}. 
These include general demographic figures 
as well as information on employment, 
education, insurance and race.

The descriptions of the regressors might 
suggest that they contain all the 
information to compute the target exactly. 
More specifically, one would expect that 
the target is exactly equal to the ratio 
of \texttt{avgdeathsperyear} and 
\texttt{popest2015}. This turns out not 
to be true at all (the adjusted $R^2$ is 
only around 0.32). The reason for this 
apparent mismatch is the fact that the 
data comes from different sources and was 
collected over similar but different time 
periods. For instance, 
\texttt{avgdeathsperyear} was computed 
over the seven-year period 2010-2016, 
together with \texttt{avganncount}, 
\texttt{incidencerate} and the target 
feature. The other features are census
estimates of 2013 and 2015.

To predict the target death rate, we keep 
all features except the two 
categorical variables \texttt{binnedinc} 
(which is just a binned version of
\texttt{medianincome}) and 
\texttt{geography} (the name of the county). 
We also remove \texttt{pctsomecol18\_24} as 
it has about 75\% missing values, and 
\texttt{studypercap} as it has over 63\% of 
zeroes. Finally, we discard the 
\texttt{avgdeathsperyear} feature, as it 
seems unnatural to predict the death rate 
using the information in this predictor. 
This leaves us with 27 explanatory 
features. Most of the features are 
completely observed, except for 
\texttt{pctemployed16\_over} that has 5\% 
of missing values, and 
\texttt{pctprivatecoveragealone} that
has 20\%.

\begin{figure}[!ht]
\centering
\vspace{-2mm}
\includegraphics[height = 6.5cm]
  {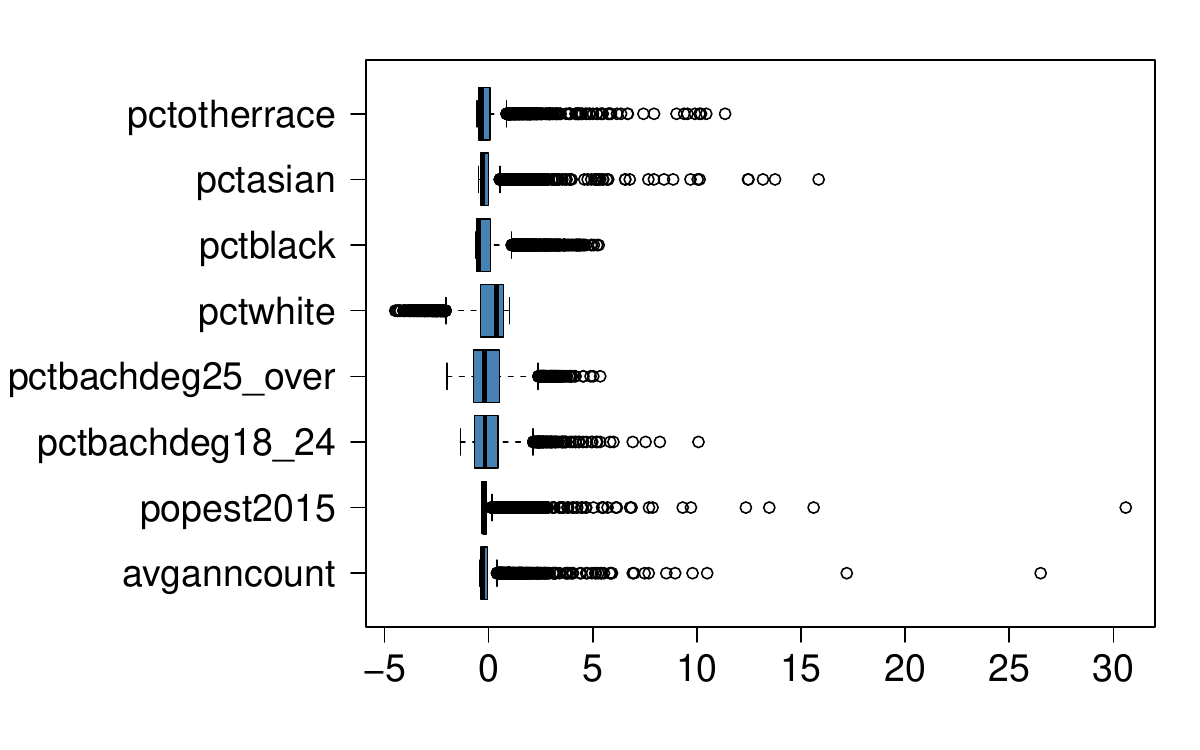}
\includegraphics[height = 6.5cm]
  {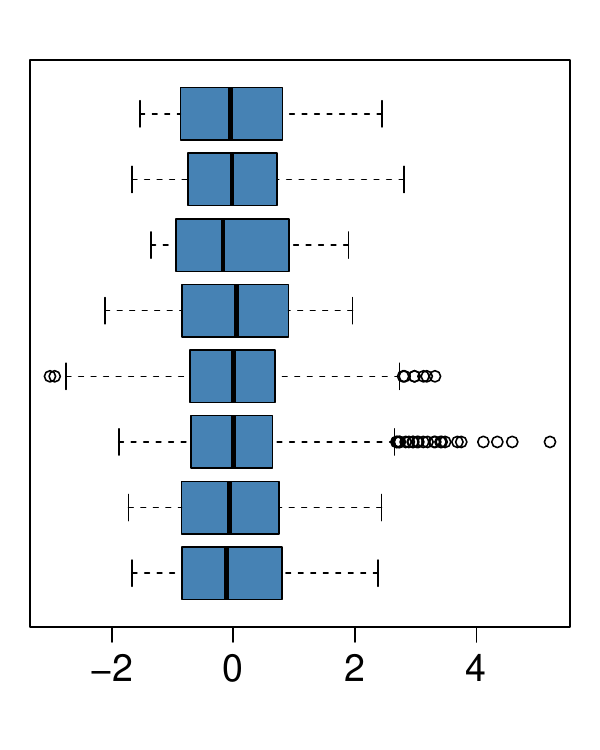}\\
\caption{The most skewed variables before 
(left) and after (right) a robust power 
transformation towards symmetry. All 
variables were scaled before plotting.}
\label{fig:usCancerlarge_transformation}
\end{figure}

Some of the explanatory features are very 
skewed. While cellLTS uses symmetrization 
to combat skewness, it can help to address 
the most extreme variables in a 
preprocessing step. Using the algorithm of 
\cite{transfo} to robustly fit a power 
transformation, we transformed the most 
skewed variables towards symmetry as seen in 
Figure~\ref{fig:usCancerlarge_transformation}.

\begin{table}[!ht]
\footnotesize
\centering
\begin{tabular}{p{4.3cm}|p{8cm}}
feature & explanation\\
\hline
\texttt{avganncount} & Average number of 
  cancer cases diagnosed annually.\\
\texttt{avgdeathsperyear} & Average number 
  of deaths due to cancer per year.\\
\texttt{incidencerate} & Incidence rate 
  of cancer.\\
\texttt{medincome} & Median income in 
  the region.\\
\texttt{popest2015} & Estimated 
  population in 2015.\\
\texttt{povertypercent} & Share of 
  population below the poverty line.\\
\texttt{studypercap} & Per capita 
  number of cancer‐related clinical trials.\\
\texttt{binnedinc} & Binned median income.\\
\texttt{medianage} & Median age in the region.\\
\texttt{medianagemale} & Median age of men.\\
\texttt{medianagefemale} & Median age of women.\\
\texttt{geograpy} & County identifier.\\
\texttt{percentmarried} & Share of people 
  who are married.\\
\texttt{pctnohs18\_24} & Share of residents 
  aged 18–24 whose highest education\\
  & is less than high school.\\
\texttt{pcths18\_24} & Share of residents 
  aged 18–24 whose highest education\\
  & is a high school diploma.\\
\texttt{pctbachdeg18\_24} & Share of 
  people aged 18–24 with a bachelor's degree.\\
\texttt{pctssomecol18\_24} & Share of people
  aged 18–24 who attended some college.\\
\texttt{pcths25\_over} & Share of people 
  aged 25+ who graduated high school.\\
\texttt{pctbachdeg25\_over} & Share of people 
  aged 25+ with a bachelor's degree.\\
\texttt{pctemployed16\_over} & Share of 
  people aged 16+ who are employed.\\
\texttt{pctunemployed16\_over} & Share of 
  people aged 16+ who are unemployed.\\
\texttt{pctprivatecoverage} & Share of 
  population with private health insurance.\\
\texttt{pctprivatecoveragealone} & Share 
  covered solely by private health insurance.\\
\texttt{pctempprivcoverage} & Share 
  covered by employer‐provided private\\ 
  & health insurance.\\
\texttt{pctpubliccoverage} & Share with 
  public health insurance.\\
\texttt{pctpubliccoveragealone} & Share 
  covered solely by public health insurance.\\
\texttt{pctwhite} & Share of the population 
  that is White.\\
\texttt{pctblack} & Share of the population 
  that is Black.\\
\texttt{pctasian} & Share of the population 
  that is Asian.\\
\texttt{pctotherrace} & Share of the 
  population identifying with other races.\\
\texttt{pctmarriedhouseholds} & Share of 
  households that are married-couple 
  households.\\
\texttt{birthrate} & Birth rate in the 
  region.\\
\end{tabular}
\caption{The regressors for 
  predicting cancer mortality. }
\label{tab:USCancerlarge_features}
\end{table}
\normalsize

\FloatBarrier

\newpage

\phantom{abc}\\
\phantom{abc}\\
\phantom{abc}\\
We run an OLS regression as well as the 
cellLTS method. For OLS, the missing values 
are imputed by the mean of the feature. 
A comparison between the coefficients of 
the two fits is given in 
Table~\ref{tab:USCancerlarge_coefficients}.
We see that many features get similar 
coefficients. There are some more substantial 
differences, such as for the 
\texttt{avganncount} and \texttt{popest2015} 
features, as well as for the features 
encoding insurance coverage. However, given 
the substantial correlation structure in the
data, it is not immediately clear whether 
these differences are due to robustness or 
to multicollinearity (or both).\\

\begin{figure}[!ht]
\centering
\includegraphics[width=0.82\linewidth]
{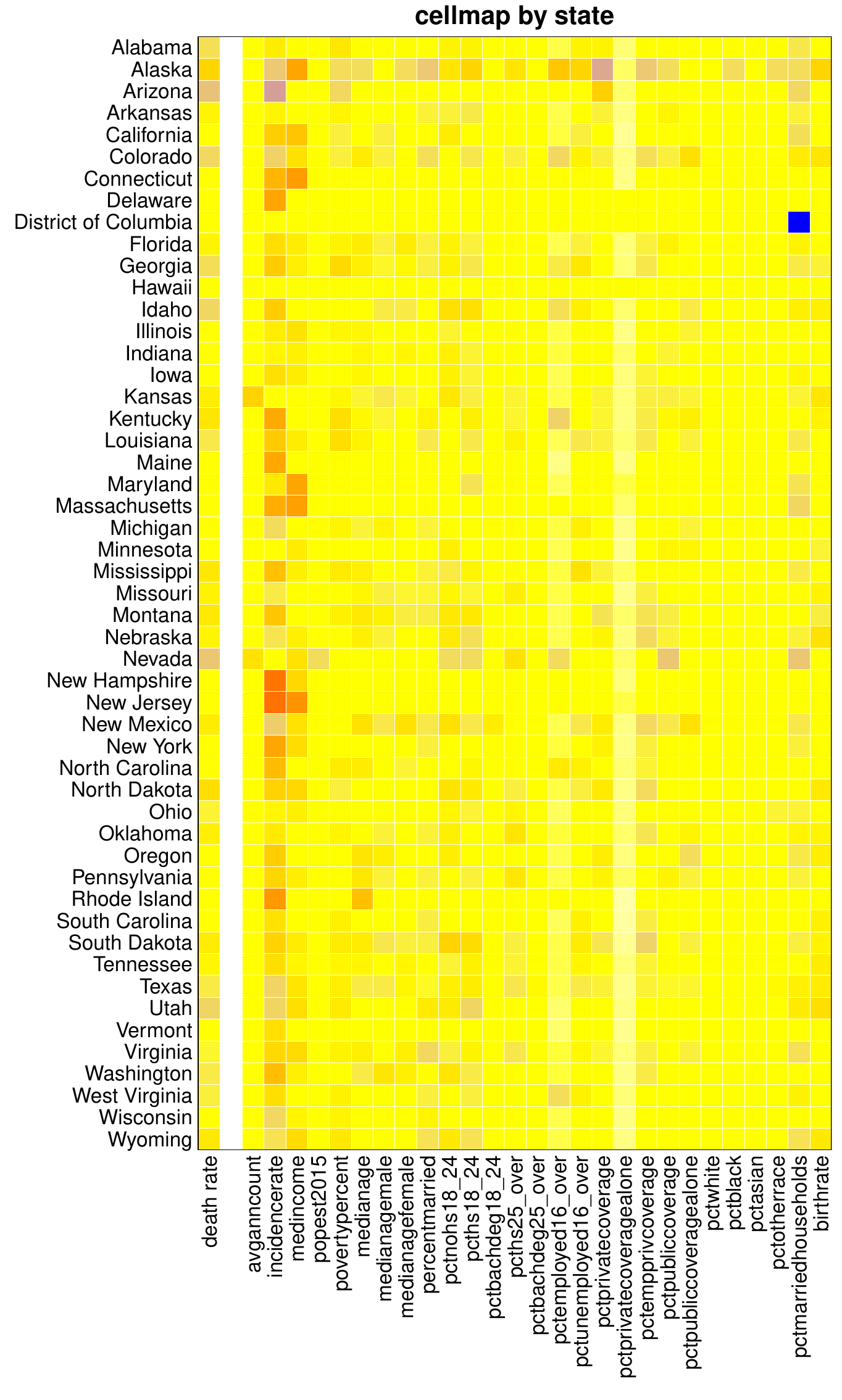}
\caption{Cellmap of the US Cancer data by state.}
\label{fig:USCancerlarge_cellmapstates}
\end{figure}

To gain more insight, we look at the cellmap 
of the data grouped per state in 
Figure~\ref{fig:USCancerlarge_cellmapstates}. 
This reveals a lot of unusual cells in the 
\texttt{incidencerate} and \texttt{medincome} 
features. Alaska appears to have the most
cellwise activity. The District of Columbia 
(Washington DC) has a bright blue cell for 
\texttt{pctmarriedhouseholds}. This district 
indeed has a very low percentage of married 
households (about 24\%, while the mean in 
the entire dataset is slightly over 50\%).

\begin{table}[ht]
\small
\centering
\begin{tabular}{rrr}
  \hline
 & OLS & cellLTS \\ 
  \hline
intercept & 139.59 & 157.82 \\ 
  avganncount & -1.88 & -12.59 \\ 
  incidencerate & 8.59 & 8.67 \\ 
  medincome & 0.26 & 0.73 \\ 
  popest2015 & 1.23 & 13.88 \\ 
  povertypercent & 0.75 & 0.52 \\ 
  medianage & -0.01 & 2.45 \\ 
  medianagemale & -2.10 & -3.22 \\ 
  medianagefemale & -1.35 & -0.97 \\ 
  percentmarried & 8.49 & 10.63 \\ 
  pctnohs18\_24 & -0.96 & -1.40 \\ 
  pcths18\_24 & 1.91 & 1.25 \\ 
  pctbachdeg18\_24 & -1.00 & -0.35 \\ 
  pcths25\_over & 3.31 & 3.32 \\ 
  pctbachdeg25\_over & -5.20 & -4.91 \\ 
  pctemployed16\_over & -3.82 & -6.52 \\ 
  pctunemployed16\_over & 0.80 & 0.63 \\ 
  pctprivatecoverage & -6.42 & -2.42 \\ 
  pctprivatecoveragealone & 0.65 & -9.04 \\ 
  pctempprivcoverage & 3.42 & 8.22 \\ 
  pctpubliccoverage & 0.34 & -4.98 \\ 
  pctpubliccoveragealone & 0.25 & 3.79 \\ 
  pctwhite & 0.02 & 1.63 \\ 
  pctblack & 3.66 & 5.43 \\ 
  pctasian & 0.17 & -0.08 \\ 
  pctotherrace & -3.42 & -3.56 \\ 
  pctmarriedhouseholds & -7.46 & -9.70 \\ 
  birthrate & -1.10 & -0.81 \\ 
   \hline
\end{tabular}
\caption{Coefficients of OLS and cellLTS on 
the complete US Cancer data. The coefficients
(except the intercept) were multiplied by the 
median absolute deviation of the corresponding 
feature, to put them on a more equal
footage.}
\label{tab:USCancerlarge_coefficients}
\end{table}
\normalsize
\FloatBarrier

\begin{figure}[!ht]
\vspace{3mm}
\centering
\includegraphics[width=1\linewidth]
{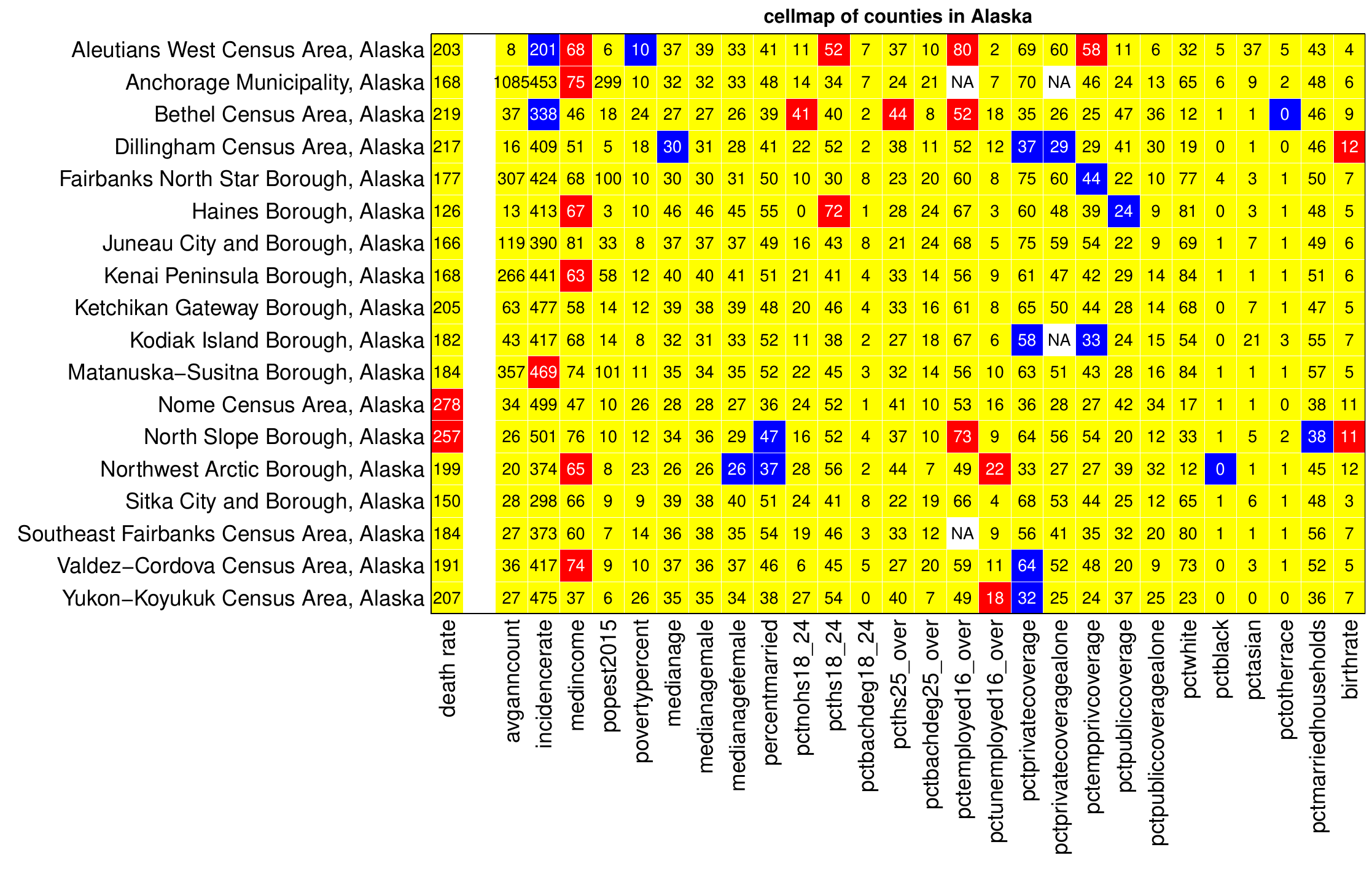}
\caption{Cellmap of Alaska.}
\label{fig:USCancerlarge_cellmapAlaska}
\end{figure}

\newpage
Figure~\ref{fig:USCancerlarge_cellmapAlaska} 
zooms in on Alaska, like we did in 
Section~\ref{sec:realdata} with fewer
regressors. It is interesting to compare 
the two cellmaps. In this one we have a 
lot more information in the regressors. 
This not only means that we might predict 
the response more accurately, but also 
that we have more information to flag 
suspicious cells.

The cellmaps in 
Figure~\ref{fig:USCancer_cellmap_Alaska} and
Figure~\ref{fig:USCancerlarge_cellmapAlaska}
have some patterns in common. The
\texttt{medincome} feature has red cells in
both, reflecting that Alaska is a 
high-income state. Interestingly, the median 
income of Juneau City, the highest in Alaska, 
now is no longer flagged as outlying. The 
likely explanation is that we now have 
additional information indicating a high 
general education, employment rate, and low 
poverty in the city, all of which are in line 
with this high median income.
We also note that the death rate of Bethel 
Census Area is no longer flagged as outlying. 
Instead, its incidence rate is now considered 
lower than expected. From the additional 
regressors we see 
that it is among the Alaskan counties with 
the lowest education levels, employment, and 
private health insurance. It also has among 
the lowest median incomes and highest poverty 
levels in Alaska. With this additional 
information its high death rate becomes 
predictable, which it wasn't previously.\\

\begin{figure}[!ht]
\centering
\includegraphics[width=1.0\linewidth]
{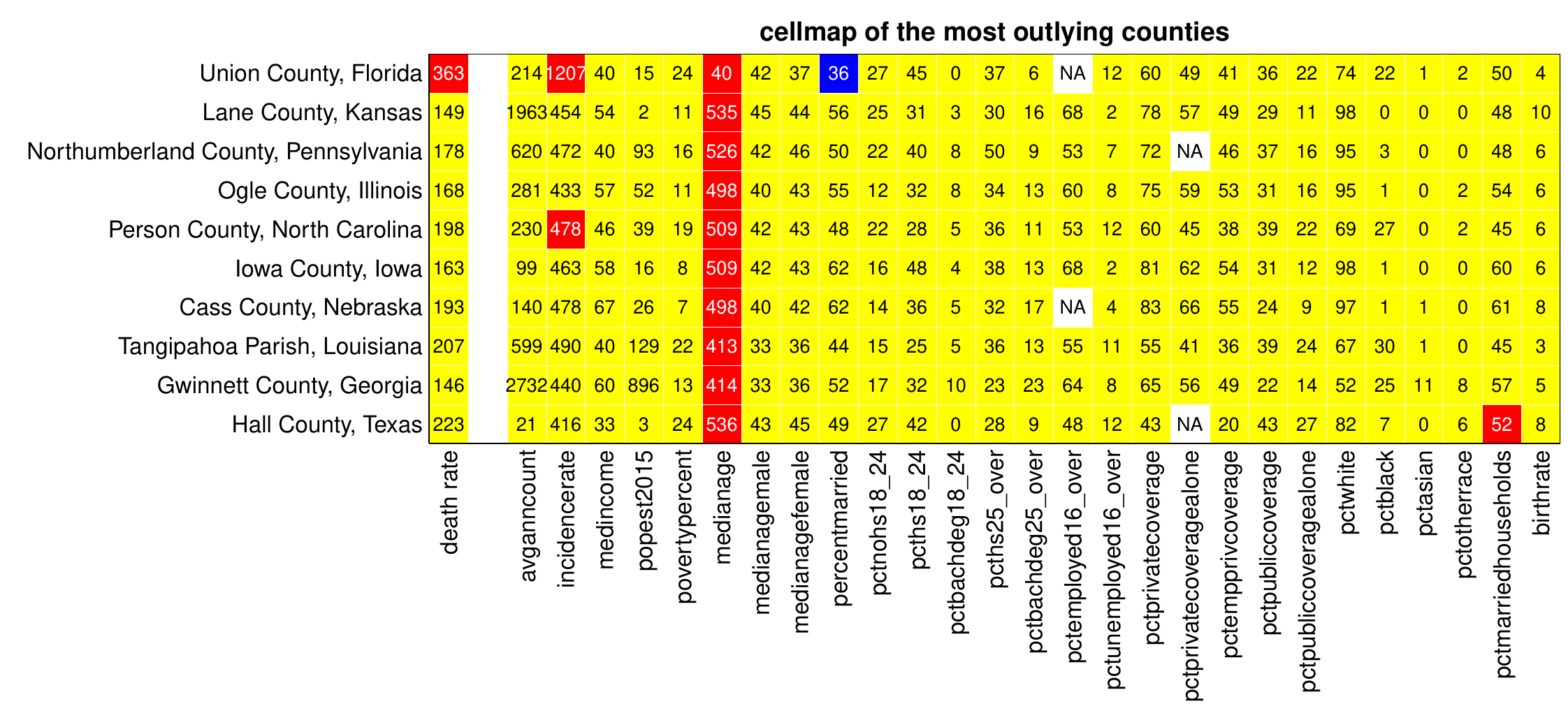}
\caption{Cellmap of the 10 counties with the highest 
sum of absolute standardized cellwise residuals.}
\label{fig:USCancerlarge_cellmaplargestresiduals}
\end{figure}

\newpage
We now inspect the 10 counties with the 
largest sum of absolute cell residuals in 
the current analysis. The cellmap of these 
counties is shown in 
Figure~\ref{fig:USCancerlarge_cellmaplargestresiduals},
whereas it was in 
Figure~\ref{fig:USCancersmall_cellmaplargestresiduals}
for the smaller set of regressors.
Union County still stands out here, with 
exceptionally high
death rate and incidence rate. We now also
see a low percentage of married households.
Overall, the highest number of flagged cells
are in the \texttt{medianage} variable, due
to its high values.

\begin{figure}[!ht]
\centering
\includegraphics[width=\linewidth]
{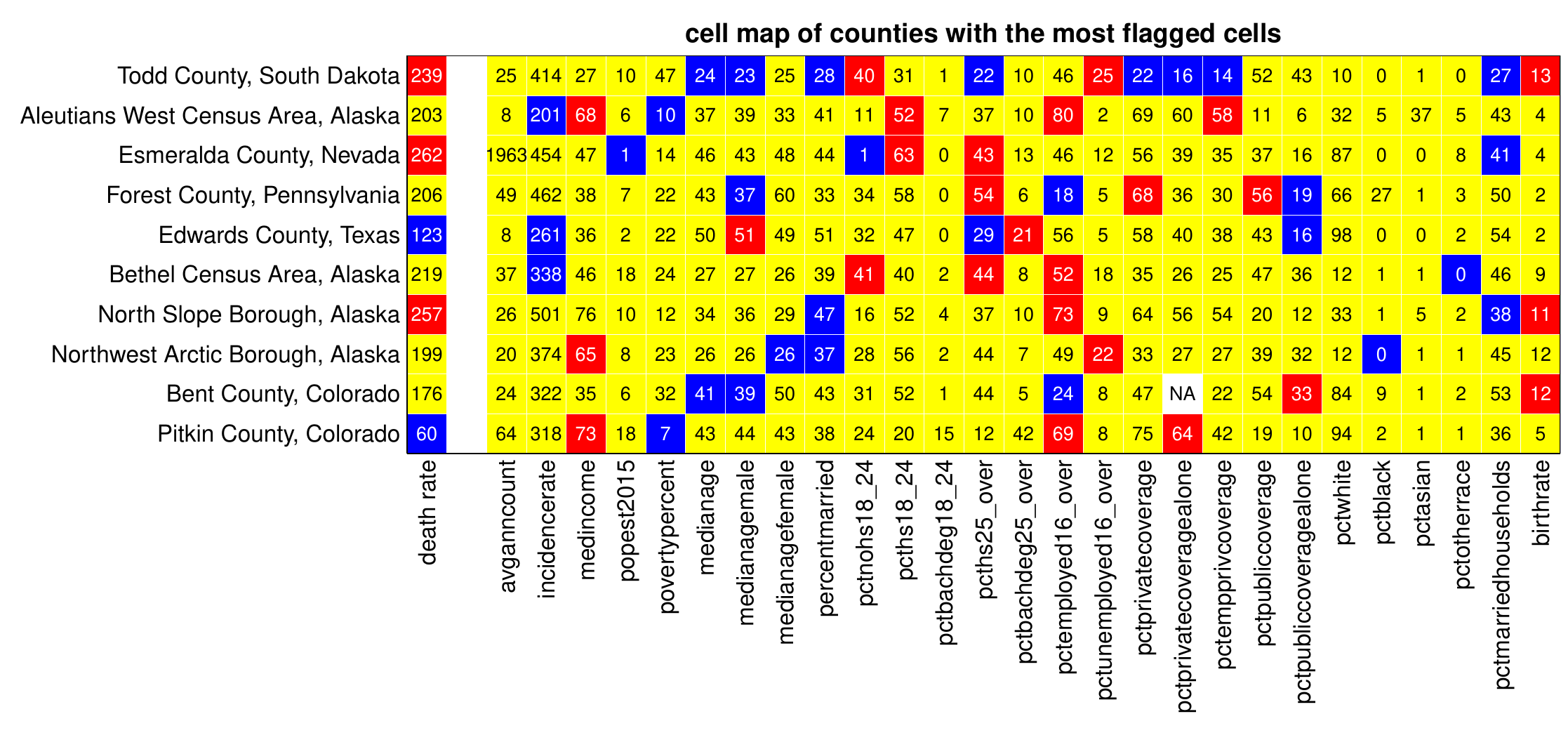}
\caption{Cellmap of the 10 counties with the 
highest number of flagged cells.}
\label{fig:USCancerlarge_cellmapmostflagged}
\end{figure}

Finally, we inspect the cellmap of the counties 
that had the highest number of unusual cells, 
shown in 
Figure~\ref{fig:USCancerlarge_cellmapmostflagged}. 
The county with the most flagged cells is Todd 
County, South Dakota. We see that it has a very 
young population, high unemployment, low 
education, and low insurance. It turns out that
over 85\% of its population are Native Americans. 
This is unlike the vast majority of counties in 
the US, which may explain its deviating behavior.

Esmeralda County, Nevada has fewer than 1000 
inhabitants, making all census data very noisy. 
Forest County in Pennsylvania has a low
median age for males (much lower than that
for females) and a low percentage of employed 
people of age 16 and over. The 
reason is that about a third of its 
population are inmates at the State 
Correctional Institution. This results in 
an over-representation of young men. 
Edwards County in Texas is another tiny 
rural county, where the data may not be very 
reliable. We see that the education levels are 
somewhat higher than expected, and its incidence 
rates and death rates are low. The incidence 
rate may be an error, as the more recent 
incidence rate for the county is 545.6 according 
to the National Cancer Institute, which is much 
more in line with the state's average. 
Bent County in Colorado is similar to Forest 
County in that it also has a large correctional 
facility with about 1500 inmates. Given its 
total population of about 6000, this is a large
fraction. The Alaskan counties in this cellmap 
have been discussed before, and show deviating 
behavior due to remoteness, high fraction of 
native population, and overall high incomes 
with low unemployment. Finally, Pitkin county 
is also an interesting case. It has the lowest 
death rate due to cancer in the dataset. Even in 
recent years, it ranks among the lowest in the 
whole of the US. There are several explanations, 
including the facts that Pitkin is generally 
wealthy with a high median income, low poverty, 
and high private insurance coverage. Moreover, 
Pitkin has a lot of seasonal homes for 
skiing, and such counties tend to have low 
cancer death rates. The reason may be that 
seriously ill people often prefer to receive 
treatment at or near their main home.

\end{document}